\documentclass[12pt,a4paper]{article}
\usepackage{amsmath,amssymb,bm,ascmac,bbm,mathtools}
\usepackage[dvipdfmx, usenames]{color}
\usepackage[dvipdfmx]{graphicx}
\usepackage{xcolor}
\usepackage{here}
\usepackage{authblk}

\newcommand{\tdif}[2]{\frac{d#1}{d#2}}

\newcommand{\tr}{\text{tr}}

\newcommand{\mcal}{\mathcal}
\newcommand{\mbb}{\mathbb}
\newcommand{\mfrak}{\mathfrak}

\newcommand{\ra}{\rangle}

\newcommand{\hmu}{{\hat\mu}}
\newcommand{\hl}{{\hat\lambda}}
\newcommand{\ho}{\hat\omega}

\newcommand{\hg}{\hat{\mathfrak g}}
\newcommand{\wf}[1]{\widehat{\mathfrak{#1}}}
\begin{document}
\title{RG flows from WZW models}
\author{Ken KIKUCHI} 
% \author{second author\footnote{email}}
\affil{Yau Mathematical Sciences Center,
Tsinghua University}
\date{}
\maketitle

\begin{abstract}
We constrain renormalization group flows from $ABCDE$ type Wess-Zumino-Witten models triggered by adjoint primaries. We propose positive Lagrangian coupling leads to massless flow and negative to massive. In the conformal phase, we prove an interface with the half-integral condition obeys the double braiding relations. Distinguishing simple and non-simple flows, we conjecture the former satisfies the half-integral condition. If the conjecture is true, some previously allowed massless flows are ruled out. For $A$ type, known mixed anomalies fix the ambiguity in identifications of Verlinde lines; an object is identified with its charge conjugate. In the massive phase, we compute ground state degeneracies.
\end{abstract}

\tableofcontents

\makeatletter
\renewcommand{\theequation}
{\arabic{section}.\arabic{equation}}
\@addtoreset{equation}{section}
\makeatother

\section{Introduction and summary}
Solving quantum field theory (QFT) typically means to explore infrared (IR) behavior of a given ultraviolet (UV) theory. One of the most famous examples is the quantum chromodynamics (QCD). The theory is defined in UV as $SU(3)$ gauge theory with fundamental matters. Since the theory is not scale-invariant, physics change as we shift energy scales. In particular, as we zoom out and explore long distance physics, it is described by hadrons (mesons and baryons). These (e.g. protons and neutrons) constitute physics at energy scales we live. Thus, we are typically interested in IR behavior (such as chiral symmetry breaking or confinement) of QCD. Concisely, we are solving renormalization group (RG) flows.

Traditionally, RG invariants have been employed to solve RG flows. Fundamental invariants are surviving symmetries and their 't Hooft anomalies \cite{tH79}. If a UV symmetry with an 't Hooft anomaly is preserved, it should be matched by possible IR theory. This logic has succeeded to reveal IR behaviors of pure gauge theory \cite{GKKS17}. Since the invariant was almost all constraints we had (in strongly coupled theories), the 't Hooft anomaly matching associated to symmetries has been the central tool to solve RG flows.

However, we now have more constraints on RG flows thanks to generalized symmetries \cite{GKSW14}. In the modern language, symmetries are generated by topological operators supported on codimension $(q+1)$ defects. They are called $q$-form symmetries. Such symmetries act on extended $q$-dimensional observables. Since we no longer impose an existence of inverse, symmetries can even be non-invertible. Non-invertible (finite) symmetries are in general described by fusion categories \cite{MS1,MS2,BT17}. Just like group symmetries have group elements and multiplication, categorical symmetries have objects and fusion.\footnote{The number of objects in a fusion category is called rank. More precisely, it counts the number of isomorphism classes of simple objects.} The fusion should satisfy consistency condition called the pentagon axiom. Solutions to the pentagon axiom are called $F$-symbols. It is mathematically proven that the set of solutions are discrete \cite{ENO}. (This property is called the Ocneanu rigidity.) The discreteness of solutions gives mathematical support of anomaly matching because $F$-symbols capture anomalies. Just like group symmetries are preserved under relevant deformation with singlets, fusion subcategories are preserved if objects commute with relevant operators \cite{G12}. Furthermore, just like an RG flow gives a homomorphism from UV symmetry group to IR symmetry group, an RG flow gives a monoidal inclusion functor from surviving symmetry category to IR symmetry category \cite{KK-ARG}. Corresponding to the pullback of the homomorphism gives anomaly matching, the functor gives matching of $F$-symbols, generalizing anomaly matching to categorical symmetries.

In rational conformal field theories (RCFTs), there is another structure, braiding. (A fusion category with braiding is called braided fusion category, or BFC for short.) Braidings obey another consistency condition called the hexagon axiom. The solutions to the condition are usually denoted by $R$ (or $c$) and called $R$-symbols. It is also mathematically proven that the solution set of $R$-symbols are discrete.

Thus, one may expect that in RG flows from RCFTs, $R$-symbols give additional invariants. The expectation is not completely correct; the $R$-symbols can jump at conformal fixed points to their opposites. Therefore, even though it is not invariant, the additional braiding structure imposes further constraints on RG flows from RCFTs. In particular, we have succeeded to explain symmetry enhancement in IR RCFTs  \cite{KK21,KKSUSY,KK22II,KK22free}.

There are two types of additional constraints on RG flows; the double braiding relation \cite{KK22II} and various ``monotonicities'' \cite{KKSUSY,KK22II,KK22free}. The double braiding relation claims that, for surviving objects $i,j$, their double braidings are the opposite in UV and IR (given some conditions explained below):
\begin{equation}
    c^\text{IR}_{j,i}c^\text{IR}_{i,j}=\left(c^\text{UV}_{j,i}c^\text{UV}_{i,j}\right)^*.\label{cIRUV}
\end{equation}
Since it is this condition that could tighten previous constraints, we elaborate on this point.

In \cite{KK22II}, we gave an intuitive explanation of the relation (\ref{cIRUV}) based on the RG interface \cite{G12}. However, the picture was too crude, and there is a `counterexample' \cite{NK22} to one naive consequence, reality condition. The example demands more accurate reason why braidings flip. We found a reason. Given a condition, we can prove the relation. The condition also reconciles the `counterexample.'

Our demonstration applies to more general setup\footnote{We thank Yuji Tachikawa for discussions on this point.} not necessarily restricted to RG interfaces. Namely, consider two theories $T_1,T_2$ separated by an interface. We assume two theories are described by modular tensor categories (MTCs), $\mcal M_1$ and $\mcal M_2$, respectively. (An MTC is a BFC with non-singular $S$-matrix. In our context, MTCs are collections of lines generating symmetries in diagonal RCFTs.) From the two MTCs, collect pairs of objects $(j_1,j_2)\in(\mcal M_1,\mcal M_2)$ which can end topologically on the interface. Such collection of objects together with fusion and braiding in, say, $\mcal M_1$ defines a braided fusion subcategory $\mcal B\subset\mcal M_1$.\footnote{The BFC $\mcal B$ does not have to be maximal. For example, one can consider two copies of identical critical Ising models separated by the trivial interface. If one takes $\mcal B$ as the rank three full MTC, then half-integral condition below is not satisfied. Thus, double braidings are identical on both sides of the trivial interface as expected. However, one can also take $\mcal B$ as the $\mbb Z_2$ braided fusion subcategory. In this case, the integrality condition is satisfied. Our proof below claims the double braidings should be opposite in $\mcal M_1$ and $\mcal M_2$. This is true because the $\mbb Z_2$ fusion subcategory has real double braidings.} By identifying\footnote{Mathematically, this means to consider a monoidal functor $m_{12}:\{j_1\}\to\{j_2\}$ together with another monoidal functor $m_{21}:\{j_2\}\to\{j_1\}$ such that for all pairs $(j_1,j_2)$, $m_{21}m_{12}(j_1)\cong j_1,m_{12}m_{21}(j_2)\cong j_2$.} $j_1$ and $j_2$, we can also view $\mcal B$ as a subcategory of $\mcal M_2$. Now, pick $j_1\in\mcal B$ along with its counterpart $j_2\cong m_{12}(j_1)$. We write their conformal dimensions $h^{\mcal M_1}_{j_1}$ and $h^{\mcal M_2}_{j_2}$, respectively. If the sum of their conformal dimensions $h^{\mcal M_1}_{j_1}+h^{\mcal M_2}_{j_2}$ are integral for all $j_1\in\mcal B$, we can prove the double braidings are the opposite. The reason is because such pairs have the opposite topological twists $\theta_j:=e^{2\pi ih_j}$:
\begin{equation}
    \theta_{j_1}=e^{2\pi ih^{\mcal M_1}_{j_1}}=e^{-2\pi ih^{\mcal M_2}_{j_2}}=\theta_{j_2}^{-1}.\label{theta12}
\end{equation}
Since double braidings are given by topological twists (with fusion coefficient $N_{ij}^k\in\mbb N$)
\begin{equation}
    c_{j,i}c_{i,j}=\sum_kN_{ij}^k\frac{\theta_k}{\theta_i\theta_j}id_k,\label{doublebraid}
\end{equation}
this shows the double braiding relation
\begin{equation}
    c^{\mcal M_1}_{j_1,i_1}c^{\mcal M_1}_{i_1,j_1}=\left(c^{\mcal M_2}_{j_2,i_2}c^{\mcal M_2}_{i_2,j_2}\right)^*\label{DBRint}
\end{equation}
if for all $(j_1,j_2)$, $h^{\mcal M_1}_{j_1}+h^{\mcal M_2}_{j_2}\in\mbb Z$. The integrality condition means the product theory $T_1\times T_2$ has extended chiral algebra generated by (bosonic) currents.

It is not hard to relax the integrality condition slightly; the proof can be repeated for more general case $h^{\mcal M_1}_{j_1}+h^{\mcal M_2}_{j_2}\in\frac12\mbb Z$. In this case, the product theory also has extended chiral algebra generated by fermionic currents. Thus, we can introduce $\mbb Z_2$ graiding in $\mcal B$. We denote the collection of ``even'' objects with integral sum as $\mcal B_e$, and the collection of ``odd'' objects with half-integral sum as $\mcal B_o$. For the $\mbb Z_2$ graiding to be well-defined, we must have
\begin{equation}
    j_e\otimes j_e\in\mcal B_e,\quad j_o\otimes j_o\in\mcal B_e,\quad j_e\otimes j_o,j_o\otimes j_e\in\mcal B_o,\label{Z2graid}
\end{equation}
where $j_e\in\mcal B_e,j_o\in\mcal B_o$. The first case was treated above. In the second case, ``odd'' objects produce the sign `mismatch' of topological twists:
\begin{equation}
    \theta_{j_1}=e^{2\pi ih^{\mcal M_1}_{j_1}}=e^{2\pi i(\frac12-h^{\mcal M_2}_{j_2})}=-\theta_{j_2}^{-1}.\label{-theta12}
\end{equation}
However, since the result of fusion consists only of ``even'' objects, the sign mismatch always cancel:
\begin{align*}
    c^{\mcal M_1}_{j_o,i_o}c^{\mcal M_1}_{i_o,j_o}&=\sum_{k_e}N_{i_oj_o}^{k_e}\frac{\theta_{k_e}}{\theta_{i_o}\theta_{j_o}}id_{k_e}\\
    &=\sum_{k_e}N_{i_oj_o}^{k_e}\frac{\theta_{k_e}^{-1}}{(-\theta_{i_o}^{-1})(-\theta_{j_o}^{-1})}id_{k_e}=\left(c^{\mcal M_2}_{j_o,i_o}c^{\mcal M_2}_{i_o,j_o}\right)^*.
\end{align*}
(Here, we identified objects in $\mcal M_1$ and $\mcal M_2$ to simplify the notation.) In this second case, the sign cancellation happened in the denominator, but in the third case, the cancellation happens in the ratio
\[ \frac{-\theta_{k_o}^{-1}}{\theta_{i_e}^{-1}(-\theta_{j_o}^{-1})}=\frac{\theta_{k_o}^{-1}}{\theta_{i_e}^{-1}\theta_{j_o}^{-1}}. \]
Therefore, we see the double braiding relations hold even in case of half-integral sum
\begin{equation}
    h^{\mcal M_1}_{j_1}+h^{\mcal M_2}_{j_2}\in\frac12\mbb Z\label{halfint}
\end{equation}
with well-defined $\mbb Z_2$ graiding. We have shown the theorem:\newline

\textbf{Theorem.}
\textit{Consider an interface separating two MTCs $\mcal M_1,\mcal M_2$. If collection $\mcal B$ of pairs $(j_1,j_2)\in(\mcal M_1,\mcal M_2)$ ending topologically on the interface satisfies the half-integral condition $h^{\mcal M_1}_{j_1}+h^{\mcal M_2}_{j_2}\in\frac12\mbb Z$ with well-defined $\mbb Z_2$ graiding, then they obey the double braiding relation
\begin{equation}
    c^{\mcal M_1}_{j_1,i_1}c^{\mcal M_1}_{i_1,j_1}=\left(c^{\mcal M_2}_{j_2,i_2}c^{\mcal M_2}_{i_2,j_2}\right)^*.\label{dbr}
\end{equation}}

This proof can reconcile the `counterexample' thanks to the half-integral condition. To explain this point, let us pick two interfaces separating three theories $T_1,T_2,T_3$. We assume the interfaces support extended current algebras. Namely, we suppose the two interfaces satisfy the half-integral condition. Now, we collide the two interfaces to get one interface. Does the resulting interface satisfy half-integral condition? In general, the answer is no. An example is the Tambara-Yamagami category surviving a sequence of RG flows $M(4,11)\to M(5,4)\to M(4,3)$.\footnote{This `counterexample' was pointed out by Yu Nakayama in \cite{NK22}. Note that the minimal model $M(4,11)$ is non-unitary, but we will see more examples of this sort in unitary theories below.} Denoting their conformal dimensions with subscripts, we have flows of objects
\[ \begin{array}{cccc}
M(4,11):&1_0&\eta_{\frac92}&N_{\frac{25}{16}}\\
&\downarrow&\downarrow&\downarrow\\
M(5,4):&1_0&\eta_{\frac32}&N_{\frac7{16}}\\
&\downarrow&\downarrow&\downarrow\\
M(4,3):&1_0&\eta_{\frac12}&N_{\frac1{16}}
\end{array}. \]
Each RG flow (or RG interface) satisfies the half-integral condition. However, if we view a sequence of RG flows as a single RG flow $M(4,11)\to M(4,3)$, the Kramers-Wannier duality line $N$ does not satisfy the half-integral condition $\frac{25+1}{16}\notin\frac12\mbb Z$. Thus, the double braidings in $M(4,11)$ and $M(4,3)$ do not have to be the opposite.\footnote{On the other hand, the $\mbb Z_2$ subcategory $\{1,\eta\}$ does satisfy the (half-)integral condition. Thus, their double braidings have to be the opposite. This is trivially obeyed because their double braidings are real.} This resolves the `counterexample.'

The observation requires us to distinguish two RG flows: simple and non-simple. We call an RG flow simple if it cannot be separated into a sequence of more than one RG flows. A sequence of RG flows is non-simple when it is viewed as a single RG flow. Examples of RG flows believed to be simple are those between consecutive unitary minimal models $M(m+1,m)\to M(m,m-1)$.

In these examples, Gaiotto found \cite{G12} half-integral conditions are satisfied. More precisely, they are necessary conditions for his algebraic construction to work because RG interface a la Gaiotto is a representation theory of the extended current algebras. The algebraic construction is believed to work in broader examples, and we expect the half-integral condition is satisfied more generally. Indeed, one can check the half-integral conditions are satisfied in RG flows between $D$- and $E$-type minimal models, $(D_4,A_6)\to(A_4,D_4),(E_6,A_{12})\to(A_{10},E_6)$, fermionic minimal models (at least for Verlinde objects) \cite{KK22II}, and also non-unitary minimal models \cite{KK22free}. Furthermore, we will see the half-integral conditions are satisfied in most (but not all\footnote{If these exceptional massless RG flows exists, they disprove our conjecture. We will comment on this point in section \ref{massless}.}) of the strongest conjectures on RG flows between $\wf{su}(r+1)_k$ WZW models. Given these infinitely many examples in different classes of RCFTs with or without unitarity, it is tempting to conjecture the half-integral conditions are satisfied in all simple RG flows:\newline

\textbf{Conjecture.}
\textit{In simple RG flows between RCFTs, surviving Verlinde objects $j$'s satisfy the half-integral condition $h^\text{UV}_j+h^\text{IR}_j\in\frac12\mbb Z$.}\label{conj}\newline

If the conjecture is true, then simple RG flows obey the double braiding relation from our theorem above. Also note that if the conjecture holds, then its contraposition gives us a convenient criterion to judge whether a given RG flow is simple or not. If the half-integral condition is unsatisfied just as in the example $M(4,11)\to M(4,3)$, then the RG flow is not simple; the RG flow can be separated into a sequence of RG flows.

Here, a reader may wonder how useful the double braiding relation is. Our demonstration claims the relation does not hold for generic non-simple RG flows. However, when we are solving RG flows, we typically assume the flows are simple. We usually do not care a putative IR CFT further flows to another CFT or a gapped phase described by a topological quantum field theory (TQFT). Therefore, in solving RG flows, we can employ the double braiding relation. This is the first type of additional constraints on RG flows satisfying the half-integral condition.

The second type of additional constraints, ``monotonicity,'' include three new conditions (beyond the $c$-theorem \cite{Z86}):
\begin{itemize}
    \item spin constraint $S_j^\text{IR}\subset S_j^\text{UV}$ \cite{KKSUSY},
    \item scaling dimension $h_j^\text{IR}\le h_j^\text{UV}$ \cite{KK22II,KK22free},
    \item global dimension $|D^\text{IR}|<|D^\text{UV}|$ \cite{KK22free}.
\end{itemize}
Each ``monotonicity'' claims the following. (We will see not all constraints are independent in examples below.)

\paragraph{Spin constraint:} For each object $j$, there is a defect Hilbert space $\mcal H_j$. Operators in the space have specific spins. We denote the set of spins as $S_j$ and call it spin content. If relevant operators are spacetime scalars, the deformation preserves rotation symmetry, and its quantum numbers (i.e., spins) are conserved. Thus, for each surviving object $j$, the spin contents should be (basically) preserved. More precisely, heavy operators may be lifted along RG flows. Therefore, for a surviving object $j$, spin content associated to it in IR $S_j^\text{IR}$ should be a subset of that in UV $S_j^\text{UV}$, i.e., $S_j^\text{IR}\subset S_j^\text{UV}$.

\paragraph{Scaling dimension:} In the Wilsonian RG flow, one integrates out heavy modes. Hence, it is intuitively expected that effective field theories have lighter and lighter operators. This intuition has been proved in minimal models \cite{KK22II,KK22free}. In diagonal RCFT, there is a one-to-one correspondence between primaries and objects (especially called Verlinde objects) generating (zero-form) symmetries. Thus, for a surviving Verlinde object $j$, there is one corresponding primary $\phi_j^\text{UV}$ in UV and another primary $\phi_j^\text{IR}$ in IR. Let us denote their conformal dimensions $h_j^\text{UV}$ and $h_j^\text{IR}$, respectively. They are ``monotonic,'' $h_j^\text{IR}\le h_j^\text{UV}$.

\paragraph{Global dimension:} Kitaev and Preskill \cite{KP05} found that an MTC gives universal contribution to entropy (see also \cite{LW05}). More precisely, an MTC with global dimension\footnote{A global dimension is defined as follows. An action of an object $j$ on the identity operator gives a real number $d_j$. The number is called quantum dimension. A global dimension of a category $\mcal C$ is defined by
\[ D^2:=\sum_{j\in\mcal C}d_j^2. \]
Alternatively, it is given by the vacuum element of the $S$-matrix:
\[ D=\frac1{S_{11}}. \]
For unitary theories, $D$ is positive, but for non-unitary theories, it can be negative.} $D$ contributes $S\ni-\ln D$ to entropy $S$.\footnote{The contribution enters free energy as $F\ni T\ln D$ at temperature $T$. Thus, we can compare various MTCs from the viewpoint of free energy. This was employed to explain emergent symmetry \cite{KK22free}.} Then, the second law of thermodynamics motivates monotonic decrease of global dimensions. The expectation $|D^\text{IR}|<|D^\text{UV}|$ was proved in RG flows among unitary and non-unitary minimal models \cite{KK22free} (in examples with index $|I|=1$ introduced in \cite{DDT00} for the latter).
\\

These additional constraints were found in minimal models, an important class of RCFTs. However, these examples are special in the following sense; nontrivial generators of (invertible) symmetries are basically identical. More precisely, they are related under charge conjugation. In unitary discrete series of bosonic minimal models, it is proven \cite{RV98} that they only have $\mbb Z_2$ or $S_3$ (invertible) symmetries. For $\mbb Z_2$, there is only one nontrivial generator. For $\mbb Z_3\subset S_3$, two nontrivial generators are swapped under charge conjugation. Accordingly, their conformal dimensions $h_j$'s are the same, and there is only one class of topological twists $\theta_j=e^{2\pi ih_j}$ (up to complex conjugation). This observation raises the following question. If there are multiple classes of topological twists, which two are related under RG flows? We would like to address this question in this paper. However, our answer should be clear now. If we assume RG interface a la Gaiotto works, then the half-integral condition is satisfied. As a result, a double braiding in UV jumps to its complex conjugate in IR. This means different classes of double braidings are `disconnected.'

A more physical explanation is the following. A topological twist $\theta_j=e^{2\pi ih_j}$ can be viewed as an amplitude for a process where an anyon $j$ is twisted once. A twist in the opposite direction gives $\theta_j^{-1}$. These phases reflect statistics of the anyons. (The topological twist is $1$ if spin $h$ is an integer and $-1$ if spin $h$ is a half-integer, but in general topological twists are complex phases.) Since we consider deformations with spacetime scalar commuting with the worldline of the anyons, it is physically expected that the statistics is not affected. More precise statement is as follows. The topological twists are coming from (half) braiding $c_{j,j}$, but it is not `gauge invariant.' To get a `gauge invariant' quantity, we take a double braiding $c_{j^*,j}c_{j,j^*}$. The fusion $j\otimes j^*$ has the identity-channel as evident from the symmetry of fusion $j\otimes1=j$. The identity-channel gives a phase $\theta_j^{-1}\theta_{j^*}^{-1}=\theta_j^{-2}$. Taking twists in the opposite direction into account, we can interpret the collection $\Theta=\{\theta_j^{\pm2}\}$ as a reflection of statistics. In particular, for surviving objects, the set should match in order not to change the statistics.\footnote{One can view this new statistics matching constraint as an analogue of the spin constraint.} Indeed, in all examples studied in \cite{KK21,KK22free}, the set is invariant as a result of the half-integral condition found there. On the other hand, the statistics matching supports our conjecture.\footnote{The half-integral condition implies $\theta_j=e^{2\pi ih_j}$ goes to $\pm\theta_j^{-1}$. Thus, the set $\{\theta_j^{\pm2}\}$ provides an invariant. On the other hand, if the set of statistics $\Theta=\{\theta_j^{\pm2}\}$ is invariant, $\theta_j$ can go to $\pm\theta_j^{\pm1}$. The possibility $+\theta_j^{+1}$ is realized by the trivial interface. If we further assume nontrivial interface changes topological twists, what is left is to rule out $-\theta_j^{+1}$. Then, the invariance of statistics implies the half-integral condition.} We will see this new constraint can rule out some previously allowed massless flows.

In order to see this phenomenon, we study another important class of RCFTs, Wess-Zumino-Witten (WZW) models. Some models have $\mbb Z_N$ symmetries with $N\ge4$ (e.g. $\wf{su}(4)_k$ model, $\wf{so}(10)_k$ model, or $\wf{su}(5)_k$ model). Consequently, they have several classes of topological twists which cannot be related by complex conjugation. The `disconnected' nature\footnote{One might support this `disconnectedness' based on the Ocneau rigidity.} of different classes of double braidings in particular indicates that symmetric RG flows between them do not exist. For example, previous conjectures based on anomaly matching allowed $\wf{su}(5)_3\to\wf{su}(5)_1$ and $\wf{su}(5)_2\to\wf{su}(5)_1$, but we claim they are unlikely because UV and IR theories have different sets $\Theta$'s of statistics.

There are also other motivations to pick WZW models. From our personal viewpoint, RG flows from WZW models are less understood than those from minimal models. (For example, there are only a few numerical studies \cite{BBLKS13,KPTT15} based on truncated conformal space approach \cite{YZ89}.) One of our goals is to fill the gap by applying additional constraints and solve RG flows from WZW models. Another motivation is the connection to experiments. Systems with $SU(N)$ spin symmetry can be experimentally realized with ultra-cold atoms \cite{WHZ03}. If they are gapless at long distances, natural candidates are $\wf{su}(N)_k$ WZW models.
These experiments would also serve as checks of additional constraints on RG flows beyond minimal models.

We study all $ABCDE$ type WZW models with nontrivial centers. Our proposal for both massless $\wf g_k\to\wf g_{k'}$ and massive scenarios together with various conditions are summarized in Table \ref{result}. When theories have center symmetries $\Gamma$ larger than $\mbb Z_2$, the half-integral condition leads to stronger conditions which cannot be seen just from anomalies in $\Gamma$. Our conjecture motivates that previously allowed massless flows violating the half-integral condition are non-simple (or even massive).

We also study another scenario of massive flows. The fundamental observable in these cases is the ground state degeneracy (GSD). In order to compute GSDs, we apply the Cardy's method \cite{C17}. The method also signals which sign of the relevant coupling leads to massless and massive phases.

\begin{table}[H]
\begin{center}
%\hspace{-30pt}
\begin{tabular}{c|c|c|c|c}
UV theory&Center $\Gamma$&Anomaly-free $\Gamma$&$h^{\wf g_{k'}}_{j'}+h^{\wf g_k}_j\in\frac12\mbb Z$&Scenario\\\hline
$\wf{su}(r+1)_k$&$\mbb Z_{r+1}$&$rk\in2\mbb Z$&$k'+k\in(r+1)\mbb Z$&$\begin{cases}(\ref{su34to2to1}),(\ref{su32})&(r=2)\\(\ref{su45to3to1}),(\ref{su43}),(\ref{su42})&(r=3)\\(\ref{su53to2}),(\ref{su56to4to1}),(\ref{su54})&(r=4)\end{cases}$\\
$\wf{so}(2r+1)_k$&$\mbb Z_2$&$k\in\mbb Z$&$k'\in\mbb Z$&(\ref{so72to1}),(\ref{so72})\\
$\wf{sp}(2r)_k$&$\mbb Z_2$&$rk\in2\mbb Z$&$r(k'+k)\in2\mbb Z$&(\ref{sp65to3to1}),(\ref{sp63})\\
$\wf{so}(4R)_k$&$\mbb Z_2\times\mbb Z_2$&$Rk\in2\mbb Z$&$R(k'+k)\in2\mbb Z$&(\ref{so125to3to1}),(\ref{so123})\\
$\wf{so}(4R+2)_k$&$\mbb Z_4$&$k\in2\mbb Z$&$k'+k\in4\mbb Z$&(\ref{so105to3to1}),(\ref{so103})\\
$(\wf{e_6})_k$&$\mbb Z_3$&$k\in\mbb Z$&$k'+k\in3\mbb Z$&(\ref{e62to1}),(\ref{e62})\\
$(\wf{e_7})_k$&$\mbb Z_2$&$k\in2\mbb Z$&$k'+k\in2\mbb Z$&(\ref{e73to1}),(\ref{e73})
\end{tabular}.
\end{center}
\caption{Summary of our proposal}\label{result}
\end{table}

\section{Massless RG flows}\label{massless}
In this section, we study RG flows from $ABCDE$ type WZW models. (For a brief review of WZW models, see the Appendix \ref{review}.) They have zero-form center symmetries. We will mainly study UV theories with anomalies. In order to preserve the symmetry and anomalies, we choose relevant operators to be singlets under the center symmetries. Concretely, we only consider relevant deformations with primaries in the adjoint $\phi_\text{adj}$ just like in \cite{L15}. Thus, IR theory cannot be trivial (in anomalous cases), and there are two possible scenarios: TQFT with $\text{GSD}>1$, or CFT. In this section, we consider the massless scenario. (Massive scenarios are studied in the next section.) We begin with well-studied $A_r$ type WZW models. Based on the experience, we then proceed to less studied $BCDE$ type WZW models.

\subsection{$A_r$ type, i.e., $\wf{su}(r+1)_k$ model}
An $\wf{su}(r+1)_k$ model has central charge
\begin{equation}
    c_k=\frac{k(r^2+2r)}{k+(r+1)}.\label{csur+1k}
\end{equation}
Its primaries are labeled by the affine dominant weights (\ref{Pk+}):
\begin{equation}
    P^k_+=\left\{\hmu\Big|0\le\mu_j\ \&\ 0\le\sum_{j=1}^r\mu_j\le k\right\}.\label{Pk+su}
\end{equation}
Their conformal dimensions can be computed from the formula (\ref{hhmu}). Some of them relevant for our purposes are given by
\begin{equation}
    h_{\widehat{[k;0,\dots,0]}}=0,\quad h_{\widehat{[0;k,0,\dots,0]}}=\frac{kr}{2(r+1)}=h_{\widehat{[0;0,\dots,0,k]}},\quad h_{\widehat{[0;0,k,0\dots,0]}}=\frac{k(r-1)}{r+1}=h_{\widehat{[0;0,\dots,0,k,0]}},\quad\dots\ .\label{sur+1h}
\end{equation}

The theory has various symmetries. One symmetry which plays the key role in our analysis is the $\mbb Z_{r+1}$ zero-form center symmetry. It is generated by Verlinde lines $\eta^l$'s associated to affine weights $A^l[k;0,0,\dots,0]$ with $l=0,1,\dots,r$. (The action of the primitive outer automorphism $A$ can be found in Table \ref{outerauto}.) Their conformal dimensions are given by (\ref{sur+1h}). Another symmetry we will use below is the vector-like $PSU(r+1)=SU(r+1)/\mbb Z_{r+1}$ symmetry. These are subgroups of the symmetry we mentioned in the Appendix \ref{review}:
\[ PSU(r+1)\times\mbb Z_{r+1}\subset\frac{SU(r+1)_L\times SU(r+1)_R}{\mbb Z_{r+1}}. \]
The symmetries act on elementary field $U_\text{UV}$, which takes values in $SU(r+1)$, as follows. The $\mbb Z_{r+1}$ center symmetry acts as $U_\text{UV}\mapsto e^{\frac{2\pi i}{r+1}}U_\text{UV}$, while the vector-like $PSU(r+1)$ symmetry acts as $U_\text{UV}\mapsto g_VU_\text{UV}g_V^\dagger$ with $g_V\in SU(r+1)_V$. These symmetries are preserved by primary operator $\phi_\text{adj}$ in the adjoint representation.

In order to solve RG flows, traditionally, we have employed anomalies associated to symmetries. An 't Hooft anomaly in Verlinde objects can be easily read off from its relation with spin \cite{NY17}. A $\mbb Z_N$ symmetry is anomaly-free iff its generating object has spin
\begin{equation}
    \mbb Z_N\text{ is anomaly-free}\iff s\in\frac1N\mbb Z.\label{anomalyfreecond}
\end{equation}
Thus, from the conformal dimensions (\ref{sur+1h}), we immediately find the $\mbb Z_{r+1}$ is anomaly-free iff
\begin{equation}
    kr\in2\mbb Z.\label{Zr+1anomfree}
\end{equation}
For $A_r$ type WZW models, more anomalies are known \cite{TS18}. They studied mixed anomalies between the center symmetry $\mbb Z_{r+1}$ and the vector-like $PSU(r+1)$ symmetry. The theory has no mixed anomaly iff
\begin{equation}
    k\in(r+1)\mbb Z.\label{mixedanomsu}
\end{equation}
When either (\ref{Zr+1anomfree}) or (\ref{mixedanomsu}) is unsatisfied, the theory has an anomaly, and an RG flow triggered by relevant operators preserving the symmetries cannot be trivial in IR. Two possibilities are TQFT with $\text{GSD}>1$, or CFT. We study massless scenarios in this section.

Previous conjectures are on specific massless flows $\wf{su}(r+1)_k\to\wf{su}(r+1)_{k'}$. In this case, first of all, the $c$-theorem imposes
\begin{equation}
    k'<k.\label{cthmsu}
\end{equation}
Furthermore, the mixed anomaly matching requires \cite{TS18} (see also \cite{YHO18}; there, this $\mbb Z_{r+1}$ number was called the Lieb-Schultz-Mattis index, or LSM index for short)
\begin{equation}
    k=k'q\quad\mod\ r+1,\label{sumixed}
\end{equation}
where $\mbb Z_{r+1}$ is assumed to act in IR as $U_\text{IR}\mapsto e^{\frac{2\pi iq}{r+1}}U_\text{IR}$ with $\gcd(q,r+1)=1$. (This transformation is generated by the Verlinde line corresponding to $A^q\ho_0$.) To the best of our knowledge, these are the strongest constraints on RG flows between $\wf{su}(r+1)$ WZW models.

We will see this constraint fixes $\mbb Z_2$ ambiguities in identifications of Verlinde objects in UV and IR. In our proposal, an object $j$ in UV always flows to its charge conjugate $j^*$ in IR. One intuitive explanation of this fact is the following. Verlinde objects are supported on oriented lines. Therefore, in the RG interface picture, a line entering the RG interface from one side goes out from the other side. Equivalently, Verlinde lines penetrating the RG interface have different orientations relative to the interface on two sides of the interface. This would explain why an object in one side of the RG interface is mapped to its charge conjugate on the other side. Since this intuitive picture seems to hold as well in $BCDE$ type WZW models, we believe an object $j$ is identified with its charge conjugate $j^*$ in these models. This should be verified by studying mixed anomalies.

Under this identification, the half-integral condition is translated to
\[ \frac{k'r}{2(r+1)}+\frac{kr}{2(r+1)}\in\frac12\mbb Z, \]
or\footnote{Note that, for $r=1$, this is basically the Haldane conjecture \cite{H82,H83}; $\wf{su}(2)_k$ with odd $k$ is gappless in IR, and the IR theory also has odd $k'$. In this sense, the half-integral condition (\ref{halfintsu}) can be viewed as a slight generalization of the Haldane conjecture.}
\begin{equation}
    k'+k\in(r+1)\mbb Z.\label{halfintsu}
\end{equation}
This means that when the $(r+1)$-ality of the level flips, the half-integral condition is satisfied. Accordingly, surviving $\mbb Z_{r+1}$ Verlinde objects obey the double braiding relations. On the other hand, if a putative massless flow does not satisfy (\ref{halfintsu}), our conjecture signals the flow is non-simple (or even massive). We will study many examples below, and support our conjecture. We start with UV theories with smaller rank $r$ and level $k$.

\subsubsection{$\wf{su}(2)_3\to\wf{su}(2)_1$}
Let us start from the well-known case, $r=1$. In this case, $\mbb Z_2$ center symmetry is anomalous iff the level is odd. An easiest way to see this is to compute conformal dimensions. The trivial and $\mbb Z_2$ Verlinde lines (denoted $1$ and $\eta$) have
\[ h_{\widehat{[k;0]}}=0,\quad h_{\widehat{[0;k]}}=\frac k4. \]
For odd $k$, $h_\eta\notin\frac12\mbb Z$, and the $\mbb Z_2$ is anomalous. Thus, an RG flow triggered by relevant operators preserving the symmetry cannot be trivial in IR. Since the center symmetry acts as (\ref{center}), we find $\hmu=[\mu_0;\mu_1]$ with even $\mu_1$ preserves the anomalous $\mbb Z_2$ symmetry. In $\wf{su}(2)_3$ model, the only nontrivial $\mbb Z_2$ singlet is $\phi_{\widehat{[1;2]}}$. It has conformal dimension $h_{\widehat{[1;2]}}=2/5$, and triggers an RG flow. Let us study the IR phase of the flow.

In $\wf{su}(2)_k$ model, we have explicit formula of $S$-matrix
\begin{equation}
    S^{\wf{su}(2)_k}_{\hmu,\hl}=\sqrt{\frac k{k+2}}\sin\left(\frac{\pi(\mu_1+1)(\lambda_1+1)}{k+2}\right).\label{SSU2k}
\end{equation}
Using the $S$-matrix, we find only the $\mbb Z_2$ braided fusion subcategory $\{1,\eta\}$ of the rank four UV MTC is preserved. In order to match the 't Hooft anomaly in the preserved $\mbb Z_2$ center symmetry, the IR WZW model (recall we are assuming conformal phase in this section) should also have odd level \cite{FO15}. For $k=3$, $k'=1$ is the only possibility consistent with the $c$-theorem. Note that the $c$-theorem also guarantees the ``monotonicity'' of conformal dimensions:
\[ h^{\wf{su}(2)_1}_1=0=h^{\wf{su}(2)_3}_1,\quad h^{\wf{su}_1}_\eta=\frac14<\frac34=h^{\wf{su}(2)_3}_\eta. \]
The scenario $\wf{su}(2)_3\to\wf{su}(2)_1$ satisfies all constraints above. The mixed anomaly matching condition (\ref{sumixed}) is satisfied because $3=1\cdot1$ mod $2$. Spin constraint is also satisfied:
\[ S^{\wf{su}(2)_1}_\eta=\{\pm\frac14\}=S^{\wf{su}(2)_3}_\eta\quad(\text{mod }1). \]
We further find this well-established example satisfies the half-integral condition (\ref{halfint}).\footnote{In fact, for all odd $k,k'$ (with $k'<k$ from $c$-theorem), one can easily see the half-integral condition is satisfied. Since the identity line trivially satisfies the integral condition, it is enough to study $\eta$. Setting $k=2K+1,k'=2K'+1$ with $K,K'\in\mbb N$, we get the sum
\[ h^{\wf{su}(2)_1}_\eta+h^{\wf{su}(2)_3}_\eta=\frac{K'+K+1}2\in\frac12\mbb Z. \]} This supports our conjecture beyond minimal models. As expected from our theorem, we also see the double braiding relations are satisfied:
\[ c^{\wf{su}(2)_1}_{j,i}c^{\wf{su}(2)_1}_{i,j}=\begin{pmatrix}id_1&id_\eta\\id_\eta&-id_1\end{pmatrix},\quad c^{\wf{su}(2)_3}_{j,i}c^{\wf{su}(2)_3}_{i,j}\Big|_{\mbb Z_2}=\begin{pmatrix}id_1&id_\eta\\id_\eta&-id_1\end{pmatrix}. \]
Since the surviving rank two BFC is a consistent MTC, no emergent symmetry is needed.

Finally, in this $\wf{su}(2)_k$ case, we can also prove the ``monotonicity'' of global dimensions. The level $k$ model has global dimension
\[ D_k=\frac1{S^{\wf{su}(2)_k}_{\widehat{[k;0]},\widehat{[k;0]}}}=\sqrt{\frac{k+2}k}\frac1{\sin\frac\pi{k+2}}. \]
One can explicitly check $D_{k+2}/D_k>1$ for $k=1,3,\dots,49$. For larger $k$, we can reliably employ the Taylor expansion to find
\[ \frac{D_{k+2}}{D_k}=1+\frac2k+O(1/k^2). \]
This shows the ``monotonicity'' $D^\text{IR}<D^\text{UV}$ of global dimensions beyond minimal models.

\subsubsection{$\wf{su}(3)_2\to\wf{su}(3)_1$}
Next, let us move to less studied examples. We start from RG flows between $\wf{su}(3)_k$ WZW models. Since $r=2$, (\ref{Zr+1anomfree}) says the $\mbb Z_3$ center symmetry is anomaly-free.\footnote{Another way to see this fact is as follows \cite{LS21}. Generic spin content of $\mbb Z_N$ topological defect line is given by
\[ s\in\frac p{N^2}+\frac1N\mbb Z. \]
The number $p$ takes $\mbb Z_N$ values $p\in H^3(\mbb Z_N,U(1))\cong\mbb Z_N$. In particular, nonzero $p$ implies an anomaly. If the topological defect line commutes with parity just as Verlinde lines, the spin contents should be invariant under parity $s\mapsto-s$. This gives a condition
\[ 2p\in N\mbb Z. \]
We see $\mbb Z_N$ symmetries with odd $N$ generated by Verlinde lines cannot have nontrivial anomalies.\label{anomspin}} This can be seen from the conformal dimensions
\[ h_{\widehat{[k;0,0]}}=0,\quad h_{\widehat{[0;k,0]}}=\frac k3=h_{\widehat{[0;0,k]}}. \]
However, we can still have nontrivial mixed anomaly (\ref{sumixed}). When the level $k$ is not a multiple of three, we have the anomaly. As the minimal choice, let us pick $k=2$. We perturb the model with its adjoint primary $\phi_\text{adj}=\phi_{\widehat{[0;1,1]}}$. The primary preserves only the $\mbb Z_3$ braided fusion subcategory of the rank six UV MTC, and $PSU(3)$ symmetry. Thus, the IR theory should match the mixed anomaly in the symmetry. With assumption of conformal phase in this section (and a natural scenario $\wf{su}(3)_2\to\wf{su}(3)_{k'}$), the $c$-theorem allows only $k'=1$. The mixed anomaly matching (\ref{sumixed}) imposes $q=2$; the UV Verlinde line $\eta$ acting as $U_\text{UV}\mapsto e^{2\pi i/3}U_\text{UV}$ should flow to the IR Verlinde line $\eta^2$ acting as $U_\text{IR}\mapsto e^{4\pi i/3}U_\text{IR}$:
\begin{equation}
\begin{array}{cccc}
\wf{su}(3)_2:&1_0&\eta_{\frac23}&\eta^2_{\frac23}\\
&\downarrow&\downarrow&\downarrow\\
\wf{su}(3)_1:&1_0&\eta^2_{\frac13}&\eta_{\frac13}
\end{array}.\label{su32to1}
\end{equation}
A priori, we cannot distinguish $\eta$ and $\eta^2$ because they obey the same fusion rules, have the same conformal dimensions and spin contents, and are swapped under charge conjugation. However, we see the detailed constraint from mixed anomaly matching enabled us to fix the ambiguity.

The matching of Verlinde lines satisfies all constraints above. The spin contents are matched as
\[ S^{\wf{su}(3)_1}_{\eta,\eta^2}=\{0,\pm\frac13\}=S^{\wf{su}(3)_2}_{\eta,\eta^2}\quad(\text{mod }1). \]
Conformal dimensions are ``monotonic''
\[ h^{\wf{su}(3)_1}_1=0=h^{\wf{su}(3)_2}_1,\quad h^{\wf{su}(3)_1}_{\eta,\eta^2}=\frac13<\frac23=h^{\wf{su}(3)_2}_{\eta,\eta^2}. \]
From the explicit $S$-matrices (\ref{Ssu}), we also observe the global dimensions are ``monotonic''
\[ D^{\wf{su}(3)_1}=\sqrt3<\sqrt3\sqrt{1+\zeta^2}=D^{\wf{su}(3)_2}, \]
where $\zeta:=\frac{1+\sqrt5}2$ is the golden ratio.

Furthermore, the (half-)integral condition (\ref{halfint}) is also satisfied. Since this massless flow $\wf{su}(3)_2\to\wf{su}(3)_1$ is also well-established, this example serves as another support of our conjecture.

Since the half-integral condition is satisfied, our theorem claims the RG flow should obey the double braiding relation. Indeed, a straightforward computation gives
\[ c^{\wf{su}(3)_1}_{j,i}c^{\wf{su}(3)_1}_{i,j}=\begin{pmatrix}id_1&id_\eta&id_{\eta^2}\\id_\eta&e^{4\pi i/3}id_{\eta^2}&e^{2\pi i/3}id_1\\id_{\eta^2}&e^{2\pi i/3}id_1&e^{4\pi i/3}id_\eta\end{pmatrix},\quad c^{\wf{su}(3)_2}_{j,i}c^{\wf{su}(3)_2}_{i,j}\Big|_{\mbb Z_3}=\begin{pmatrix}id_1&id_\eta&id_{\eta^2}\\id_\eta&e^{2\pi i/3}id_{\eta^2}&e^{4\pi i/3}id_1\\id_{\eta^2}&e^{4\pi i/3}id_1&e^{2\pi i/3}id_\eta\end{pmatrix}. \]
We see the double braiding relations (\ref{dbr}) are beautifully satisfied. Note that as the surviving rank three BFC is a consistent MTC, no emergent symmetry is needed. Since all known constraints are satisfied, the scenario $\wf{su}(3)_2\to\wf{su}(3)_1$ is allowed. These additional consistency checks further support the proposal in \cite{L15}.

\subsubsection{$\wf{su}(3)_4\to\wf{su}(3)_2$}
Let us also study a UV theory with larger level. We pick $k=4$ for the next nontrivial example. The $\mbb Z_3$ center symmetry alone is again anomaly-free, but it has the mixed anomaly. The adjoint primary $\phi_\text{adj}=\phi_{\widehat{[2;1,1]}}$ preserves the symmetry $PSU(3)\times\mbb Z_3$, and we can use the anomaly. (The adjoint primary preserves only the $\mbb Z_3$ braided fusion subcategory of the rank 15 UV MTC.) The anomaly matching imposes
\[ 4=k'q\quad\text{mod }3. \]
Taking the $c$-theorem into account, we only have two possibilities, $(k',q)=(1,1),(2,2)$. The first possibility says $\eta_\text{UV}$ flows to $\eta_\text{IR}$, and the second claims $\eta_\text{UV}$ flows to $\eta_\text{IR}^2$. The $k'=1$ scenario does not satisfy the half-integral condition. If such a massless flow exists, we interpret it as non-simple. On the other hand, the second possibility does satisfy the (half-)integral condition $\frac43+\frac23\in\mbb Z$. Indeed, the scenario satisfies all constraints above. The spin constraint is satisfied because
\[ S^{\wf{su}(3)_2}_{\eta,\eta^2}=\{0,\pm\frac13\}=S^{\wf{su}(3)_4}_{\eta,\eta^2}\quad(\text{mod }1). \]
The ``monotonicity'' of conformal dimensions is also satisfied thanks to the $c$-theorem:
\[ h^{\wf{su}_2}_1=0=h^{\wf{su}_4}_1,\quad h^{\wf{su}(3)_4}_{\eta,\eta^2}=\frac23<\frac43=h^{\wf{su}_4}_{\eta,\eta^2}. \]
We also see the global dimensions are ``monotonic''
\[ 3.29\approx\sqrt3\sqrt{1+\zeta^2}=D^{\wf{su}(3)_2}<D^{\wf{su}(3)_4}\approx10.3. \]
Since the (half-)integral condition is satisfied, we also observe the double braiding relations are obeyed:
\[ c^{\wf{su}(3)_2}_{j,i}c^{\wf{su}(3)_2}_{i,j}\Big|_{\mbb Z_3}=\begin{pmatrix}id_1&id_\eta&id_{\eta^2}\\id_\eta&e^{2\pi i/3}id_{\eta^2}&e^{4\pi i/3}id_1\\id_{\eta^2}&e^{4\pi i/3}id_1&e^{2\pi i/3}id_\eta\end{pmatrix},\quad c^{\wf{su}(3)_4}_{j,i}c^{\wf{su}(3)_4}_{i,j}\Big|_{\mbb Z_3}=\begin{pmatrix}id_1&id_\eta&id_{\eta^2}\\id_\eta&e^{4\pi i/3}id_{\eta^2}&e^{2\pi i/3}id_1\\id_{\eta^2}&e^{2\pi i/3}id_1&e^{4\pi i/3}id_\eta\end{pmatrix}. \]
Our additional consistency checks suggest $\wf{su}(3)_4\to\wf{su}(3)_2$ is allowed and simple. Together with the previous example, we propose a sequence of simple RG flows:
\begin{equation}
\begin{array}{cccc}
\wf{su}(3)_4:&1_0&\eta^2_{\frac43}&\eta_{\frac43}\\
&\downarrow&\downarrow&\downarrow\\
\wf{su}(3)_2:&1_0&\eta_{\frac23}&\eta^2_{\frac23}\\
&\downarrow&\downarrow&\downarrow\\
\wf{su}(3)_1:&1_0&\eta^2_{\frac13}&\eta_{\frac13}
\end{array}.\label{su34to2to1}
\end{equation}
Note that the mixed anomaly is matched even if we view the sequence of RG flows as a single RG flow because $\eta$ in $k=4$ flows to $\eta$ in $k'=1$. This example supports the contraposition of our conjecture; since the half-integral condition is unsatisfied, it signals the flow is non-simple.

Before we move to the next example, let us briefly comment on emergent symmetries. The scenario says there are three emergent symmetry objects. In the spirit of \cite{KK21}, we have to explain why they appear. First, we notice that the surviving rank three BFC has non-singular $S$-matrix and it is actually modular. Furthermore, its fusion rules have $SU(3)_1$ realization \cite{GK94}. The rank three MTC has central charge $c=2$ mod $4$, which can be smaller than $c_\text{UV}=\frac{24}7\approx3.4$. If these are all consistencies, we cannot explain why the symmetries emerge. Then why the symmetry enhances? The only possible reason we are aware of is the half-integral condition. If we are searching for simple RG flows between RCFTs and if we assume the conjecture is true, then the IR RCFT is constrained to satisfy the half-integral condition. (This reasoning succeeded to fix IR symmetry category in \cite{KK22free}.) The putative IR theory $\wf{su}(3)_1$ does \textit{not} satisfy the condition, and may not be consistent. The next smallest (in the sense of global dimension) $\wf{su}(3)_{k'}$ model is the $\wf{su}(3)_2$. This would explain why three symmetry lines emerge. We view this observation encouraging for our conjecture.

\subsubsection{$\wf{su}(4)_3\to\wf{su}(4)_1$}
Next, let us study examples with anomalous center symmetries larger than $\mbb Z_2$. The minimal example is $\wf{su}(4)_k$. The conformal dimensions of $\mbb Z_4$ Verlinde lines are given by
\[ h_{[k;0,0,0]}=0,\quad h_{\widehat{[0;k,0,0]}}=\frac{3k}8=h_{\widehat{[0;0,0,k]}},\quad h_{\widehat{[0;0,k,0]}}=\frac k2. \]
The $\mbb Z_2$ subgroup generated by $\eta^2$ is anomaly-free for any integer $k$, while the full $\mbb Z_4$ center symmetry is anomalous for odd $k$. As a minimal anomalous example, we pick $k=3$, and consider its relevant deformation.

We take the adjoint primary $\phi_\text{adj}=\phi_{\widehat{[1;1,0,0,1]}}$ as the relevant operator. This operator preserves $PSU(4)\times\mbb Z_4$. (The primary preserves only the $\mbb Z_4$ braided fusion subcategory of the rank 20 UV MTC.) Thus, we can employ two anomaly matching conditions --- one from mixed anomaly and one from $\mbb Z_4$ alone --- to constrain the RG flow triggered by the relevant operator.

The second anomaly requires the IR level $k'$ to be odd.\footnote{This anomaly is matched by any odd $k'$. To see this, recall the generic spin content in the footnote \ref{anomspin}. For odd $k$, the anomaly $p$ is given by $[p]=[6k]=[2]\in\mbb Z_4$. Therefore, the $\mbb Z_4$ anomaly is matched as long as the level $k'$ is odd.} Together with the $c$-theorem, the only possibility is $k'=1$. The mixed anomaly then imposes
\[ 3=q\quad\text{mod }4. \]
The only possibility is $q=3$. Again, we see the mixed anomaly matching fixes the ambiguity in identifications of Verlinde lines:
\begin{equation}
\begin{array}{ccccc}
\wf{su}(4)_3:&1_0&\eta_{\frac98}&\eta^2_{\frac32}&\eta^3_{\frac98}\\
&\downarrow&\downarrow&\downarrow&\downarrow\\
\wf{su}(4)_1:&1_0&\eta^3_{\frac38}&\eta^2_{\frac12}&\eta_{\frac38}
\end{array}.\label{su43to1}
\end{equation}

This scenario passes all constraints above. The spin constraint is satisfied because
\begin{align*}
    S^{\wf{su}(4)_1}_{\eta,\eta^3}=&\{\pm\frac18,\pm\frac38\}=S^{\wf{su}(4)_3}_{\eta,\eta^3}\quad(\text{mod }1),\\
    S^{\wf{su}(4)_1}_{\eta^2}=&\{0,\pm\frac12\}=S^{\wf{su}(4)_3}_{\eta^2}\quad(\text{mod }1).
\end{align*}
The ``monotonicity'' of conformal dimensions should be clear from (\ref{su43to1}). The global dimensions are also ``monotonic''
\[ D^{\wf{su}(4)_1}=2<\frac{\sqrt7}{\sin\frac\pi{14}}=D^{\wf{su}(4)_3}. \]
Since the half-integral condition is satisfied, we see the double braiding relations are also obeyed:
\[ \hspace{-30pt}c^{\wf{su}(4)_1}_{j,i}c^{\wf{su}(4)_1}_{i,j}=\begin{pmatrix}id_1&id_\eta&id_{\eta^2}&id_{\eta^3}\\id_\eta&-i\cdot id_{\eta^2}&-id_{\eta^3}&i\cdot id_1\\id_{\eta^2}&-id_{\eta^3}&id_1&-id_\eta\\id_{\eta^3}&i\cdot id_1&-id_\eta&-i\cdot id_{\eta^2}\end{pmatrix},\quad c^{\wf{su}(4)_3}_{j,i}c^{\wf{su}(4)_3}_{i,j}\Big|_{\mbb Z_4}=\begin{pmatrix}id_1&id_\eta&id_{\eta^2}&id_{\eta^3}\\id_\eta&i\cdot id_{\eta^2}&-id_{\eta^3}&-i\cdot id_1\\id_{\eta^2}&-id_{\eta^3}&id_1&-id_\eta\\id_{\eta^3}&-i\cdot id_1&-id_\eta&i\cdot id_{\eta^2}\end{pmatrix}. \]
These checks suggest the massless flow $\wf{su}(4)_3\to\wf{su}(4)_1$ is allowed and simple. Since the rank four MTC is a consistent category, no emergent symmetry is needed.

\subsubsection{$\wf{su}(4)_2\to\text{TQFT}$}
In order to support our conjecture, let us also study $k=2$. Since the level is even, the $\mbb Z_4$ center symmetry alone is anomaly-free. However, since the level is not a multiple of four, we still have the mixed anomaly. We perturb the UV theory with the primary operator in the adjoint $\phi_\text{adj}=\phi_{\widehat{[0;1,0,0, 1]}}$. Since the operator preserves the product symmetry $PSU(4)\times\mbb Z_4$, the IR theory cannot be trivial. With the assumption of conformal phase, let us see what we can draw.

A natural scenario is $\wf{su}(4)_2\to\wf{su}(4)_{k'}$. From the $c$-theorem, the only possibility is $k'=1$. Then, the mixed anomaly matching imposes
\[ 2=q\quad\text{mod }4. \]
The only possibility is $q=2$. This means $\eta$ in $k=2$ flows to $\eta^2$ in $k'=1$. Notice that powers of $\eta^2$ cannot generate the whole $\mbb Z_4$ symmetry in the $\wf{su}(4)_1$ model. In other words, this scenario requires emergent symmetries at the cost of free energy. Furthermore, although the putative flow seems the `shortest' possible one, we also notice that the identifications of Verlinde lines do not satisfy the half-integral condition.\footnote{The primitive line $\eta$ in $k=2$ has $h=3/4$ while $\eta^2$ in $k'=1$ has $h=1/2$. Their sum is not half-integral $\frac34+\frac12\notin\frac12\mbb Z$.} Relatedly, the statistics sets do not match. In UV, the surviving objects have $\Theta=\{\pm1\}$, while in IR the set is given by $\Theta=\{1\}$ because the surviving objects only span $\{1,\eta^2\}$. From these observations, we find the massless scenario unlikely. Therefore, we propose the RG flow is massive leading to a TQFT with $\text{GSD}>1$ to match the mixed anomaly.

\subsubsection{$\wf{su}(4)_5\to\wf{su}(4)_3$}
Let us also study one example with level larger than $r+1=4$. As a minimal example, we pick $k=5$. The model has anomalous $\mbb Z_4$ center symmetry. The primary operator in the adjoint $\phi_\text{adj}=\phi_{\widehat{[3;1,0,1]}}$ preserves (only) the $\mbb Z_4$ braided fusion subcategory of the rank 56 UV MTC. Thus, the RG flow triggered by the relevant operator cannot be trivial in IR. We study massless scenario in this section.

In order to saturate the anomaly in $\mbb Z_4$, the level in IR should also be odd (if we assume a natural scenario $\wf{su}(4)_5\to\wf{su}(4)_{k'}$). Taking the $c$-theorem into account, we can only have $k'=3,1$. Since the adjoint primary also preserves $PSU(4)$ symmetry, we can further employ the mixed anomaly in $PSU(4)\times\mbb Z_4$. The anomaly matching imposes
\[ 5=k'q\quad\text{mod }4. \]
Thus, the only possibilities are $(k',q)=(3,3),(1,1)$. The first option satisfies the (half-)integral condition, while the second does not. We interpret the first RG flow is simple, while the second (if exists) is non-simple. Indeed, the first scenario satisfies all the constraints above.

The spin constraint is satisfied as
\begin{align*}
    S^{\wf{su}(4)_3}_{\eta,\eta^3}&=\{\pm\frac18,\pm\frac38\}=S^{\wf{su}(4)_5}_{\eta,\eta^3}\quad(\text{mod }1),\\
    S^{\wf{su}(4)_3}_{\eta^2}&=\{0,\pm\frac12\}=S^{\wf{su}(4)_5}_{\eta^2}\quad(\text{mod }1).
\end{align*}
The ``monotonicity'' of conformal dimensions follow as a byproduct of the $c$-theorem. The global dimensions are also ``monotonic''
\[ 11.89\approx D^{\wf{su}(4)_3}<D^{\wf{su}(4)_5}\approx58.94. \]
Since the (half-)integral condition is satisfied, we see the double braiding relations are beautifully obeyed:
\[ \hspace{-40pt}c^{\wf{su}(4)_3}_{j,i}c^{\wf{su}(4)_3}_{i,j}\Big|_{\mbb Z_4}=\begin{pmatrix}id_1&id_\eta&id_{\eta^2}&id_{\eta^3}\\id_\eta&i\cdot id_{\eta^2}&-id_{\eta^3}&-i\cdot id_1\\id_{\eta^2}&-id_{\eta^3}&id_1&-id_\eta\\id_{\eta^3}&-i\cdot id_1&-id_\eta&i\cdot id_{\eta^2}\end{pmatrix}\quad c^{\wf{su}(4)_5}_{j,i}c^{\wf{su}(4)_5}_{i,j}\Big|_{\mbb Z_4}=\begin{pmatrix}id_1&id_\eta&id_{\eta^2}&id_{\eta^3}\\id_\eta&-i\cdot id_{\eta^2}&-id_{\eta^3}&i\cdot id_1\\id_{\eta^2}&-id_{\eta^3}&id_1&-id_\eta\\id_{\eta^3}&i\cdot id_1&-id_\eta&-i\cdot id_{\eta^2}\end{pmatrix}. \]
These consistency checks support our massless flow $\wf{su}(4)_5\to\wf{su}(4)_3$. Together with the previous example, we propose a sequence of simple RG flows
\begin{equation}
\begin{array}{ccccc}
    \wf{su}(4)_5:&1_0&\eta^3_{\frac{15}8}&\eta^2_{\frac52}&\eta_{\frac{15}8}\\
    &\downarrow&\downarrow&\downarrow&\downarrow\\
    \wf{su}(4)_3:&1_0&\eta_{\frac98}&\eta^2_{\frac32}&\eta^3_{\frac98}\\
    &\downarrow&\downarrow&\downarrow&\downarrow\\
    \wf{su}(4)_1:&1_0&\eta^3_{\frac38}&\eta^2_{\frac12}&\eta_{\frac38}
\end{array}.\label{su45to3to1}
\end{equation}
Note that, as in the $\wf{su}(3)_k$ case, the mixed anomaly is matched even if we view the sequence of RG flows as a single RG flow because $\eta$ in $k=5$ flows to $\eta$ in $k'=1$. This example also supports the contraposition of our conjecture. The emergence of 16 Verlinde objects would be explained from the half-integral condition as in $\wf{su}(3)_4\to\wf{su}(3)_2$.

\subsubsection{$\wf{su}(5)_2\to\text{TQFT}$}
Finally, let us also provide examples which could disprove our conjecture. (Conversely, if we could find the examples obey our conjecture in future, they give strong supports of it.) They are given by models with $r=4$. Since $r$ is even, the $\mbb Z_5$ symmetry alone is anomaly-free. However, for level $k$ not a multiple of five, we still have a mixed anomaly in $PSU(5)\times\mbb Z_5$. We consider relevant deformations preserving the product symmetry. Concretely, we consider perturbations with the adjoint primary. We start from a minimal example, $k=2$. This is the smallest level the model has the adjoint.

Since the anomaly has to be matched, the IR theory cannot be trivial. We consider a massless scenario in this section. With the $c$-theorem, a natural IR candidate is $\wf{su}(5)_1$. The mixed anomaly matching imposes
\[ 2=q\quad\text{mod }5. \]
Thus, in this scenario $\wf{su}(5)_2\to\wf{su}(5)_1$, $\eta$ in $k=2$ has to flow to $\eta^2$ in $k'=1$. Indeed, this scenario satisfies all the constraints above but one.

The spin constraint is satisfied as
\[ S^{\wf{su}(5)_1}_{\eta,\eta^2,\eta^3,\eta^4}=\{0,\pm\frac15,\pm\frac25\}=S^{\wf{su}(5)_2}_{\eta,\eta^2,\eta^3,\eta^4}\quad\text{mod }1. \]
The ``monotonicity'' of conformal dimensions are obeyed as
\[ h^{\wf{su}(5)_1}_1=0=h^{\wf{su}(5)_2}_1,\quad h^{\wf{su}(5)_1}_{\eta,\eta^4}=\frac25<\frac65=h^{\wf{su}(5)_2}_{\eta^2,\eta^3},\quad h^{\wf{su}(5)_1}_{\eta^2,\eta^3}=\frac35<\frac45=h^{\wf{su}(5)_2}_{\eta,\eta^4}. \]
The global dimensions are also ``monotonic''
\[ D^{\wf{su}(5)_1}=\sqrt5<\sqrt{\frac{35}{2\left(1-\sin\frac{3\pi}{14}\right)}}=D^{\wf{su}(5)_2}. \]

Let us look at the conformal dimensions in more detail. The identifications forced by the mixed anomaly matching do \textit{not} satisfy the half-integral condition:
\[ \frac25+\frac65,\ \frac35+\frac45\notin\frac12\mbb Z. \]
Related to this, the sets of statistics do not match. In UV, it is given by $\Theta=\{1,e^{\pm\frac{4\pi i}5}\}$, while in IR, it is given by $\Theta=\{1,e^{\pm\frac{2\pi i}5}\}$. This fact can be interpreted in several ways. It could imply the flow is non-simple. However, it seems the flow is the `shortest' possible one, and we believe this possibility is unlikely. Relatedly, if this massless flow existed, it could be a `counterexample' to our conjecture (or the statistics matching condition) because the flow seems simple. This would mean the algebraic construction of RG interface a la Gaiotto, which seems to work in other examples, mysteriously fails to work in this case. We believe this possibility is also unlikely. Therefore, a natural logical possibility is that the flow is massive. (If the flow is massive, IR theory should have $\text{GSD}>1$ to saturate the mixed anomaly.) Since we know our scenario is arguable, further investigation, say numerical study, is desirable.

\subsubsection{$\wf{su}(5)_3\to\wf{su}(5)_2$}
Let us continue our study. Next, we pick $k=3$. The model has a mixed $PSU(5)\times\mbb Z_5$ anomaly. Thus, an RG flow triggered by singlets under the symmetry cannot be trivial. We consider deformation with adjoint primary.

With assumption of massless flow $\wf{su}(5)_3\to\wf{su}(5)_{k'}$ in this section, the mixed anomaly matching requires
\[ 3=k'q\quad\text{mod }5. \]
The only possibilities are $(k',q)=(2,4),(1,3)$. We find both scenarios satisfy\footnote{The spin constraint is trivially satisfied because, for $k=1,2,3$, defect Hilbert spaces associated to nontrivial $\eta^l$ have spin contents
\[ S^{\wf{su}(5)_k}_{\eta^l}=\{0,\pm\frac15,\pm\frac25\}\quad\text{mod }1\quad(l=1,2,3,4). \]
The ``monotonicity'' of conformal dimensions follows as is clear from explicit values
\begin{align*}
    &h^{\wf{su}(5)_3}_1=0,\quad h^{\wf{su}(5)_3}_\eta=\frac65=h^{\wf{su}(5)_3}_{\eta^4},\quad h^{\wf{su}(5)_3}_{\eta^2}=\frac95=h^{\wf{su}(5)_3}_{\eta^3},\\
    &h^{\wf{su}(5)_2}_1=0,\quad h^{\wf{su}(5)_2}_\eta=\frac45=h^{\wf{su}(5)_2}_{\eta^4},\quad h^{\wf{su}(5)_2}_{\eta^2}=\frac65=h^{\wf{su}(5)_2}_{\eta^3},\\
    &h^{\wf{su}(5)_1}_1=0,\quad h^{\wf{su}(5)_1}_\eta=\frac25=h^{\wf{su}(5)_1}_{\eta^4},\quad h^{\wf{su}(5)_1}_{\eta^2}=\frac35=h^{\wf{su}(5)_1}_{\eta^3}.
\end{align*}
Finally, the global dimensions are also ``monotonic'' because
\[ D^{\wf{su}(5)_1}=\sqrt5<D^{\wf{su}(5)_2}=\sqrt{\frac{35}{2\left(1-\sin\frac{3\pi}{14}\right)}}<D^{\wf{su}(5)_3}=4\sqrt5(\sqrt2+1). \]} the ``monotonicities,'' however, only the first scenario satisfies the (half-)integral condition:\footnote{As a result, the sets of statistics also match:
\[ \Theta^\text{IR}=\{1,e^{\pm\frac{4\pi i}5}\}=\Theta^\text{UV}. \]}
\begin{equation}
\begin{array}{cccccc}
    \wf{su}(5)_3:&1_0&\eta_{\frac65}&\eta^2_{\frac95}&\eta^3_{\frac95}&\eta^4_{\frac65}\\
    &\downarrow&\downarrow&\downarrow&\downarrow&\downarrow\\
    \wf{su}(5)_2:&1_0&\eta^4_{\frac45}&\eta^3_{\frac65}&\eta^2_{\frac65}&\eta_{\frac45}
\end{array}.\label{su53to2}
\end{equation}
Accordingly, we see the double braiding relations are obeyed:
\[ \hspace{-50pt}\scalebox{0.8}{$c^{\wf{su}(5)_2}_{j,i}c^{\wf{su}(5)_2}_{i,j}\Big|_{\mbb Z_5}=\begin{pmatrix}id_1&id_\eta&id_{\eta^2}&id_{\eta^3}&id_{\eta^4}\\id_\eta&e^{-\frac{4\pi i}5}id_{\eta^2}&e^{\frac{2\pi i}5}id_{\eta^3}&e^{-\frac{2\pi i}5}id_{\eta^4}&e^{\frac{4\pi i}5}id_1\\id_{\eta^2}&e^{\frac{2\pi i}5}id_{\eta^3}&e^{\frac{4\pi i}5}id_{\eta^4}&e^{-\frac{4\pi i}5}id_1&e^{-\frac{2\pi i}5}id_\eta\\id_{\eta^3}&e^{-\frac{2\pi i}5}id_{\eta^4}&e^{-\frac{4\pi i}5}id_1&e^{\frac{4\pi i}5}id_\eta&e^{\frac{2\pi i}5}id_{\eta^2}\\id_{\eta^4}&e^{\frac{4\pi i}5}id_1&e^{-\frac{2\pi i}5}id_\eta&e^{\frac{2\pi i}5}id_{\eta^2}&e^{-\frac{4\pi i}5}id_{\eta^2}\end{pmatrix},\quad c^{\wf{su}(5)_3}_{j,i}c^{\wf{su}(5)_3}_{i,j}\Big|_{\mbb Z_5}=\begin{pmatrix}id_1&id_\eta&id_{\eta^2}&id_{\eta^3}&id_{\eta^4}\\id_\eta&e^{\frac{4\pi i}5}id_{\eta^2}&e^{-\frac{2\pi i}5}id_{\eta^3}&e^{\frac{2\pi i}5}id_{\eta^4}&e^{-\frac{4\pi i}5}id_1\\id_{\eta^2}&e^{-\frac{2\pi i}5}id_{\eta^3}&e^{-\frac{4\pi i}5}id_{\eta^4}&e^{\frac{4\pi i}5}id_1&e^{\frac{2\pi i}5}id_\eta\\id_{\eta^3}&e^{\frac{2\pi i}5}id_{\eta^4}&e^{\frac{4\pi i}5}id_1&e^{-\frac{4\pi i}5}id_\eta&e^{-\frac{2\pi i}5}id_{\eta^2}\\id_{\eta^4}&e^{-\frac{4\pi i}5}id_1&e^{\frac{2\pi i}5}id_\eta&e^{-\frac{2\pi i}5}id_{\eta^2}&e^{\frac{4\pi i}5}id_{\eta^2}\end{pmatrix}.$} \]
Therefore, the first scenario $\wf{su}(5)_3\to\wf{su}(5)_2$ is an allowed simple RG flow. The emergence of 10 Verlinde objects would be explained from the half-integral condition as in $\wf{su}(3)_4\to\wf{su}(3)_2$. We believe the second massless flow does not exist because it violates the statistics matching, but if it existed, we propose it would be non-simple.

\subsubsection{$\wf{su}(5)_4\to\wf{su}(5)_1$}
For completeness, we also study $k=4$. Its relevant deformation with adjoint primary preserves $PSU(5)\times\mbb Z_5$ symmetry, and we can employ the mixed anomaly matching to constrain the IR theory. With assumption of massless phase in this section, natural IR candidates are $\wf{su}(5)_{k'}$'s. The anomaly matching imposes
\[ 4=k'q\quad\text{mod }5. \]
Taking the $c$-theorem into account, we only have $(k',q)=(3,3),(2,2),(1,4)$. Among the three possibilities, we find only the last one satisfies the (half-)integral condition
\begin{equation}
\begin{array}{cccccc}
    \wf{su}(5)_4:&1_0&\eta_{\frac85}&\eta^2_{\frac{12}5}&\eta^3_{\frac{12}5}&\eta^4_{\frac85}\\
    &\downarrow&\downarrow&\downarrow&\downarrow&\downarrow\\
    \wf{su}(5)_1:&1_0&\eta^4_{\frac25}&\eta^3_{\frac35}&\eta^2_{\frac35}&\eta_{\frac25}
\end{array}.\label{su54to1}
\end{equation}
Indeed, this scenario $\wf{su}(5)_4\to\wf{su}(5)_1$ satisfies all the constraints above.

The sets of statistics match:
\[ \Theta^\text{IR}=\{1,e^{\pm\frac{2\pi i}5}\}=\Theta^\text{UV}. \]
The spin constraint is satisfied as
\[ S^{\wf{su}(5)_1}_{\eta,\eta^2,\eta^3,\eta^4}=\{0,\pm\frac15,\pm\frac25\}=S^{\wf{su}(5)_4}_{\eta,\eta^2,\eta^3,\eta^4}\quad\text{mod }1. \]
The ``monotonicity'' of conformal dimensions are obvious from (\ref{su54to1}). The global dimensions are also ``monotonic''\footnote{The analytic expression is given by
\[ \frac{54\sqrt5(-1)^{2/3}}{-3+8(-1)^{1/9}-8(-1)^{2/9}+4(-1)^{4/9}-4(-1)^{5/9}-4(-1)^{7/9}+4(-1)^{8/9}+3\sqrt3i}\approx65.8943. \]}
\[ D^{\wf{su}(5)_1}=\sqrt5<D^{\wf{su}(5)_4}\approx65.8943. \]
Since the (half-)integral condition is satisfied, we also see the double braiding relations are beautifully obeyed:
\[ \hspace{-50pt}\scalebox{0.83}{$c^{\wf{su}(5)_1}_{j,i}c^{\wf{su}(5)_1}_{i,j}=\begin{pmatrix}id_1&id_\eta&id_{\eta^2}&id_{\eta^3}&id_{\eta^4}\\id_\eta&e^{-\frac{2\pi i}5}id_{\eta^2}&e^{-\frac{4\pi i}5}id_{\eta^3}&e^{\frac{4\pi i}5}id_{\eta^4}&e^{\frac{2\pi i}5}id_1\\id_{\eta^2}&e^{-\frac{4\pi i}5}id_{\eta^3}&e^{\frac{2\pi i}5}id_{\eta^4}&e^{-\frac{2\pi i}5}id_1&e^{\frac{4\pi i}5}id_\eta\\id_{\eta^3}&e^{\frac{4\pi i}5}id_{\eta^4}&e^{-\frac{2\pi i}5}id_1&e^{\frac{2\pi i}5}id_\eta&e^{-\frac{4\pi i}5}id_{\eta^2}\\id_{\eta^4}&e^{\frac{2\pi i}5}id_1&e^{\frac{4\pi i}5}id_\eta&e^{-\frac{4\pi i}5}id_{\eta^2}&e^{-\frac{2\pi i}5}id_{\eta^2}\end{pmatrix},\quad c^{\wf{su}(5)_4}_{j,i}c^{\wf{su}(5)_4}_{i,j}\Big|_{\mbb Z_5}=\begin{pmatrix}id_1&id_\eta&id_{\eta^2}&id_{\eta^3}&id_{\eta^4}\\id_\eta&e^{\frac{2\pi i}5}id_{\eta^2}&e^{\frac{4\pi i}5}id_{\eta^3}&e^{-\frac{4\pi i}5}id_{\eta^4}&e^{-\frac{2\pi i}5}id_1\\id_{\eta^2}&e^{\frac{4\pi i}5}id_{\eta^3}&e^{-\frac{2\pi i}5}id_{\eta^4}&e^{\frac{2\pi i}5}id_1&e^{-\frac{4\pi i}5}id_\eta\\id_{\eta^3}&e^{-\frac{4\pi i}5}id_{\eta^4}&e^{\frac{2\pi i}5}id_1&e^{-\frac{2\pi i}5}id_\eta&e^{\frac{4\pi i}5}id_{\eta^2}\\id_{\eta^4}&e^{-\frac{2\pi i}5}id_1&e^{-\frac{4\pi i}5}id_\eta&e^{\frac{4\pi i}5}id_{\eta^2}&e^{\frac{2\pi i}5}id_{\eta^2}\end{pmatrix}.$} \]
These consistency checks support our flow $\wf{su}(5)_4\to\wf{su}(5)_1$ is allowed and simple. Since the surviving rank five BFC is a consistent MTC, no emergent Verlinde object is needed.

\subsubsection{$\wf{su}(5)_6\to\wf{su}(5)_4$}
To draw universal pattern, we proceed one step further; we also study $k=6$. Since the level is not a multiple of five, the model has the mixed anomaly in $PSU(5)\times\mbb Z_5$. The primary operator in the adjoint $\phi_\text{adj}=\phi_{\widehat{[4;1,0,0,1]}}$ preserves the product symmetry. Hence, the RG flow triggered by the operator cannot be trivial in IR. We focus on the massless scenario in this section.

Natural IR candidates are $\wf{su}(5)_{k'}$ with $k'=4,3,2,1$. The mixed anomaly matching imposes
\[ 6=k'q\quad\text{mod }5. \]
The possible solutions are $(k',q)=(4,4),(3,2),(2,3),(1,1)$. We find only the first scenario satisfies the (half-)integral condition:
\begin{equation}
\begin{array}{cccccc}
    \wf{su}(5)_6:&1_0&\eta_{\frac{12}5}&\eta^2_{\frac{18}5}&\eta^3_{\frac{18}5}&\eta^4_{\frac{12}5}\\
    &\downarrow&\downarrow&\downarrow&\downarrow&\downarrow\\
    \wf{su}(5)_4:&1_0&\eta^4_{\frac85}&\eta^3_{\frac{12}5}&\eta^2_{\frac{12}5}&\eta_{\frac85}
\end{array}.\label{su56to4}
\end{equation}
Indeed, this scenario $\wf{su}(5)_6\to\wf{su}(5)_4$ satisfies all the constraints above.

The sets of statistics match:
\[ \Theta^\text{IR}=\{1,e^{\pm\frac{2\pi i}5}\}=\Theta^\text{UV}. \]
The spin constraint is satisfied as
\[ S^{\wf{su}(5)_4}_{\eta,\eta^2,\eta^3,\eta^4}=\{0,\pm\frac15,\pm\frac25\}=S^{\wf{su}(5)_6}_{\eta,\eta^2,\eta^3,\eta^4}\quad\text{mod }1. \]
The ``monotonicity'' of conformal dimensions are clear from (\ref{su56to4}). The global dimensions are also ``monotonic''
\[ 65.8943\approx D^{\wf{su}(5)_4}<D^{\wf{su}(5)_6}\approx510.827. \]
Since the (half-)integral condition is satisfied, we observe the double braiding relations are obeyed:
\[ \hspace{-50pt}\scalebox{0.8}{$c^{\wf{su}(5)_4}_{j,i}c^{\wf{su}(5)_4}_{i,j}\Big|_{\mbb Z_5}=\begin{pmatrix}id_1&id_\eta&id_{\eta^2}&id_{\eta^3}&id_{\eta^4}\\id_\eta&e^{\frac{2\pi i}5}id_{\eta^2}&e^{\frac{4\pi i}5}id_{\eta^3}&e^{-\frac{4\pi i}5}id_{\eta^4}&e^{-\frac{2\pi i}5}id_1\\id_{\eta^2}&e^{\frac{4\pi i}5}id_{\eta^3}&e^{-\frac{2\pi i}5}id_{\eta^4}&e^{\frac{2\pi i}5}id_1&e^{-\frac{4\pi i}5}id_\eta\\id_{\eta^3}&e^{-\frac{4\pi i}5}id_{\eta^4}&e^{\frac{2\pi i}5}id_1&e^{-\frac{2\pi i}5}id_\eta&e^{\frac{4\pi i}5}id_{\eta^2}\\id_{\eta^4}&e^{-\frac{2\pi i}5}id_1&e^{-\frac{4\pi i}5}id_\eta&e^{\frac{4\pi i}5}id_{\eta^2}&e^{\frac{2\pi i}5}id_{\eta^2}\end{pmatrix},\quad c^{\wf{su}(5)_6}_{j,i}c^{\wf{su}(5)_6}_{i,j}\Big|_{\mbb Z_5}=\begin{pmatrix}id_1&id_\eta&id_{\eta^2}&id_{\eta^3}&id_{\eta^4}\\id_\eta&e^{-\frac{2\pi i}5}id_{\eta^2}&e^{-\frac{4\pi i}5}id_{\eta^3}&e^{\frac{4\pi i}5}id_{\eta^4}&e^{\frac{2\pi i}5}id_1\\id_{\eta^2}&e^{-\frac{4\pi i}5}id_{\eta^3}&e^{\frac{2\pi i}5}id_{\eta^4}&e^{-\frac{2\pi i}5}id_1&e^{\frac{4\pi i}5}id_\eta\\id_{\eta^3}&e^{\frac{4\pi i}5}id_{\eta^4}&e^{-\frac{2\pi i}5}id_1&e^{\frac{2\pi i}5}id_\eta&e^{-\frac{4\pi i}5}id_{\eta^2}\\id_{\eta^4}&e^{\frac{2\pi i}5}id_1&e^{\frac{4\pi i}5}id_\eta&e^{-\frac{4\pi i}5}id_{\eta^2}&e^{-\frac{2\pi i}5}id_{\eta^2}\end{pmatrix}.$} \]
These consistency checks support our flow $\wf{su}(5)_6\to\wf{su}(5)_4$ is allowed and simple. Together with the previous example, we propose a sequence of simple RG flows
\begin{equation}
\begin{array}{cccccc}
    \wf{su}(5)_6:&1_0&\eta_{\frac{12}5}&\eta^2_{\frac{18}5}&\eta^3_{\frac{18}5}&\eta^4_{\frac{12}5}\\
    &\downarrow&\downarrow&\downarrow&\downarrow&\downarrow\\
    \wf{su}(5)_4:&1_0&\eta^4_{\frac85}&\eta^3_{\frac{12}5}&\eta^2_{\frac{12}5}&\eta_{\frac85}\\
    &\downarrow&\downarrow&\downarrow&\downarrow&\downarrow\\
    \wf{su}(5)_1:&1_0&\eta_{\frac25}&\eta^2_{\frac35}&\eta^3_{\frac35}&\eta^4_{\frac25}
\end{array}.\label{su56to4to1}
\end{equation}
Note that as in $\wf{su}(3)_k$ or $\wf{su}(4)_k$ cases, the mixed anomaly is matched even if we view the sequence of RG flows as a single RG flow because $\eta$ in $k=6$ flows to $\eta$ in $k'=1$. This example further supports the contraposition of our conjecture. The emergence of 65 Verlinde objects would be explained from the half-integral condition as usual.

\subsection{$B_r$ type, i.e., $\wf{so}(2r+1)_k$ model}
An $\wf{so}(2r+1)_k$ model has central charge
\begin{equation}
    c_k=\frac{k(2r^2+r)}{k+(2r-1)}.\label{cso2r+1k}
\end{equation}
Its primaries are labeled by the affine dominant weights (\ref{Pk+}):
\begin{equation}
    P^k_+=\left\{\hmu\Big|0\le\mu_j\ \&\ 0\le\mu_1+\mu_r+2\sum_{j=2}^{r-1}\mu_j\le k\right\}.\label{Pk+soodd}
\end{equation}
Their conformal dimensions can be computed from the formula (\ref{hhmu}). Three of them important for our purposes are given by
\begin{equation}
    h_{\widehat{[k;0,\dots,0]}}=0,\quad h_{\widehat{[0;k,0,\dots,0]}}=\frac k2,\quad h_{\widehat{[k-2;0,1,0,\dots,0]}}=\frac{2r-1}{k+2r-1}.\label{so2r+1h}
\end{equation}
The first two are conformal dimensions of Verlinde lines $\{1,\eta\}$ generating the $\mbb Z_2$ center symmetry of the model. The third one is the conformal dimension of the primary operator in the adjoint.

The theory has various symmetries. The $\mbb Z_2$ center symmetry mentioned above is part of the categorical symmetry. From the conformal dimension, it is clear that the $\mbb Z_2$ symmetry is anomaly-free for integer $k$. The $\mbb Z_2$ symmetry would also be free of mixed anomalies as implied by an existence of symmetry-preserving boundary state \cite{KY19} for $k\in\mbb Z$. Therefore, as far as we know, there is no obstruction for IR theory to be trivial.\footnote{Note, however, this does \textit{not} mean IR phase is always gapped. An example is the unitary discrete series of (bosonic) minimal models. The models do not have 't Hooft anomalies \cite{CW20}. An easy way to see this fact is to compute spins. Since $\mbb Z_3$ symmetries are anomaly-free, it is sufficient to study models with $\mbb Z_2$ symmetry. In an $M(m+1,m)$ model, a Verlinde line labeled with Kac index $(r,s)=(m-1,1)$ generates the $\mbb Z_2$ symmetry. The line has conformal dimension $h=\frac{(m-1)(m-2)}4$. Since $h\in\frac12\mbb Z$, the $\mbb Z_2$ symmetry is anomaly-free. However, it is known that its integrable deformation (with positive Lagrangian coupling) triggers a massless flow $M(m+1,m)+\phi_{1,3}\to M(m,m-1)$. These examples indicate that, even in absence of anomaly, fine-tuning can lead to conformal phases.} (Indeed, we will see the negative Lagrangian coupling leads to a trivially gapped phase in the next section.) Nevertheless, for completeness, we also study RG flows from $B_r$ type WZW models.

We study massless scenarios $\wf{so}(2r+1)_k\to\wf{so}(2r+1)_{k'}$ in this section. Since there is only one nontrivial generator of the center symmetry, we can identify them. Then, the half-integral condition is translated to
\[ \frac{k'}2+\frac k2\in\frac12\mbb Z, \]
or
\begin{equation}
    k'\in\mbb Z.\label{halfintsoodd}
\end{equation}

\subsubsection{$\wf{so}(7)_2\to\wf{so}(7)_1$?}
We study the minimal example, $r=3,k=2$. The model has a primary operator in the adjoint, $\phi_\text{adj}=\phi_{\widehat{[0;0,1,0]}}$. We consider an RG flow triggered by the relevant operator.

The adjoint primary preserves only the $\mbb Z_2$ braided fusion subcategory of the rank seven UV MTC. If the flow is massless, a natural candidate for IR CFT is $\wf{so}(7)_1$. Indeed, this scenario $\wf{so}(7)_2\to\wf{so}(7)_1$ passes all constraints above.

The spin constraint is satisfied as
\[ S^{\wf{so}(7)_1}_\eta=\{0,\pm\frac12\}=S^{\wf{so}(7)_2}_\eta\quad(\text{mod }1). \]
The ``monotonicity'' of conformal dimensions follows as a byproduct of the $c$-theorem. The global dimensions are also ``monotonic''
\[ D^{\wf{so}(7)_1}=2<2\sqrt7=D^{\wf{so}(7)_2}. \]
Finally, this scenario
\begin{equation}
\begin{array}{ccc}
    \wf{so}(7)_2:&1_0&\eta_1\\
    &\downarrow&\downarrow\\
    \wf{so}(7)_1:&1_0&\eta_{\frac12}
\end{array}\label{so72to1}
\end{equation}
also satisfies the half-integral condition (\ref{halfint}).\footnote{As a result, the sets of statistics also match:
\[\Theta^\text{IR}=\{1\}=\Theta^\text{UV}. \]} Accordingly, we see the double braiding relations are obeyed:
\[ c^{\wf{so}(7)_1}_{j,i}c^{\wf{so}(7)_1}_{i,j}\Big|_{\mbb Z_2}=\begin{pmatrix}id_1&id_\eta\\id_\eta&id_1\end{pmatrix},\quad c^{\wf{so}(7)_2}_{j,i}c^{\wf{so}(7)_2}_{i,j}\Big|_{\mbb Z_2}=\begin{pmatrix}id_1&id_\eta\\id_\eta&id_1\end{pmatrix}. \]
These consistency checks allow our massless scenario $\wf{so}(7)_2\to\wf{so}(7)_1$.

Note that if this massless flow is realized, there is an emergent symmetry object in IR. In the spirit of \cite{KK21}, we have to explain why the symmetry emerges. Luckily, the reason is known, modularity. The surviving BFC has (unnormalized) topological $S$-matrix
\[ \widetilde S_\text{top}=\begin{pmatrix}1&1\\1&1\end{pmatrix}. \]
Since the matrix is singular, we need emergent object to make it modular if the IR theory is described by an MTC. Can one emergent object make the surviving BFC modular? Yes. According to \cite{GK94}, there is only one rank three MTC enlarging the $\mbb Z_2$ BFC. The MTC has $SU(2)_2$ realization, and fusion rules are those of $\wf{so}(7)_1$. Since the rank three MTC can have central charge smaller than $c_\text{UV}=6$, we conclude the rank three MTC is consistent.

\subsection{$C_r$ type, i.e., $\wf{sp}(2r)_k$ model}
An $\wf{sp}(2r)_k$ model has central charge
\begin{equation}
    c_k=\frac{k(2r^2+r)}{k+(r+1)}.\label{csp2rk}
\end{equation}
Its primaries are labeled by the affine dominant weights (\ref{Pk+}):
\begin{equation}
    P^k_+=\left\{\hmu\Big|0\le\mu_j\ \&\ 0\le\sum_{j=1}^r\mu_j\le k\right\}.\label{Pk+sp}
\end{equation}
Their conformal dimensions can be computed from the formula (\ref{hhmu}). Three of them we use below are given by
\begin{equation}
    h_{\widehat{[k;0,,\dots,0]}}=0,\quad h_{\widehat{[0;0,\dots,0,k]}}=\frac{kr}4,\quad h_{\widehat{[k-2;2,0,\dots,0]}}=\frac{r+1}{k+r+1}.\label{sp2rh}
\end{equation}
The first two are conformal dimensions of Verlinde lines $\{1,\eta\}$ generating the $\mbb Z_2$ center symmetry of the model. The third is the conformal dimension of the adjoint primary. From the conformal dimension, we immediately learn the $\mbb Z_2$ symmetry is anomalous if $kr$ is odd. In order to have nontrivial IR theory, we consider these cases. As a minimal example, we pick $r=3$ and odd $k$.

Before we dive into details, let us comment on the half-integral condition. In the massless scenario $\wf{sp}(2r)_k\to\wf{sp}(2r)_{k'}$, we can identify the only nontrivial generators. Under the identification, the half-integral condition is translated to
\[ \frac{k'r}4+\frac{kr}4\in\frac12\mbb Z, \]
or
\begin{equation}
    r(k'+k)\in2\mbb Z.\label{halfintsp}
\end{equation}

\subsubsection{$\wf{sp}(6)_3\to\wf{sp}(6)_1$}
We start from $k=3$. The adjoint primary preserves only the $\mbb Z_2$ braided fusion subcategory of the rank 20 UV MTC. Since the anomalous symmetry is preserved, the IR theory cannot be trivial. Assuming massless phase in this section, we get a natural scenario $\wf{sp}(6)_3\to\wf{sp}(6)_1$. Indeed, this scenario satisfies all the constraints above.

The spin constraint is satisfied as
\[ S^{\wf{sp}(6)_1}_\eta=\{\pm\frac14\}=S^{\wf{sp}(6)_3}_\eta\quad(\text{mod }1). \]
The ``monotonicity'' of the conformal dimensions follow as a byproduct of the $c$-theorem. The global dimensions are also ``monotonic''
\[ D^{\wf{sp}(6)_1}=\sqrt2\sqrt{1+\zeta^2}<\frac{7\sqrt{14}}{\cos\frac\pi{14}-3\cos\frac{3\pi}{14}+5\sin\frac\pi7}=D^{\wf{sp}(6)_3}, \]
where $\zeta=\frac{1+\sqrt5}2$ is the golden ratio. Finally, our scenario
\begin{equation}
\begin{array}{ccc}
    \wf{sp}(6)_3&1_0&\eta_{\frac94}\\
    &\downarrow&\downarrow\\
    \wf{sp}(6)_1&1_0&\eta_{\frac34}
\end{array}\label{sp63to1}
\end{equation}
satisfies the (half-)integral condition. As a result, we observe the double braiding relations are obeyed:
\[ c^{\wf{sp}(6)_1}_{j,i}c^{\wf{sp}(6)_1}_{i,j}\Big|_{\mbb Z_2}=\begin{pmatrix}id_1&id_\eta\\id_\eta&-id_1\end{pmatrix},\quad c^{\wf{sp}(6)_3}_{j,i}c^{\wf{sp}(6)_3}_{i,j}\Big|_{\mbb Z_2}=\begin{pmatrix}id_1&id_\eta\\id_\eta&-id_1\end{pmatrix}. \]
These consistency checks support our massless scenario $\wf{sp}(6)_3\to\wf{sp}(6)_1$. 

Let us briefly comment on emergent symmetry. The scenario claims there are two emergent Verlinde objects in IR. Why they appear? We first notice that the surviving rank two BFC has a non-singular $S$-matrix, and it is actually modular. According to \cite{GK94}, the rank two MTC has $SU(2)_1$ realization and central charge $c=1$ mod $2$, which can be smaller than $c_\text{UV}=9$. From these observations, one might think a putative massless flow to $\wf{su}(2)_1$ plausible, however, we have to recall the fact that the adjoint primary also preserves continuous symmetry. For a given rank, $Sp$ group is larger than $SU$ group, and the presence of additional continuous symmetry makes natural to have $C_r$ type IR theory. Since the only lower (in the sense of central charge and global dimension) $C_r$ type theory is $\wf{sp}(6)_1$, this would explain why the two symmetry objects emerge.

\subsubsection{$\wf{sp}(6)_5\to\wf{sp}(6)_3$ or $\wf{sp}(6)_1$}
We study one more example, $k=5$. The model also has anomalous $\mbb Z_2$ center symmetry. The adjoint primary preserves only the $\mbb Z_2$ braided fusion subcategory of the rank 56 UV MTC. Thus, the RG flow triggered by the operator cannot be trivial.

A natural massless scenario is $\wf{sp}(6)_5\to\wf{sp}(6)_{k'}$ with $k'=3,1$ to saturate the anomaly matching and the $c$-theorem. We find both scenarios satisfy the half-integral condition. Hence, together with the previous example, we propose a sequence of simple RG flows:
\begin{equation}
\begin{array}{ccc}
    \wf{sp}(6)_5&1_0&\eta_{\frac{15}4}\\
    &\downarrow&\downarrow\\
    \wf{sp}(6)_3&1_0&\eta_{\frac94}\\
    &\downarrow&\downarrow\\
    \wf{sp}(6)_1&1_0&\eta_{\frac34}
\end{array}.\label{sp65to3to1}
\end{equation}
Indeed, the proposal $\wf{sp}(6)_5\to\wf{sp}(6)_3$ satisfies all the constraints above.

The spin constraint is satisfied as
\[ S^{\wf{sp}(6)_3}_\eta=\{\pm\frac14\}=S^{\wf{sp}(6)_5}_\eta\quad(\text{mod }1). \]
The ``monotonicity'' of conformal dimensions is clear from (\ref{sp65to3to1}). The global dimensions are also ``monotonic''
\[ D^{\wf{sp}(6)_3}=\frac{7\sqrt{14}}{\cos\frac\pi{14}-3\cos\frac{3\pi}{14}+5\sin\frac\pi7}<\frac{9\sqrt2}{\sin\frac\pi9-2\sin\frac{2\pi}9+\cos\frac\pi{18}}=D^{\wf{sp}(6)_5}. \]
The double braiding relations are also obeyed:
\[ c^{\wf{sp}(6)_3}_{j,i}c^{\wf{sp}(6)_3}_{i,j}\Big|_{\mbb Z_2}=\begin{pmatrix}id_1&id_\eta\\id_\eta&-id_1\end{pmatrix},\quad c^{\wf{sp}(6)_5}_{j,i}c^{\wf{sp}(6)_5}_{i,j}\Big|_{\mbb Z_2}=\begin{pmatrix}id_1&id_\eta\\id_\eta&-id_1\end{pmatrix}. \]
These consistency checks allow our massless scenario $\wf{sp}(6)_5\to\wf{sp}(6)_3$.

Note that the scenario has 18 emergent Verlinde objects. They increase the free energy. Furthermore, the half-integral condition is satisfied in the putative massless flow $\wf{sp}(6)_5\to\wf{sp}(6)_1$. Therefore, we also leave the direct flow as a candidate.

\subsection{$D_r$ type with $r=2R$, i.e., $\wf{so}(4R)_k$ model}
An $\wf{so}(2r)_k$ model has central charge
\begin{equation}
    c_k=\frac{k(2r^2-r)}{k+(2r-2)}.\label{cso2rk}
\end{equation}
Its primaries are labeled by the affine dominant weights (\ref{Pk+}):
\begin{equation}
    P^k_+=\left\{\hmu\Big|0\le\mu_j\ \&\ 0\le\mu_1+\mu_{r-1}+\mu_r+2\sum_{j=2}^{r-2}\mu_j\le k\right\}.\label{Pk+soeven}
\end{equation}
Their conformal dimensions can be computed from the formula (\ref{hhmu}). Five of them relevant for our purposes are given by
\begin{equation}
    h_{\widehat{[k;0,\dots,0]}}=0,\quad h_{\widehat{[0;k,0,\dots,0]}}=\frac k2,\quad h_{\widehat{[0;0,\dots,0,k,0]}}=\frac{kr}8=h_{\widehat{[0;0,\dots,0,k]}},\quad h_{\widehat{[k-2;0,1,0,\dots,0]}}=\frac{2r-2}{k+2r-2}.\label{so2rh}
\end{equation}
The first four are conformal dimensions of Verlinde lines generating invertible symmetries, $\mbb Z_2\times\mbb Z_2$ for even $r$, and $\mbb Z_4$ for odd $r$. The last is the conformal dimension of the adjoint primary. In this subsection, we consider even $r=2R$. Thus, the center symmetry is given by $\mbb Z_2\times\mbb Z_2$. Since there are three different $\mbb Z_2$ subgroups, we label them as $\mbb Z_2^A\times\mbb Z_2^{\widetilde A}$ according to outer automorphism elements. Namely, the first $\mbb Z_2$ is generated by $kA\ho_0$, the second by $k\widetilde A\ho_0$, and the diagonal $\mbb Z_2$ by $kA\widetilde A\ho_0$. The conformal dimensions (\ref{so2rh}) say $\mbb Z_2^A$ is anomaly-free for any $k\in\mbb Z$, while the other $\mbb Z_2$ subgroups are anomalous iff $kr\notin4\mbb Z$, or $kR\notin2\mbb Z$. Since we are interested in nontrivial cases, we study models with odd $kR$. Taking $r\ge4$ or $R\ge2$ into account, the minimal example is $R=3$, or $\wf{so}(12)_k$ with odd $k$.

In the massless scenario $\wf{so}(4R)_k\to\wf{so}(4R)_{k'}$, for models we study, we can identify $\eta_A$'s. The other two generators $\eta_{\widetilde A},\eta_{A\widetilde A}$ are swapped under charge conjugation, and there is a $\mbb Z_2$ ambiguity in their identifications. However, the half-integral condition is unaffected by the ambiguity. It is given by
\[ \frac{k'}2+\frac k2\in\frac12\mbb Z,\quad\frac{k'R}4+\frac{kR}4\in\frac12\mbb Z, \]
or
\begin{equation}
    R(k'+k)\in2\mbb Z.\label{halfintso4R}
\end{equation}

\subsubsection{$\wf{so}(12)_3\to\wf{so}(12)_1$}
We start from $k=3$. Since two $\mbb Z_2$ subgroups generated by $k\widetilde A\ho_0$ and $kA\widetilde A\ho_0$ are anomalous, an RG flow triggered by singlets under the symmetries cannot be trivial in IR. In particular, the primary in the adjoint preserves (only) the $\mbb Z_2^A\times\mbb Z_2^{\widetilde A}$ braided fusion subcategory of the rank 32 UV MTC. Thus, the RG flow cannot be trivial, and the IR theory is either TQFT with $\text{GSD}>1$, or CFT. We focus on massless scenario in this section.

A natural IR candidate is the $\wf{so}(12)_1$ WZW model. Indeed, the scenario $\wf{so}(12)_3\to\wf{so}(12)_1$ satisfies all constraints above. The spin constraint is satisfied as
\begin{align*}
    S^{\wf{so}(12)_1}_{\eta_A}&=\{0,\pm\frac12\}=S^{\wf{so}(12)_3}_{\eta_A}\quad(\text{mod }1),\\
    S^{\wf{so}(12)_1}_{\eta_{\widetilde A}}&=\{\pm\frac14\}=S^{\wf{so}(12)_3}_{\eta_{\widetilde A}}\quad(\text{mod }1),\\
    S^{\wf{so}(12)_1}_{\eta_{A\widetilde A}}&=\{\pm\frac14\}=S^{\wf{so}(12)_3}_{\eta_{A\widetilde A}}\quad(\text{mod }1).
\end{align*}
The ``monotonicity'' of conformal dimensions follows as a byproduct of the $c$-theorem. The global dimension is also ``monotonic''
\[ D^{\wf{so}(12)_1}=2<\frac{13}{1-2\cos\frac\pi{13}+2\cos\frac{3\pi}{13}-\sin\frac\pi{26}}=D^{\wf{so}(12)_3}. \]
Although we cannot fix the ambiguity in identifications of Verlinde lines, both
\[
\begin{array}{ccccc}
    \wf{so}(12)_3:&1_0&(\eta_A)_{\frac32}&(\eta_{\widetilde A})_{\frac94}&(\eta_{A\widetilde A})_{\frac94}\\
    &\downarrow&\downarrow&\downarrow&\downarrow\\
    \wf{so}(12)_1:&1_0&(\eta_A)_{\frac12}&(\eta_{\widetilde A})_{\frac34}&(\eta_{A\widetilde A})_{\frac34}
\end{array},
\]
or
\[
\begin{array}{ccccc}
    \wf{so}(12)_3:&1_0&(\eta_A)_{\frac32}&(\eta_{\widetilde A})_{\frac94}&(\eta_{A\widetilde A})_{\frac94}\\
    &\downarrow&\downarrow&\downarrow&\downarrow\\
    \wf{so}(12)_1:&1_0&(\eta_A)_{\frac12}&(\eta_{A\widetilde A})_{\frac34}&(\eta_{\widetilde A})_{\frac34}
\end{array},
\]
satisfy the (half-)integral condition. Consequently, we see the double braiding relations are obeyed:
\[ \hspace{-40pt}c^{\wf{so}(12)_1}_{j,i}c^{\wf{so}(12)_1}_{i,j}=\begin{pmatrix}id_1&id_{\eta_A}&id_{\eta_{A\widetilde A}}&id_{\eta_{\widetilde A}}\\id_{\eta_A}&id_1&-id_{\eta_{\widetilde A}}&-id_{\eta_{A\widetilde A}}\\id_{\eta_{A\widetilde A}}&-id_{\eta_{\widetilde A}}&-id_1&id_{\eta_A}\\id_{\eta_{\widetilde A}}&-id_{\eta_{A\widetilde A}}&id_{\eta_A}&-id_1\end{pmatrix},\quad c^{\wf{so}(12)_3}_{j,i}c^{\wf{so}(12)_3}_{i,j}\Big|_{\mbb Z_2\times\mbb Z_2}=\begin{pmatrix}id_1&id_{\eta_A}&id_{\eta_{A\widetilde A}}&id_{\eta_{\widetilde A}}\\id_{\eta_A}&id_1&-id_{\eta_{\widetilde A}}&-id_{\eta_{A\widetilde A}}\\id_{\eta_{A\widetilde A}}&-id_{\eta_{\widetilde A}}&-id_1&id_{\eta_A}\\id_{\eta_{\widetilde A}}&-id_{\eta_{A\widetilde A}}&id_{\eta_A}&-id_1\end{pmatrix}. \]
These consistency checks support our massless scenario $\wf{so}(12)_3\to\wf{so}(12)_1$.

Let us briefly comment on the consistency of the category. The surviving rank four BFC is actually an MTC. Furthermore, according to \cite{GK94}, the rank four MTC has $SO(8)_1$ realization with central charge $0$ mod $4$, which can be smaller than $c_\text{UV}=\frac{198}{13}$. Therefore, the rank four MTC is consistent and requires no emergent symmetry. This is consistent with the absence of emergent Verlinde object in our scenario.

\subsubsection{$\wf{so}(12)_5\to\wf{so}(12)_3$ or $\wf{so}(12)_1$}
We study one more example with $k=5$. As in the previous example, two $\mbb Z_2$ subgroups generated by $k\widetilde A\ho_0$ and $kA\widetilde A\ho_0$ are anomalous. In order to preserve the anomaly, we consider relevant deformation with the adjoint. Then, IR theories cannot be trivial.

Assuming massless scenario in this section, we obtain a natural flow $\wf{so}(12)_5\to\wf{so}(12)_{k'}$ with $k'=3,1$ to saturate the anomaly. We find both scenarios satisfy the half-integral condition. Let us see the fist scenario $\wf{so}(12)_5\to\wf{so}(12)_3$ further satisfies all constraints above.

The spin constraint is satisfied as
\begin{align*}
    S^{\wf{so}(12)_3}_{\eta_A}&=\{0,\pm\frac12\}=S^{\wf{so}(12)_5}_{\eta_A}\quad(\text{mod }1),\\
    S^{\wf{so}(12)_3}_{\eta_{\widetilde A}}&=\{\pm\frac14\}=S^{\wf{so}(12)_5}_{\eta_{\widetilde A}}\quad(\text{mod }1),\\
    S^{\wf{so}(12)_3}_{\eta_{A\widetilde A}}&=\{\pm\frac14\}=S^{\wf{so}(12)_5}_{\eta_{A\widetilde A}}\quad(\text{mod }1).
\end{align*}
The ``monotonicity'' of conformal dimensions follows as a byproduct of the $c$-theorem. The global dimensions are also ``monotonic''
\[ D^{\wf{so}(12)_3}=\frac{13}{1-2\cos\frac\pi{13}+2\cos\frac{3\pi}{13}-\sin\frac\pi{26}}<\frac{120}{9-\sqrt5-\sqrt{6(5+\sqrt5)}}=D^{\wf{so}(12)_5}. \]
These consistency checks allow our simple massless flow $\wf{so}(12)_5\to\wf{so}(12)_3$. This scenario requires 28 emergent Verlinde objects. They increase free energy. Therefore, we also leave the direct flow $\wf{so}(12)_5\to\wf{so}(12)_1$ as a candidate.

Although we cannot fix the ambiguity in identifications of Verlinde objects, if we assume the universal pattern we found in the $A$ type  WZW models, we get a sequence of simple RG flows with identifications
\begin{equation}
\begin{array}{ccccc}
    \wf{so}(12)_5:&1_0&(\eta_A)_{\frac52}&(\eta_{A\widetilde A})_{\frac{15}4}&(\eta_{\widetilde A})_{\frac{15}4}\\
    &\downarrow&\downarrow&\downarrow&\downarrow\\
    \wf{so}(12)_3:&1_0&(\eta_A)_{\frac32}&(\eta_{\widetilde A})_{\frac94}&(\eta_{A\widetilde A})_{\frac94}\\
    &\downarrow&\downarrow&\downarrow&\downarrow\\
    \wf{so}(12)_1:&1_0&(\eta_A)_{\frac12}&(\eta_{A\widetilde A})_{\frac34}&(\eta_{\widetilde A})_{\frac34}
\end{array}.\label{so125to3to1}
\end{equation}

\subsection{$D_r$ type with $r=2R+1$, i.e., $\wf{so}(4R+2)_k$ model}
An $\wf{so}(4R+2)_k$ model has central charge (\ref{cso2rk}), affine dominant weights (\ref{Pk+soeven}), and some conformal dimensions (\ref{so2rh}). The difference from the last subsection is its center symmetry; the model for odd $r=2R+1$ has $\mbb Z_4$ symmetry. Since the generating Verlinde line $\eta$ has conformal dimension $h=\frac{k(2R+1)}8$, the symmetry is anomalous if $k$ is odd. To have nontrivial IR theory, we consider these cases.

For these cases, in the massless scenario $\wf{so}(4R+2)_k\to\wf{so}(4R+2)_{k'}$, $\eta$ is identified with either $\eta$ or $\eta^3$. Both identifications yield the half-integral condition
\[ \frac{k'(2R+1)}8+\frac{k(2R+1)}8\in\frac12\mbb Z, \]
or
\begin{equation}
    k'+k\in4\mbb Z.\label{halfintso4R+2}
\end{equation}

\subsubsection{$\wf{so}(10)_3\to\wf{so}(10)_1$}
As a minimal example, let us pick $R=2,k=3$. The $\mbb Z_4$ symmetry is anomalous, and RG flows triggered by singlets cannot be trivial in IR. In particular, adjoint primary of the model preserves (only) the $\mbb Z_4$ braided fusion subcategory of the rank 28 UV MTC. We consider the flow.

Since the IR theory is nontrivial, there are two possibilities: a TQFT with $\text{GSD}>1$, or CFT. In this section, we consider the massless scenario. Then, a natural IR candidate is $\wf{so}(10)_1$. Indeed, the scenario $\wf{so}(10)_3\to\wf{so}(10)_1$ satisfies all constraints above.

The spin constraint is satisfied as
\[ S^{\wf{so}(10)_1}_\eta=\{\pm\frac18,\pm\frac38\}=S^{\wf{so}(10)_3}_\eta\quad(\text{mod }1). \]
The ``monotonicity'' of conformal dimensions follow as a byproduct of the $c$-theorem. The global dimensions are also ``monotonic''
\[ D^{\wf{so}(10)_1}=2<\frac{11\sqrt{11}}{1+9\sin\frac\pi{22}+2\sin\frac{3\pi}{22}-2\sin\frac{5\pi}{22}-2\cos\frac\pi{11}+2\cos\frac{2\pi}{11}}=D^{\wf{so}(10)_3}. \]
Finally, although we cannot fix the ambiguity in identifications of Verlinde lines, both
\[\begin{array}{ccccc}
    \wf{so}(10)_3:&1_0&\eta^2_{\frac32}&\eta^3_{\frac{15}8}&\eta_{\frac{15}8}\\
    &\downarrow&\downarrow&\downarrow&\downarrow\\
    \wf{so}(10)_1:&1_0&\eta^2_{\frac12}&\eta^3_{\frac58}&\eta_{\frac58}
\end{array},\]
or
\[\begin{array}{ccccc}
    \wf{so}(10)_3:&1_0&\eta^2_{\frac32}&\eta^3_{\frac{15}8}&\eta_{\frac{15}8}\\
    &\downarrow&\downarrow&\downarrow&\downarrow\\
    \wf{so}(10)_1:&1_0&\eta^2_{\frac12}&\eta_{\frac58}&\eta^3_{\frac58}
\end{array},\]
satisfy the (half-)integral condition. Accordingly, we see the double braiding relations are beautifully obeyed:
\[ \hspace{-30pt}c^{\wf{so}(10)_1}_{j,i}c^{\wf{so}(10)_1}_{i,j}=\begin{pmatrix}id_1&id_{\eta^2}&id_{\eta^3}&id_\eta\\id_{\eta^2}&id_1&-id_\eta&-id_{\eta^3}\\id_{\eta^3}&-id_\eta&i\cdot id_{\eta^2}&-i\cdot id_1\\id_\eta&-id_{\eta^3}&-i\cdot id_1&i\cdot id_{\eta^2}\end{pmatrix},\quad c^{\wf{so}(10)_3}_{j,i}c^{\wf{so}(10)_3}_{i,j}\Big|_{\mbb Z_4}=\begin{pmatrix}id_1&id_{\eta^2}&id_{\eta^3}&id_\eta\\id_{\eta^2}&id_1&-id_\eta&-id_{\eta^3}\\id_{\eta^3}&-id_\eta&-i\cdot id_{\eta^2}&i\cdot id_1\\id_\eta&-id_{\eta^3}&i\cdot id_1&-i\cdot id_{\eta^2}\end{pmatrix}. \]
These consistency checks support our massless scenario $\wf{so}(10)_3\to\wf{so}(10)_1$. Since the surviving rank four BFC is a consistent MTC, no emergent Verlinde object is needed.

\subsubsection{$\wf{so}(10)_5\to\wf{so}(10)_3$}
We proceed one step further; we study $k=5$. The model also has anomalous $\mbb Z_4$, and relevant deformation with singlets cannot be trivial. We consider perturbation with the adjoint $\phi_\text{adj}=\phi_{\widehat{[3;0,1,0,0,0]}}$. The primary operator preserves only the $\mbb Z_4$ braided fusion subcategory of the rank 108 UV MTC.

Assuming conformal phase in this section, we get a natural scenario $\wf{so}(10)_5\to\wf{so}(10)_{k'}$ with $k'=3,1$ to saturate the anomaly (together with the $c$-theorem). We find only the first possibility satisfies the (half-)integral condition. Indeed, we see the scenario $\wf{so}(10)_5\to\wf{so}(10)_3$ satisfies all the constraints above.

The spin constraint is satisfied as
\[ S^{\wf{so}(10)_3}_\eta=\{\pm\frac18,\pm\frac38\}=S^{\wf{so}(10)_5}_\eta\quad(\text{mod }1). \]
The ``monotonicity'' of conformal dimensions follow as a byproduct of the $c$-theorem. The global dimensions are also ``monotonic''
\[ 23.3048\approx D^{\wf{so}(10)_3}<D^{\wf{so}(10)_5}\approx341.841. \]
Although we cannot fix the ambiguity in the identifications of Verlinde lines, both\[\begin{array}{ccccc}
    \wf{so}(10)_5:&1_0&\eta^2_{\frac52}&\eta^3_{\frac{25}8}&\eta_{\frac{25}8}\\
    &\downarrow&\downarrow&\downarrow&\downarrow\\
    \wf{so}(10)_3:&1_0&\eta^2_{\frac32}&\eta^3_{\frac{15}8}&\eta_{\frac{15}8}
\end{array},\]
or
\[\begin{array}{ccccc}
    \wf{so}(10)_5:&1_0&\eta^2_{\frac52}&\eta^3_{\frac{25}8}&\eta_{\frac{25}8}\\
    &\downarrow&\downarrow&\downarrow&\downarrow\\
    \wf{so}(10)_3:&1_0&\eta^2_{\frac32}&\eta_{\frac{15}8}&\eta^3_{\frac{15}8}
\end{array},\]
obey the double braiding relations. A straightforward computation yields
\[ \hspace{-40pt}c^{\wf{so}(10)_3}_{j,i}c^{\wf{so}(10)_3}_{i,j}\Big|_{\mbb Z_4}=\begin{pmatrix}id_1&id_{\eta^2}&id_{\eta^3}&id_\eta\\id_{\eta^2}&id_1&-id_\eta&-id_{\eta^3}\\id_{\eta^3}&-id_\eta&-i\cdot id_{\eta^2}&i\cdot id_1\\id_\eta&-id_{\eta^3}&i\cdot id_1&-i\cdot id_{\eta^2}\end{pmatrix},\quad c^{\wf{so}(10)_5}_{j,i}c^{\wf{so}(10)_5}_{i,j}\Big|_{\mbb Z_4}=\begin{pmatrix}id_1&id_{\eta^2}&id_{\eta^3}&id_\eta\\id_{\eta^2}&id_1&-id_\eta&-id_{\eta^3}\\id_{\eta^3}&-id_\eta&i\cdot id_{\eta^2}&-i\cdot id_1\\id_\eta&-id_{\eta^3}&-i\cdot id_1&i\cdot id_{\eta^2}\end{pmatrix}. \]
This scenario requires 24 emergent objects. They would be explained from the half-integral condition as in $\wf{su}(3)_4\to\wf{su}(3)_2$.

Although we cannot fix the ambiguity in the identifications of Verlinde objects, assuming the universal patter we found in $A$ type WZW models, we propose a sequence of simple RG flows
\begin{equation}
\begin{array}{ccccc}
    \wf{so}(10)_5:&1_0&\eta^2_{\frac52}&\eta^3_{\frac{25}8}&\eta_{\frac{25}8}\\
    &\downarrow&\downarrow&\downarrow&\downarrow\\
    \wf{so}(10)_3:&1_0&\eta^2_{\frac32}&\eta_{\frac{15}8}&\eta^3_{\frac{15}8}\\
    &\downarrow&\downarrow&\downarrow&\downarrow\\
    \wf{so}(10)_1:&1_0&\eta^2_{\frac12}&\eta^3_{\frac58}&\eta_{\frac58}
\end{array}.\label{so105to3to1}
\end{equation}

\subsection{$E_6$ type, i.e., $(\wf{e_6})_k$ model}
An $(\wf{e_6})_k$ model has central charge
\begin{equation}
    c_k=\frac{78k}{k+12}.\label{ce6k}
\end{equation}
Its primaries are labeled by affine dominant weights (\ref{Pk+}):
\begin{equation}
    P^k_+=\left\{\hmu\Big|0\le\mu_j\ \&\ 0\le\mu_1+2\mu_2+3\mu_3+2\mu_4+\mu_5+2\mu_6\le k\right\}.\label{Pk+e6}
\end{equation}

With $S$-matrices in the Appendix \ref{Smatrix} at hand, we can also study RG flows from $E$ type WZW models in detail. We start from $(\wf{e_6})_k$ WZW models. The model has $\mbb Z_3$ center symmetry. They are generated by Verlinde lines with conformal dimensions
\[ h_{\widehat{[k;0,0,0,0,0,0]}}=0,\quad h_{\widehat{[0;k,0,0,0,0,0]}}=\frac{2k}3=h_{\widehat{[0;0,0,0,0,k,0]}}. \]
As expected, the symmetry is anomaly-free. However, absence of invariant Cardy state for $k\notin3\mbb Z$ signals a mixed anomaly \cite{KY19} in $\mbb Z_3$. At least, from the $S$-submatrix in Appendix \ref{Smatrix}, we know the corresponding three-dimensional Chern-Simons (CS) theory $(E_6)_k$ has anomalous $\mbb Z_3$ one-form symmetry for $k\notin3\mbb Z$. Thus, we consider RG flows from $k\notin3\mbb Z$ triggered by relevant operators preserving the symmetry. As a minimal example, we consider $k=2$.

In massless scenario $(\wf{e_6})_k\to(\wf{e_6})_{k'}$, the generator $\eta$ of the $\mbb Z_3$ center symmetry is identified with either $\eta$ or $\eta^2$. Both identifications give the half-integral condition
\[ \frac{2k'}3+\frac{2k}3\in\frac12\mbb Z, \]
or
\begin{equation}
    k'+k\in3\mbb Z.\label{halfinte6}
\end{equation}

\subsubsection{$(\wf{e_6})_2\to(\wf{e_6})_1$}
The UV theory $(\wf{e_6})_2$ has an adjoint primary $\phi_\text{adj}=\phi_{\widehat{[0;0,0,0,0,0,1]}}$. It has conformal dimension $h=6/7$, and triggers an RG flow. The relevant operator preserves only the $\mbb Z_3$ braided fusion subcategory of the rank nine UV MTC. If we assume massless scenario, a natural candidate is $(\wf{e_6})_1$. (We study the massive scenario in the next section.) Indeed, we find this scenario $(\wf{e_6})_2\to(\wf{e_6})_1$ satisfies all constraints above.

The spin constraint is satisfied as
\[ S^{(\wf{e_6})_1}_{\eta,\eta^2}=\{0,\pm\frac13\}=S^{(\wf{e_6})_2}_{\eta,\eta^2}\quad(\text{mod }1). \]
The ``monotonicity'' of conformal dimensions follow as a result of the $c$-theorem. From the $S$-matrices, one can also check the ``monotonicity'' of global dimensions
\[ D^{(\wf{e_6})_1}=\sqrt3<\frac{\sqrt{21}}{2\sin\frac\pi7}=D^{(\wf{e_6})_2}. \]
Although we cannot fix the ambiguity in the identification of Verlinde lines, both possibilities, either
\[
\begin{array}{cccc}
(\wf{e_6})_2:&1_0&\eta_{\frac43}&\eta^2_{\frac43}\\
&\downarrow&\downarrow&\downarrow\\
(\wf{e_6})_1:&1_0&\eta_{\frac23}&\eta^2_{\frac23}
\end{array},
\]
or
\begin{equation}
\begin{array}{cccc}
(\wf{e_6})_2:&1_0&\eta_{\frac43}&\eta^2_{\frac43}\\
&\downarrow&\downarrow&\downarrow\\
(\wf{e_6})_1:&1_0&\eta^2_{\frac23}&\eta_{\frac23}
\end{array},\label{e62to1}
\end{equation}
do satisfy the (half-)integral condition. (If we assume the universal patter we found in $A$ type WZW models, the latter is chosen.) Consequently, the double braiding relations are also beautifully obeyed:
\[ c^{(\wf{e_6})_1}_{j,i}c^{(\wf{e_6})_1}_{i,j}=\begin{pmatrix}id_1&id_{\eta^2}&id_\eta\\id_{\eta^2}&e^{2\pi i/3}id_\eta&e^{4\pi i/3}id_1\\id_\eta&e^{4\pi i/3}id_1&e^{2\pi i/3}id_{\eta^2}\end{pmatrix},\quad c^{(\wf{e_6})_2}_{j,i}c^{(\wf{e_6})_2}_{i,j}\Big|_{\mbb Z_3}=\begin{pmatrix}id_1&id_{\eta^2}&id_\eta\\id_{\eta^2}&e^{4\pi i/3}id_\eta&e^{2\pi i/3}id_1\\id_\eta&e^{2\pi i/3}id_1&e^{4\pi i/3}id_{\eta^2}\end{pmatrix}. \]
These consistency checks support our massless flow $(\wf{e_6})_2\to(\wf{e_6})_1$. Since the surviving rank three BFC is a consistent MTC, no emergent Verlinde object is needed.

\subsection{$E_7$ type, i.e., $(\wf{e_7})_k$ model}
An $(\wf{e_7})_k$ model has central charge
\begin{equation}
    c_k=\frac{133k}{k+18}.\label{ce7k}
\end{equation}
Its primaries are labeled by dominant affine weights (\ref{Pk+}):
\begin{equation}
    P^k_+=\left\{\hmu\Big|0\le\mu_j\ \&\ 0\le2\mu_1+3\mu_2+4\mu_3+3\mu_4+2\mu_5+\mu_6+2\mu_7\le k\right\}.\label{Pk+e7}
\end{equation}
The model has $\mbb Z_2$ center symmetry. It is generated by Verlinde lines with conformal dimensions
\[ h_{\widehat{[k;0,0,0,0,0,0,0]}}=0,\quad h_{\widehat{[0;0,0,0,0,0,k,0]}}=\frac{3k}4. \]
For odd $k$, we have $h_\eta\notin\frac12\mbb Z$. Thus, the symmetry is anomalous. (Correspondingly, three-dimensional CS theory $(E_7)_k$ has anomalous $\mbb Z_2$ one-form symmetry for $k\notin2\mbb Z$.) We will consider this case. Then, RG flows triggered by relevant operators preserving the symmetry cannot be trivial. As a minimal example, we pick $k=3$.

In the massless scenario $(\wf{e_7})_k\to(\wf{e_7})_{k'}$, we can identify the only nontrivial generators. The identification gives the half-integral condition
\[ \frac{3k'}4+\frac{3k}4\in\frac12\mbb Z, \]
or
\begin{equation}
    k'+k\in2\mbb Z.\label{halfinte7}
\end{equation}

\subsubsection{$(\wf{e_7})_3\to(\wf{e_7})_1$}
The UV theory $(\wf{e_7})_3$ has an adjoint primary $\phi_\text{adj}=\phi_{\widehat{[1;1,0,0,0,0,0,0]}}$. It has conformal dimension $h=6/7$, and triggers an RG flow. The $S$-matrix (\ref{Se7}) shows the relevant operator preserves only the $\mbb Z_2$ braided fusion subcategory of the rank 12 UV MTC. If we assume massless scenario in this section, a natural IR candidate is the $(\wf{e_7})_1$ WZW model. Indeed, this scenario satisfies all constraints above.

The spin constraint is satisfied because
\[ S^{(\wf{e_7})_1}_\eta=\{\pm\frac14\}=S^{(\wf{e_7})_3}_\eta\quad(\text{mod }1). \]
The ``monotonicity'' of conformal dimensions follows as a byproduct of the $c$-theorem. The global dimensions are also ``monotonic''
\[ D^{(\wf{e_7})_1}=\sqrt2<\sqrt{21(5+\sqrt{21})}=D^{(\wf{e_7})_3}. \]
Finally, this scenario
\begin{equation}
\begin{array}{ccc}
(\wf{e_7})_3:&1_0&\eta_{\frac94}\\
&\downarrow&\downarrow\\
(\wf{e_7})_1:&1_0&\eta_{\frac34}
\end{array},\label{e73to1}
\end{equation}
satisfies the (half-)integral condition. As a result, we see the double braiding relations are obeyed:
\[ c^{(\wf{e_7})_1}_{j,i}c^{(\wf{e_7})_1}_{i,j}=\begin{pmatrix}id_1&id_\eta\\id_\eta&-id_1\end{pmatrix},\quad c^{(\wf{e_7})_3}_{j,i}c^{(\wf{e_7})_3}_{i,j}\Big|_{\mbb Z_2}=\begin{pmatrix}id_1&id_\eta\\id_\eta&-id_1\end{pmatrix}. \]
These consistency checks support our massless scenario $(\wf{e_7})_3\to(\wf{e_7})_1$. Since the surviving rank two BFC is a consistent MTC, no emergent Verlinde object is needed.

\section{Massive RG flows}\label{massive}
In this section, we study massive scenarios. For that purpose, we apply the Cardy's method. The method tells us GSDs. It also signals which sign of the relevant coupling triggers massless and massive RG flows. (However, as Cardy himself mentioned, it should be viewed as an approximation in the massless case. In those cases, we combine our previous analysis to support conformal phases.) We start from a brief review of the method.

The setup is the same as in the truncated conformal space approach. We quantize a UV RCFT on a cylinder. A time slice is given by a circle. It supports a Hilbert space. On the space, the UV Hamiltonian $H_\text{UV}$ acts. We consider deformation problem of the UV RCFT. The deformation operators modify the Hamiltonian to\footnote{We fix the sign of the couplings $\lambda_j$'s in the Lagrangian formalism.}
\[ H=H_\text{UV}-\sum_j\lambda_j\int_{\mbb S^1}\phi_j(x). \]
Our goal is to find ground states of this modified Hamiltonian $H$. Cardy \cite{C17} suggested an ansatz for the ground states. His proposal is a superposition of smeared Cardy states. The ansatz claims a Cardy state $|a\ra$ on a time slice has energy (with rescaled couplings)
\begin{equation}
    E_a=\frac{\pi c}{24(2\tau_a)^2}-\sum_{j\neq0}\frac{S_{ja}}{S_{0a}}\frac{\lambda_j}{(2\tau_a)^{\Delta_j}},\label{Ea}
\end{equation}
where $c$ is the UV central charge, $\tau_a$ a variational parameter, and $\Delta_j$ scaling dimension of $\phi_j$. From this formula, we can read off which Cardy state gives the minimal energy. Below, we apply this method to the deformation problem\footnote{The method was recently applied to $\wf{su}(2)_{2K}$ WZW models in \cite{LHYO22}.} with a single relevant operator, adjoint.

\subsection{$\wf{su}(r+1)_k$}
In the $\wf{su}(r+1)_k$ models, a restriction on GSD is known \cite{YHO18}. Their derivation based on mixed anomaly claims
\begin{equation}
    \text{GSD}\in\frac{r+1}{\gcd(I_{r+1},r+1)}\mbb N,\label{LSMGSD}
\end{equation}
where $I_{r+1}$ is the LSM index. For us, it is just the UV level $k$ mod $(r+1)$:
\[ I_{r+1}=k\quad\text{mod }r+1. \]
See around (\ref{sumixed}).

\subsubsection{$\wf{su}(3)_2+\phi_{\widehat{[0;1,1]}}$}
We start from a simple example, $\wf{su}(3)_2$.
The energies (\ref{Ea}) are given by (with rescaled coupling and $\zeta=\frac{1+\sqrt5}2$)
\[ E_a=\begin{cases}{\displaystyle\frac{16\pi/5}{24(2\tau_a)^2}-\zeta\frac{\lambda_\text{adj}}{(2\tau_a)^{6/5}}}&(\hmu=2\ho_0,2\ho_1,2\ho_2),\\
{\displaystyle\frac{16\pi/5}{24(2\tau_a)^2}+\zeta^{-1}\frac{\lambda_\text{adj}}{(2\tau_a)^{6/5}}}&(\hmu=\widehat{[1;1,0]},\widehat{[1;0,1]},\widehat{[0;1,1]}).\end{cases} \]
Therefore, for $\lambda_\text{adj}>0$, the first set of Cardy states $|\hmu\ra$ minimizes the energy. Recalling the correspondence between Cardy states, primary operators, and Verlinde lines, this signals massless RG flow. The sign is consistent with the proposal in \cite{L15}.

For the other sign $\lambda_\text{adj}<0$, the second set of Cardy states minimizes the energy. This suggests TQFT with ground state degeneracy three. The degeneracy is consistent with (\ref{LSMGSD}). It turns out that the ground state degeneracy is a consequence of spontaneously broken $\mbb Z_3$ center symmetry. To understand this point, recall our relevant operator preserves $PSU(3)\times\mbb Z_3$ and their mixed anomaly. Since the continuous symmetry $PSU(3)$ cannot be spontaneously broken in two dimensions, one way to match the mixed anomaly is to spontaneously break the discrete $\mbb Z_3$. Indeed, the center symmetry acts faithfully on the ground states
\begin{table}[H]
\begin{center}
\begin{tabular}{c|c|c|c}
&$\left|\widehat{[1;1,0]}\right\ra$&$\left|\widehat{[1;0,1]}\right\ra$&$\left|\widehat{[0;1,1]}\right\ra$\\\hline
$1$&$\left|\widehat{[1;1,0]}\right\ra$&$\left|\widehat{[1;0,1]}\right\ra$&$\left|\widehat{[0;1,1]}\right\ra$\\\hline
$\eta$&$\left|\widehat{[0;1,1]}\right\ra$&$\left|\widehat{[1;1,0]}\right\ra$&$\left|\widehat{[1;1,0]}\right\ra$
\end{tabular}.
\end{center}
\end{table}
\hspace{-17pt}Thus, the $\mbb Z_3$ center symmetry is spontaneously broken.

Our proposal is summarized as follows:
\begin{equation}
    \wf{su}(3)_2+\phi_{\widehat{[0;1,1]}}=\begin{cases}\wf{su}(3)_1&(\lambda_\text{adj}>0),\\\text{TQFT w/ GSD}=3&(\lambda_\text{adj}<0).\end{cases}\label{su32}
\end{equation}

\subsubsection{$\wf{su}(4)_3+\phi_{\widehat{[1;1,0,1]}}$}
Our next example is $r=3$. We start from $k=3$. The Cardy's method gives
\[ E_a=\begin{cases}{\displaystyle\frac{45\pi/7}{24(2\tau_a)^2}-4.04892 \frac{\lambda_\text{adj}}{(2\tau_a)^{8/7}}}&(\hmu=4\ho_0,4\ho_1,4\ho_2,4\ho_3),\\
{\displaystyle\frac{45\pi/7}{24(2\tau_a)^2}+0.692021\frac{\lambda_\text{adj}}{(2\tau_a)^{8/7}}}&(\hmu=\widehat{[1;1,1,0]},\widehat{[1;1,0,1]},\widehat{[1;0,1,1]},\widehat{[0;1,1,1]}).\end{cases} \]
For $\lambda_\text{adj}>0$, the first line gives the lowest energy. Since the ground states contain the identity $4\ho_0$ and they obey the same fusion rules as $\wf{su}(4)_1$, the Cardy's method signals massless flow $\wf{su}(4)_3\to\wf{su}(4)_1$ for this sign. On the other hand, for $\lambda_\text{adj}<0$, the second line minimizes the energy. The result suggests a TQFT with $\text{GSD}=4$. The degeneracy is consistent with (\ref{LSMGSD}). The four-fold degeneracy is a result of spontaneously broken $\mbb Z_4$ center symmetry. Indeed, we see the symmetry acts faithfully on the ground states:
\begin{table}[H]
\begin{center}
\begin{tabular}{c|c|c|c|c}
&$\left|\widehat{[1;1,1,0]}\right\ra$&$\left|\widehat{[1;1,0,1]}\right\ra$&$\left|\widehat{[1;0,1,1]}\right\ra$&$\left|\widehat{[0;1,1,1]}\right\ra$\\\hline
$1$&$\left|\widehat{[1;1,1,0]}\right\ra$&$\left|\widehat{[1;1,0,1]}\right\ra$&$\left|\widehat{[1;0,1,1]}\right\ra$&$\left|\widehat{[0;1,1,1]}\right\ra$\\\hline
$\eta$&$\left|\widehat{[0;1,1,1]}\right\ra$&$\left|\widehat{[1;1,1,0]}\right\ra$&$\left|\widehat{[1;1,0,1]}\right\ra$&$\left|\widehat{[1;0,1,1]}\right\ra$
\end{tabular}.
\end{center}
\end{table}

Our proposal is summarized as follows:
\begin{equation}
    \wf{su}(4)_3+\phi_{\widehat{[1;1,0,1]}}=\begin{cases}\wf{su}(4)_1&(\lambda_\text{adj}>0),\\\text{TQFT w/ GSD}=4&(\lambda_\text{adj}<0).\end{cases}\label{su43}
\end{equation}

\subsubsection{$\wf{su}(4)_2+\phi_{\widehat{[0;1,0,1]}}$}
Next, let us also study an example with nontrivial greatest common divisor in (\ref{LSMGSD}). We pick $k=2$. The Cardy's method gives energies
\[ E_a=\begin{cases}{\displaystyle\frac{5\pi}{24(2\tau_a)^2}-2\frac{\lambda_\text{adj}}{(2\tau_a)^{4/3}}}&(\hmu=2\ho_0,2\ho_1,2\ho_2,2\ho_3),\\
{\displaystyle\frac{5\pi}{24(2\tau_a)^2}}&(\hmu=\widehat{[1;1,0,0]},\widehat{[1;0,0,1]},\widehat{[0;1,1,0]},\widehat{[0;0,1,1]}),\\
{\displaystyle\frac{5\pi}{24(2\tau_a)^2}+\frac{\lambda_\text{adj}}{(2\tau_a)^{4/3}}}&(\hmu=\widehat{[1;0,1,0]},\widehat{[0;1,0,1]}).\end{cases} \]
For $\lambda_\text{adj}>0$, the first line gives the minimal energy. The ground states contain the identity $2\ho_0$ and they obey the fusion rules of $\wf{su}(4)_1$. Hence, one might think the positive relevant coupling triggers a massless flow $\wf{su}(4)_2\to\wf{su}(4)_1$. However, as we saw in section \ref{massless}, this scenario requires two emergent Verlinde lines. Furthermore, the identifications of Verlinde lines imposed by the mixed anomaly do not satisfy the half-integral condition or statistics matching. Therefore, we find the massless scenario unlikely, and we propose the flow is massive with four-fold degenerated vacua. Note that the degeneracy $\text{GSD}=4$ is also consistent with (\ref{LSMGSD}). The degeneracy can be understood as a consequence of spontaneously broken $\mbb Z_4$:
\begin{table}[H]
\begin{center}
\begin{tabular}{c|c|c|c|c}
&$\left|\widehat{[2;0,0,0]}\right\ra$&$\left|\widehat{[0;2,0,0]}\right\ra$&$\left|\widehat{[0;0,2,0]}\right\ra$&$\left|\widehat{[0;0,0,2]}\right\ra$\\\hline
$1$&$\left|\widehat{[2;0,0,0]}\right\ra$&$\left|\widehat{[0;2,0,0]}\right\ra$&$\left|\widehat{[0;0,2,0]}\right\ra$&$\left|\widehat{[0;0,0,2]}\right\ra$\\\hline
$\eta$&$\left|\widehat{[0;2,0,0]}\right\ra$&$\left|\widehat{[0;0,2,0]}\right\ra$&$\left|\widehat{[0;0,0,2]}\right\ra$&$\left|\widehat{[2;0,0,0]}\right\ra$
\end{tabular}.
\end{center}
\end{table}
\hspace{-17pt}For the other sign $\lambda_\text{adj}<0$, the third line gives the minimal energy. The result suggests a TQFT with $\text{GSD}=2$. It is consistent with (\ref{LSMGSD}). The degeneracy is also a consequence of the spontaneously broken $\mbb Z_4$ symmetry:
\begin{table}[H]
\begin{center}
\begin{tabular}{c|c|c}
&$\left|\widehat{[1;0,1,0]}\right\ra$&$\left|\widehat{[0;1,0,1]}\right\ra$\\\hline
$1$&$\left|\widehat{[1;0,1,0]}\right\ra$&$\left|\widehat{[0;1,0,1]}\right\ra$\\\hline
$\eta$&$\left|\widehat{[0;1,0,1]}\right\ra$&$\left|\widehat{[1;0,1,0]}\right\ra$
\end{tabular}.
\end{center}
\end{table}

Our proposal is summarized as follows:
\begin{equation}
    \wf{su}(4)_2+\phi_{\widehat{[0;1,0,1]}}=\begin{cases}\text{TQFT w/ GSD}=4&(\lambda_\text{adj}>0),\\\text{TQFT w/ GSD}=2&(\lambda_\text{adj}<0).\end{cases}\label{su42}
\end{equation}

\subsubsection{$\wf{su}(5)_4+\phi_{\widehat{[2;1,0,0,1]}}$}
Finally, let us also study one example with $r=4$. For $k=4$, the Cardy's method yields
\[ E_a=\begin{cases}{\displaystyle\frac{32\pi/3}{24(2\tau_a)^2}-7.291\frac{\lambda_\text{adj}}{(2\tau_a)^{10/9}}}&(\hmu=4\ho_0,4\ho_1,4\ho_2,4\ho_3,4\ho_4),\\
{\displaystyle\frac{32\pi/3}{24(2\tau_a)^2}+0.8794\frac{\lambda_\text{adj}}{(2\tau_a)^{10/9}}}&(\hmu=\widehat{[2;0,1,1,0]},\widehat{[1;1,0,2,0]},\widehat{[1;0,2,0,1]},\widehat{[0;2,0,1,1]},\widehat{[0;1,1,0,2]}).\end{cases} \]
For $\lambda_\text{adj}>0$, the first set of Cardy states minimizes the energy. The presence of the identity $4\ho_0$ and the matching of fusion rules with those of $\wf{su}(5)_1$ signals our massless flow $\wf{su}(5)_4\to\wf{su}(5)_1$. For the other sign $\lambda_\text{adj}<0$, the second set of Cardy states minimizes the energy. It suggests a TQFT with $\text{GSD}=5$. The degeneracy is consistent with (\ref{LSMGSD}). It is a consequence of spontaneously broken $\mbb Z_5$ center symmetry:
\begin{table}[H]
\begin{center}
\begin{tabular}{c|c|c|c|c|c}
&$\left|\widehat{[2;0,1,1,0]}\right\ra$&$\left|\widehat{[1;1,0,2,0]}\right\ra$&$\left|\widehat{[1;0,2,0,1]}\right\ra$&$\left|\widehat{[0;2,0,1,1]}\right\ra$&$\left|\widehat{[0;1,1,0,2]}\right\ra$\\\hline
$1$&$\left|\widehat{[2;0,1,1,0]}\right\ra$&$\left|\widehat{[1;1,0,2,0]}\right\ra$&$\left|\widehat{[1;0,2,0,1]}\right\ra$&$\left|\widehat{[0;2,0,1,1]}\right\ra$&$\left|\widehat{[0;1,1,0,2]}\right\ra$\\\hline
$\eta$&$\left|\widehat{[0;2,0,1,1]}\right\ra$&$\left|\widehat{[0;1,1,0,2]}\right\ra$&$\left|\widehat{[1;1,0,2,0]}\right\ra$&$\left|\widehat{[1;0,2,0,1]}\right\ra$&$\left|\widehat{[2;0,1,1,0]}\right\ra$
\end{tabular}.
\end{center}
\end{table}

Our proposal is summarized as follows:
\begin{equation}
    \wf{su}(5)_4+\phi_{\widehat{[2;1,0,0,1]}}=\begin{cases}\wf{su}(5)_1&(\lambda_\text{adj}>0),\\\text{TQFT w/ GSD}=5&(\lambda_\text{adj}<0).\end{cases}\label{su54}
\end{equation}

\subsection{$\wf{so}(2r+1)_k$}

\subsubsection{$\wf{so}(7)_2+\phi_{\widehat{[0;0,1,0]}}$}
Under the adjoint deformation, the Cardy's method gives some energies
\[ E_a=\begin{cases}{\displaystyle\frac{6\pi}{24(2\tau_a)^2}-2\frac{\lambda_\text{adj}}{(2\tau_a)^{10/7}}}&(\hmu=2\ho_0,2\ho_1),\\
{\displaystyle\frac{6\pi}{24(2\tau_a)^2}+2\cos\frac\pi7\frac{\lambda_\text{adj}}{(2\tau_a)^{10/7}}}&(\hmu=\ho_2).\end{cases} \]
For $\lambda_\text{adj}>0$, the first line gives the minimal energy. The ground states contain the identity $2\ho_0$, and they obey the fusion rules of primaries in $\wf{so}(7)_1$. Thus, if the IR phases is gapless, a natural candidate is the $\wf{so}(7)_1$. On the other hand, for $\lambda_\text{adj}<0$, we see only one Cardy state gives the minimal energy. This signals trivially gapped vacuum without degeneracy. Indeed, the surviving $\mbb Z_2$ center symmetry acts unfaithfully on the Cardy state. This means the $\mbb Z_2$ symmetry is preserved.

Therefore, our proposal is summarized as
\begin{equation}
    \wf{so}(7)_2+\phi_{\widehat{[0;0,1,0]}}=\begin{cases}\wf{so}(7)_1&(\lambda_\text{adj}>0),\\
    \text{trivially gapped}&(\lambda_\text{adj}<0).\end{cases}\label{so72}
\end{equation}

\subsection{$\wf{sp}(2r)_k$}
\subsubsection{$\wf{sp}(6)_3+\phi_{\widehat{[1;2,0,0]}}$}
Under the adjoint deformation, the Cardy's method yields some energies
\[ \hspace{-50pt}E_a=\begin{cases}{\displaystyle\frac{9\pi}{24(2\tau_a)^2}-\frac{4(4\cos\frac\pi7-3)(1-\sin\frac\pi{14})}{6\sin\frac\pi{14}-1}\frac{\lambda_\text{adj}}{(2\tau_a)^{8/7}}}&(\hmu=3\ho_0,3\ho_3),\\
{\displaystyle\frac{9\pi}{24(2\tau_a)^2}+\frac{41-86\sin\frac\pi{14}+102\sin\frac{3\pi}{14}-94\cos\frac\pi7}{64\sin^5\frac\pi{14}(1+2\cos\frac\pi7)^2(1+2\sin\frac\pi{14})}\frac{\lambda_\text{adj}}{(2\tau_a)^{8/7}}}&(\hmu=\widehat{[2;0,0,1]},\widehat{[1;1,0,1]},\widehat{[1;0,1,1]},\widehat{[1;0,0,2]}).\end{cases} \]
For $\lambda_\text{adj}>0$, the first line gives the minimal energy. The ground states contain the identity $3\ho_0$, and they obey the fusion rules of two primaries in $\wf{sp}(6)_1$. This suggests the IR theory is gapless and described by $\wf{sp}(6)_1$. On the other hand, for $\lambda_\text{adj}<0$, the second line gives the minimal energy. The result suggests a TQFT with $\text{GSD}=4$. Since the surviving $\mbb Z_2$ acts faithfully on the ground states, it is spontaneously broken. This explains two-fold degeneracy. However, interestingly, the Cardy's method signals further two-fold degeneracy. Since the only surviving Verlinde lines are $\mbb Z_2$, we do not know where this additional degeneracy is coming from. It is desirable to understand its origin.

To sum up, our proposal is given as
\begin{equation}
    \wf{sp}(6)_3+\phi_{\widehat{[1;2,0,0]}}=\begin{cases}\wf{sp}(6)_1&(\lambda_\text{adj}>0),\\
    \text{TQFT w/ GSD}=4&(\lambda_\text{adj}<0).\end{cases}\label{sp63}
\end{equation}
The surviving Verlinde lines act on the ground states as follows:
\begin{table}[H]
\begin{center}
\begin{tabular}{c|c|c|c|c}
&$\left|\widehat{[2;0,0,1]}\right\ra$&$\left|\widehat{[1;1,0,1]}\right\ra$&$\left|\widehat{[1;0,1,1]}\right\ra$&$\left|\widehat{[1;0,0,2]}\right\ra$\\\hline
$1$&$\left|\widehat{[2;0,0,1]}\right\ra$&$\left|\widehat{[1;1,0,1]}\right\ra$&$\left|\widehat{[1;0,1,1]}\right\ra$&$\left|\widehat{[1;0,0,2]}\right\ra$\\\hline
$\eta$&$\left|\widehat{[1;0,0,2]}\right\ra$&$\left|\widehat{[1;0,1,1]}\right\ra$&$\left|\widehat{[1;1,0,1]}\right\ra$&$\left|\widehat{[2;0,0,1]}\right\ra$
\end{tabular}.
\end{center}
\end{table}

\subsection{$\wf{so}(4R)_k$}

\subsubsection{$\wf{so}(12)_3+\phi_{\widehat{[1;0,1,0,0,0,0]}}$}
Under the adjoint deformation, some relevant energies are given by\footnote{The analytic expressions for the coefficients are given by
\begin{align*}
    2\sin\frac{5 \pi}{26}\left(1+2\sin\frac{5\pi}{26}\right)\left(1+2\sin\frac{\pi}{26}-2\sin\frac{3\pi}{26}+2\sin\frac{5\pi}{26}+2\cos\frac{2\pi}{13}-2\cos\frac{3\pi}{13}\right)&\approx4.7128,\\
    \frac{44755+87674\sin\frac\pi{26}-75344\sin\frac{3\pi}{26}+61206\sin\frac{5\pi}{26}-82572\cos\frac\pi{13}+67630\cos\frac{2\pi}{13}-57568\cos\frac{3\pi}{13}}{390133+775158\sin\frac\pi{26}-741124\sin\frac{3\pi}{26}+702610\sin\frac{5\pi}{26}-761020\cos\frac\pi{13}+720048\cos\frac{2\pi}{13}-692776\cos\frac{3\pi}{13}}&\approx-1.18296.
\end{align*}}
\[ E_a=\begin{cases}{\displaystyle\frac{198\pi/13}{24(2\tau_a)^2}-4.7128\frac{\lambda_\text{adj}}{(2\tau_a)^{20/13}}}&(\hmu=3\ho_0,3\ho_1,3\ho_5,3\ho_6),\\
{\displaystyle\frac{198\pi/13}{24(2\tau_a)^2}+1.1830\frac{\lambda_\text{adj}}{(2\tau_a)^{20/13}}}&(\hmu=\ho_0+\ho_3,\ho_1+\ho_3,\ho_3+\ho_5,\ho_3+\ho_6).\end{cases} \]
For $\lambda_\text{adj}>0$, the first line gives the minimal energy. Since the ground states contain the identity $3\ho_0$ and they obey the fusion rules of primaries in $\wf{so}(12)_1$, this result suggests the massless scenario $\wf{so}(12)_3\to\wf{so}(12)_1$. For $\lambda_\text{adj}<0$, the second line gives the minimal energy. This suggests a TQFT with four-fold degenerated vacua. Since the four surviving Verlinde lines act faithfully on the four Cardy states as
\begin{table}[H]
\begin{center}
\begin{tabular}{c|c|c|c|c}
    &$\left|\widehat{[1;0,0,1,0,0,0]}\right\ra$&$\left|\widehat{[0;1,0,1,0,0,0]}\right\ra$&$\left|\widehat{[0;0,0,1,0,1,0]}\right\ra$&$\left|\widehat{[0;0,0,1,0,0,1]}\right\ra$\\\hline
    $1$&$\left|\widehat{[1;0,0,1,0,0,0]}\right\ra$&$\left|\widehat{[0;1,0,1,0,0,0]}\right\ra$&$\left|\widehat{[0;0,0,1,0,1,0]}\right\ra$&$\left|\widehat{[0;0,0,1,0,0,1]}\right\ra$\\\hline
    $\eta_A$&$\left|\widehat{[0;1,0,1,0,0,0]}\right\ra$&$\left|\widehat{[1;0,0,1,0,0,0]}\right\ra$&$\left|\widehat{[0;0,0,1,0,0,1]}\right\ra$&$\left|\widehat{[0;0,0,1,0,1,0]}\right\ra$\\\hline
    $\eta_{\widetilde A}$&$\left|\widehat{[0;0,0,1,0,0,1]}\right\ra$&$\left|\widehat{[0;0,0,1,0,1,0]}\right\ra$&$\left|\widehat{[0;1,0,1,0,0,0]}\right\ra$&$\left|\widehat{[1;0,0,1,0,0,0]}\right\ra$\\\hline
    $\eta_{A\widetilde A}$&$\left|\widehat{[0;0,0,1,0,1,0]}\right\ra$&$\left|\widehat{[0;0,0,1,0,0,1]}\right\ra$&$\left|\widehat{[1;0,0,1,0,0,0]}\right\ra$&$\left|\widehat{[0;1,0,1,0,0,0]}\right\ra$
\end{tabular},
\end{center}
\end{table}
\hspace{-17pt}the four-fold degeneracy is a result of spontaneously broken $\mbb Z_2\times\mbb Z_2$ symmetry.

Our proposal is summarized as follows:
\begin{equation}
    \wf{so}(12)_3+\phi_{\widehat{[1;0,1,0,0,0,0]}}=\begin{cases}\wf{so}(12)_1&(\lambda_\text{adj}>0),\\
    \text{TQFT w/ GSD}=4&(\lambda_\text{adj}<0).\end{cases}\label{so123}
\end{equation}

\subsection{$\wf{so}(4R+2)_k$}

\subsubsection{$\wf{so}(10)_3+\phi_{\widehat{[1;0,1,0,0,0]}}$}
Under the adjoint deformation, some relevant energies are given by\footnote{The analytic expressions for the coefficients are given by
\begin{align*}
    \left(1-2\sin\frac\pi{22}\right)\left(1+2\sin\frac{3\pi}{22}\right)\left(1+2\sin\frac{3\pi}{22}+2\cos\frac{2\pi}{11}\right)&\approx4.60149,\\
    \frac{1691-3332\sin\frac\pi{22}+3014\sin\frac{3\pi}{22}-2742\sin\frac{5\pi}{22}-3196\cos\frac\pi{11}+2844\cos\frac{2\pi}{11}}{-1369+2656\sin\frac\pi{22}-2148\sin\frac{3\pi}{22}+1728\sin\frac{5\pi}{22}+2436\cos\frac\pi{11}-1884\cos\frac{2\pi}{11}}&\approx-1.14055.
\end{align*}}
\[ E_a=\begin{cases}{\displaystyle\frac{135\pi/11}{24(2\tau_a)^2}-4.6015\frac{\lambda_\text{adj}}{(2\tau_a)^{16/11}}}&(\hmu=3\ho_0,3\ho_1,3\ho_4,3\ho_5),\\
{\displaystyle\frac{135\pi/11}{24(2\tau_a)^2}+1.1406\frac{\lambda_\text{adj}}{(2\tau_a)^{16/11}}}&(\hmu=\ho_0+\ho_3,\ho_1+\ho_3,\ho_2+\ho_4,\ho_2+\ho_5).\end{cases} \]
For $\lambda_\text{adj}>0$, the first line gives the minimal energy. Since the ground states contain the identity $3\ho_0$ and they obey the fusion rules of the primaries in $\wf{so}(10)_1$, the result suggests the massless scenario $\wf{so}(10)_3\to\wf{so}(10)_1$. For $\lambda_\text{adj}<0$, the second line gives the minimal energy. This suggests a TQFT with four-fold degenerated vacua. The ground state degeneracy turns out to be a consequence of spontaneously broken $\mbb Z_4$ symmetry. Indeed, the $\mbb Z_4$ symmetry acts faithfully on the ground states:
\begin{table}[H]
\begin{center}
\begin{tabular}{c|c|c|c|c}
    &$\left|\widehat{[1;0,0,1,0,0]}\right\ra$&$\left|\widehat{[0;1,0,1,0,0]}\right\ra$&$\left|\widehat{[0;0,1,0,1,0]}\right\ra$&$\left|\widehat{[0;0,1,0,0,1]}\right\ra$\\\hline
    $1$&$\left|\widehat{[1;0,0,1,0,0]}\right\ra$&$\left|\widehat{[0;1,0,1,0,0]}\right\ra$&$\left|\widehat{[0;0,1,0,1,0]}\right\ra$&$\left|\widehat{[0;0,1,0,0,1]}\right\ra$\\\hline
    $\eta$&$\left|\widehat{[0;0,1,0,0,1]}\right\ra$&$\left|\widehat{[0;0,1,0,1,0]}\right\ra$&$\left|\widehat{[1;0,0,1,0,0]}\right\ra$&$\left|\widehat{[0;1,0,1,0,0]}\right\ra$
\end{tabular}.
\end{center}
\end{table}
\hspace{-17pt}Thus, the $\mbb Z_4$ symmetry is spontaneously broken.

Our proposal is summarized as follows:
\begin{equation}
    \wf{so}(10)_3+\phi_{\widehat{[1;0,1,0,0,0]}}=\begin{cases}\wf{so}(10)_1&(\lambda_\text{adj}>0),\\
    \text{TQFT w/ GSD}=4&(\lambda_\text{adj}<0).\end{cases}\label{so103}
\end{equation}

\subsection{$(\wf{e_6})_k$}
\subsubsection{$(\wf{e_6})_2+\phi_{\widehat{[0;0,0,0,0,0,1]}}$}
Since we only computed $S$-matrices up to $k=2$, we apply the Cardy's method to $(\wf{e_6})_2$ WZW model. The energies are given by (with rescaled coupling)
\[ E_a=\begin{cases}{\displaystyle\frac{78\pi/7}{24(2\tau_a)^2}-2\cos\frac\pi7\frac{\lambda_\text{adj}}{(2\tau_a)^{12/7}}}&(\hmu=2\ho_0,2\ho_1,2\ho_5),\\
{\displaystyle\frac{78\pi/7}{24(2\tau_a)^2}-2\sin\frac\pi{14}\frac{\lambda_\text{adj}}{(2\tau_a)^{12/7}}}&(\hmu=\ho_0+\ho_1,\ho_0+\ho_5,\ho_1+\ho_5),\\
{\displaystyle\frac{78\pi/7}{24(2\tau_a)^2}+2\sin\frac{3\pi}{14}\frac{\lambda_\text{adj}}{(2\tau_a)^{12/7}}}&(\hmu=\ho_2,\ho_4,\ho_6).\end{cases} \]
For $\lambda_\text{adj}>0$, the first line gives the smallest energy. The three Cardy states contain the identity $2\ho_0$. Furthermore, three primaries corresponding to the three Cardy states obey the same fusion rules as those in $(\wf{e_6})_1$. This suggests the IR theory is gapless and described by $(\wf{e_6})_1$. On the other hand, for $\lambda_\text{adj}<0$, the third line gives the smallest energy. This signals a TQFT with ground state degeneracy three. The three-fold degeneracy of ground states is a result of spontaneously broken $\mbb Z_3$ symmetry. Indeed, the $\mbb Z_3$ symmetry acts faithfully on ground states:
\begin{table}[H]
\begin{center}
\begin{tabular}{c|c|c|c}
    &$\left|\widehat{[0;0,1,0,0,0,0]}\right\ra$&$\left|\widehat{[0;0,0,0,1,0,0]}\right\ra$&$\left|\widehat{[0;0,0,0,0,0,1]}\right\ra$\\\hline
    $1$&$\left|\widehat{[0;0,1,0,0,0,0]}\right\ra$&$\left|\widehat{[0;0,0,0,1,0,0]}\right\ra$&$\left|\widehat{[0;0,0,0,0,0,1]}\right\ra$\\\hline
    $\eta$&$\left|\widehat{[0;0,0,0,0,0,1]}\right\ra$&$\left|\widehat{[0;0,1,0,0,0,0]}\right\ra$&$\left|\widehat{[0;0,0,0,1,0,0]}\right\ra$
\end{tabular}.
\end{center}
\end{table}

Our proposal is summarized as follows:
\begin{equation}
    (\wf{e_6})_2+\phi_{\widehat{[0;0,0,0,0,0,1]}}=\begin{cases}(\wf{e_6})_1&(\lambda_\text{adj}>0),\\
    \text{TQFT w/ GSD}=3&(\lambda_\text{adj}<0).\end{cases}\label{e62}
\end{equation}

\subsection{$(\wf{e_7})_k$}
\subsubsection{$(\wf{e_7})_3+\phi_{\widehat{[1;1,0,0,0,0,0,0]}}$}
We apply the Cardy's method to the deformation problem of $(\wf{e_7})_3$ WZW model. The relevant energies are given by (with rescaled coupling)
\[ E_a=\begin{cases}{\displaystyle\frac{19\pi}{24(2\tau_a)^2}-\frac{3+\sqrt{21}}2\frac{\lambda_\text{adj}}{(2\tau_a)^{12/7}}}&(\hmu=3\ho_0,3\ho_6),\\
{\displaystyle\frac{19\pi}{24(2\tau_a)^2}+\frac{2a}{3+\sqrt{21}}\frac{\lambda_\text{adj}}{(2\tau_a)^{12/7}}}&(\hmu=\ho_2,\ho_4).\end{cases} \]
The number $a$ in the numerator is given at (\ref{Se7}). For $\lambda_\text{adj}>0$, the first line gives the minimal energy. The Cardy states contain the identity $3\ho_0$, and their fusion rules are identical to those of $(\wf{e_7})_1$. This suggests that the positive relevant coupling triggers the massless flow $(\wf{e_7})_3\to(\wf{e_7})_1$. On the other hand, for $\lambda_\text{adj}<0$, the second line gives the minimal energy. The result signals a TQFT with two degenerated vacua. The degeneracy is a consequence of spontaneously broken $\mbb Z_2$ symmetry. To see this, we compute the action of the $\mbb Z_2$ symmetry on the ground states:
\begin{table}[H]
\begin{center}
\begin{tabular}{c|c|c}
    &$\left|\widehat{[0;0,1,0,0,0,0,0]}\right\ra$&$\left|\widehat{[0;0,0,0,1,0,0,0]}\right\ra$\\\hline
    $1$&$\left|\widehat{[0;0,1,0,0,0,0,0]}\right\ra$&$\left|\widehat{[0;0,0,0,1,0,0,0]}\right\ra$\\\hline
    $\eta$&$\left|\widehat{[0;0,0,0,1,0,0,0]}\right\ra$&$\left|\widehat{[0;0,1,0,0,0,0,0]}\right\ra$
\end{tabular}.
\end{center}
\end{table}
\hspace{-17pt}Since the symmetry acts faithfully, it is spontaneously broken. This explains the two-fold degeneracy in the massive phase.

Our proposal is summarized as follows:
\begin{equation}
    (\wf{e_7})_3+\phi_{\widehat{[1;1,0,0,0,0,0,0]}}=\begin{cases}(\wf{e_7})_1&(\lambda_\text{adj}>0),\\\text{TQFT w/ GSD}=2&(\lambda_\text{adj}<0).\end{cases}\label{e73}
\end{equation}

\section{Discussion}
We studied RG flows from WZW models, but left many interesting questions. In this section, we list some of them.

\begin{itemize}
    \item Charge conjugation:\\
    In massless RG flows between $A_r$ type WZW models, we fixed the ambiguity in identifications of surviving objects with the help of known mixed anomaly. Concretely, in healthy scenarios satisfying the half-integral conditions, we always saw an object $j$ in UV is identified with its charge conjugate $j^*$ in IR. In the RG interface picture, this means the interface effectively performs charge conjugation. In other words, after the folding trick, the product object $jj^*$ can end topologically on the RG boundary. In search of invariant boundary state, a ``mysterious'' appearance of charge conjugation is known \cite{NY17}. It is interesting to see whether their appearance is a coincidence or there is a deeper reason.
    \item Mixed anomaly:\\
    Related to the point above, it is desirable to study mixed anomalies in $BCDE$ type WZW models following \cite{TS18}. They will fix ambiguities in the identification of Verlinde objects.
    \item Conjecture:\\
    We provided a concrete example --- $\wf{su}(5)_2+\phi_{\widehat{[0;1,0,0,1]}}$ --- which could disprove or strongly support our conjecture. Since the conformal dimension of the relevant operator is smaller than $3/4$, the truncated conformal space approach is applicable to this problem.
    \item Origin of further two-fold degeneracy:\\
    In the deformation problem $\wf{sp}(6)_3+\phi_{\widehat{1;2,0,0}}$, we found four-fold degeneracy for $\lambda_\text{adj}<0$. The two-fold degeneracy can be understood as a result of spontaneously broken $\mbb Z_2$ center symmetry, but the Cardy's method suggests further two-fold degeneracy. It is desirable to understand the origin of the ``mysterious'' degeneracy.
    \item More general relevant deformation:\\
    In this paper, we only considered relevant deformations with adjoint primaries. In all examples we studied, we found just invertible symmetry objects of UV MTCs are preserved. It is interesting to study more general deformations. They would preserve non-invertible objects, and possibly lead to interesting physics. For example, $\phi_{\widehat{[0;0,2]}}$ in $\wf{su}(3)_2$ preserves the Fibonacci object, and the IR cannot be trivial. A natural scenario is a TQFT with $\text{GSD}=2$.
    \item Gauging/fermionization:\\
    In some examples, we had anomaly-free subgroups. For instance, the $\mbb Z_4$ center symmetry of $\wf{su}(4)_k$ always has anomaly-free $\mbb Z_2$ subgroup. Similarly, $\mbb Z_2^A\subset\mbb Z_2^A\times\mbb Z_2^{\widetilde A}$ of $\wf{so}(4R)_k$ and $\mbb Z_2\subset\mbb Z_4$ of $\wf{so}(4R+2)_k$ are always anomaly-free. Since there is no obstruction, we can gauge them. (Generically, gauging leads to categorical quantum symmetry \cite{V89,BT17}.) Furthermore, because our relevant operators are singlets under the subgroups, the gauging and the relevant deformation commute (see, e.g. \cite{KK22II}). Therefore, our scenarios immediately lead to massless scenarios in gauged theories:
    \begin{align*}
        \wf{su}(4)_3/\mbb Z_2&\to\wf{su}(4)_1/\mbb Z_2,\\
        \wf{so}(12)_5/\mbb Z_2^A\to\wf{so}(12)_3&/\mbb Z_2^A\to\wf{so}(12)_1/\mbb Z_2^A,\\
        \wf{so}(10)_5/\mbb Z_2\to\wf{so}(10)_3&/\mbb Z_2\to\wf{so}(10)_1/\mbb Z_2.
    \end{align*}
    Relatedly, one can use the anomaly-free $\mbb Z_2$ subgroups to fermionize \cite{T18} the WZW models. Our scenarios also lead to massless flows between fermionic WZW models. It would be interesting to study RG flows in gauged/fermionized theories in the context of emergent SUSY \cite{KKSUSY}.
\end{itemize}
We would like to report progresses on these problems in a future.

\section*{Acknowledgement}
We thank Yu Nakayama and Yuji Tachikawa for helpful discussions. We also appreciate comments from Yu Nakayama on the draft.

\appendix
\setcounter{section}{0}
\renewcommand{\thesection}{\Alph{section}}
\setcounter{equation}{0}
\renewcommand{\theequation}{\Alph{section}.\arabic{equation}}

\section{Review of WZW models}\label{review}
In this appendix, we provide some backgrounds on WZW models to understand our results. We also collect some data on simple Lie algebras.

A $\hg_k$ WZW model with simple Lie algebra $\mfrak g$ is defined with the action integral
\begin{equation}
    S_{\hg_k}[U]=\frac k{8\pi}\int_\Sigma d^2x\tr\left(\partial_\mu U\partial^\mu U^\dagger\right)-\frac{ik}{12\pi}\int_Bd^3y\epsilon^{\mu\nu\rho}\tr\left(U^\dagger\partial_\mu U\cdot U^\dagger\partial_\nu U\cdot U^\dagger\partial_\rho U\right),\label{WZWaction}
\end{equation}
where the elementary field $U$ takes values in group manifold $G$ associated with the Lie algebra $\mfrak g$. Originally, it is defined on $\Sigma$, but in the second term, it is extended to $B$ bounded by $\Sigma$, i.e., $\partial B=\Sigma$. For the model to be independent of the extension, $k$ has to be integral (for the models we consider). The integer is called the level. We take the level $k$ positive.

The model has global symmetry $(G_L\times G_R)/\Gamma$. $\Gamma$ is the center of $G$. The symmetry acts on $U$ as $U\mapsto g_LUg_R^\dagger$ with $(g_L,g_R)\in G_L\times G_R$. Since elements of the center $\Gamma$ commute with $U$, faithful part is given by the quotient $(G_L\times G_R)/\Gamma$.

It is known that the model is a diagonal RCFT. Its central charge is given by
\begin{equation}
    c=\frac{k\dim\mfrak g}{k+g}.\label{gkcentralcharge}
\end{equation}
Here, $g$ is the dual Coxeter number. Some data of simple Lie algebras are summarized in the following table:\footnote{It seems the Appendix 13. A. of \cite{FMS} has typos in the highest roots of $\mfrak e_8,\mfrak g_2$.}
\begin{table}[H]
%\begin{center}
\hspace{-40pt}
\begin{tabular}{c|c|c|c|c|c|c}
Type&$\mfrak g$&$\dim\mfrak g$&Rank&Dual Coxeter number $g$&Highest root $\theta$&Center $\Gamma$\\\hline
$A_{r\ge2}$&$\mfrak{su}(r+1)$&$r^2+2r$&$r$&$r+1$&$(1,0,\dots,0,1)$&$\mbb Z_{r+1}$\\\hline
$B_{r\ge3}$&$\mfrak{so}(2r+1)$&$2r^2+r$&$r$&$2r-1$&$(0,1,0,\dots,0)$&$\mbb Z_2$\\\hline
$C_{r\ge2}$&$\mfrak{sp}(2r)$&$2r^2+r$&$r$&$r+1$&$(2,0,\dots,0)$&$\mbb Z_2$\\\hline
$D_{r\ge4}$&$\mfrak{so}(2r)$&$2r^2-r$&$r$&$2r-2$&$(0,1,0,\dots,0)$&$\left\{\begin{array}{cc}\mbb Z_2\times\mbb Z_2&(\text{even }r)\\\mbb Z_4&(\text{odd }r)\end{array}\right.$\\\hline
$E_6$&$\mfrak e_6$&78&6&12&$(0,0,0,0,0,1)$&$\mbb Z_3$\\\hline
$E_7$&$\mfrak e_7$&133&7&18&$(1,0,0,0,0,0,0)$&$\mbb Z_2$\\\hline
$E_8$&$\mfrak e_8$&248&8&30&$(0,0,0,0,0,0,1,0)$&$\mbb Z_1$\\\hline
$F_4$&$\mfrak f_4$&52&4&9&$(1,0,0,0)$&$\mbb Z_1$\\\hline
$G_2$&$\mfrak g_2$&14&2&4&$(0,1)$&$\mbb Z_1$
\end{tabular}.
%\end{center}
\caption{Some data of simple Lie algebras}\label{Liedata}
\end{table}
\hspace{-17pt}One immediate conclusion we can draw from the central charge is the monotonicity of the level
\begin{equation}
    k'<k\label{kmonotone}
\end{equation}
under massless flow $\hg_k\to\hg_{k'}$.\footnote{A quick way to show this is to formally compute the derivative of $c$ with respect to $k$:
\[ \tdif ck=\frac{g\dim\mfrak g}{(k+g)^2}. \]
Since this is positive, $k'<k$ is required by the $c$-theorem.}

The primaries of the model are labeled by affine weights $\hmu$. In the basis of fundamental weights $\ho_j$, an affine weight is expanded as
\[ \hmu=\sum_{j=0}^r\mu_j\ho_j, \]
where $r$ is the rank of $\mfrak g$. We also write
\begin{equation}
    \hmu=[\mu_0;\mu_1,\dots,\mu_r].\label{affineweight}
\end{equation}
Its finite counterpart, weight, is obtained by removing the zero-th component $\mu_0$, $\mu=(\mu_1,\mu_2,\dots,\mu_r)$. The integers $\mu_j$'s are called (affine) Dynkin labels. They are constrained by the level
\[ k=\mu_0+\sum_{j=1}^ra_j^\vee\mu_j, \]
where $a_j^\vee$'s are comarks (a.k.a. dual Kac labels).\footnote{Comarks for each simple Lie algebra can be found in Appendix 13. A. of \cite{FMS}. With the comarks, the dual Coxeter number is defined as $g:=\sum_{j=1}^ra_j^\vee+1$.} Thus, for a finite level $k$, distinct primaries are labeled by a finite set of affine dominant weights
\begin{equation}
    P_+^k:=\left\{\hmu\Big|0\le\mu_j\ \&\ 0\le\sum_{j=1}^ra_j^\vee\mu_j\le k\right\}.\label{Pk+}
\end{equation}
A primary with affine weight $\hmu$ has conformal dimension
\begin{equation}
    h_\hmu=\frac{(\mu,\mu+2\rho)}{2(k+g)},\label{hhmu}
\end{equation}
where $\rho=(1,1,\dots,1)$ is the Weyl vector (a.k.a. principal vector). The scalar product in the numerator is given by
\[ (\mu,\lambda)=\sum_{i,j=1}^r\mu_i\lambda_jF_{ij}, \]
where the quadratic form matrix
\begin{equation}
    F_{ij}:=(\omega_i,\omega_j)=F_{ji}\label{Fij}
\end{equation}
for each simple Lie algebra, in the basis $\{\omega_1,\omega_2,\dots,\omega_r\}$, is given as follows (see e.g. Appendix 13. A. of \cite{FMS}):\newpage
\begin{equation}
\begin{split}
    F_{\mfrak{su}(r+1)}&=\frac1{r+1}\begin{pmatrix}r&r-1&r-2&\dots&2&1\\r-1&2(r-1)&2(r-2)&\dots&4&2\\r-2&2(r-2)&3(r-2)&\dots&6&3\\\vdots&\vdots&\vdots&\ddots&\vdots&\vdots\\2&4&6&\cdots&2(r-1)&r-1\\1&2&3&\cdots&r-1&r\end{pmatrix},\\
    F_{\mfrak{so}(2r+1)}&=\frac12\begin{pmatrix}2&2&2&\dots&2&1\\2&4&4&\dots&4&2\\2&4&6&\dots&6&3\\\vdots&\vdots&\vdots&\ddots&\vdots&\vdots\\2&4&6&\cdots&2(r-1)&r-1\\1&2&3&\cdots&r-1&\frac r2\end{pmatrix},\\
    F_{\mfrak{sp}(2r)}&=\frac12\begin{pmatrix}1&1&1&\dots&1&1\\1&2&2&\dots&2&2\\1&2&3&\dots&3&3\\\vdots&\vdots&\vdots&\ddots&\vdots&\vdots\\1&2&3&\cdots&r-1&r-1\\1&2&3&\cdots&r-1&r\end{pmatrix},\\
    F_{\mfrak{so}(2r)}&=\frac12\begin{pmatrix}2&2&2&\dots&2&1&1\\2&4&4&\dots&4&2&2\\2&4&6&\dots&6&3&3\\\vdots&\vdots&\vdots&\ddots&\vdots&\vdots&\vdots\\2&4&6&\cdots&2(r-2)&r-2&r-2\\1&2&3&\cdots&r-2&\frac r2&\frac{r-2}2\\1&2&3&\cdots&r-2&\frac{r-2}2&\frac r2\end{pmatrix},\\
    F_{\mfrak e_6}&=\frac13\begin{pmatrix}4&5&6&4&2&3\\5&10&12&8&4&6\\6&12&18&12&6&9\\4&8&12&10&5&6\\2&4&6&5&4&3\\3&6&9&6&3&6\end{pmatrix},\\
    F_{\mfrak e_7}&=\frac12\begin{pmatrix}4&6&8&6&4&2&4\\6&12&16&12&8&4&8\\8&16&24&18&12&6&12\\6&12&18&15&10&5&9\\4&8&12&10&8&4&6\\2&4&6&5&4&3&3\\4&8&12&9&6&3&7\end{pmatrix}.
\end{split}\label{quadraticF}
\end{equation}
A primary operator important for our purposes is the adjoint. It is labeled by the highest root $\theta$. The highest root has scalar product (see equation (13.128) of \cite{FMS})
\[ (\theta,\theta+2\rho)=2g. \]
Thus, the primary in the adjoint has conformal dimension
\begin{equation}
    h_{\hat\theta}=\frac g{k+g}.\label{hadjoint}
\end{equation}
It is relevant for positive $k$. We have also included the highest root in the Table \ref{Liedata}. As one notices from the Dynkin labels, adjoints do not exist for small levels. (For example, $\wf{su}(3)_1$ has no adjoint.)

When $\hg$ has nontrivial center $\Gamma$, it acts on primaries. The primitive element of the center acts as
\begin{equation}
    \phi_\hmu\mapsto e^{-2\pi i(A\ho_0,\mu)}\phi_\hmu,\label{center}
\end{equation}
where $A$ is an outer automorphism $A\in\mcal O(\hg)\simeq\Gamma$. The action of primitive outer automorphisms are summarized in the following table:
\begin{table}[H]
\begin{center}
%\hspace{-30pt}
\begin{tabular}{c|c|c}
Type&$\mcal O(\hg)\simeq\Gamma$&$A\hmu$\\\hline
$A_r$&$\mbb Z_{r+1}$&$A[\mu_0;\mu_1,\cdots,\mu_{r-1},\mu_r]=[\mu_r;\mu_0,\cdots,\mu_{r-2},\mu_{r-1}]$\\
$B_r$&$\mbb Z_2$&$A[\mu_0;\mu_1,\mu_2,\cdots,\mu_r]=[\mu_1;\mu_0,\mu_2,\cdots,\mu_r]$\\
$C_r$&$\mbb Z_2$&$A[\mu_0;\mu_1,\cdots,\mu_r]=[\mu_r;\mu_{r-1},\cdots,\mu_0]$\\
$D_{r=2R}$&$\mbb Z_2^A\times\mbb Z_2^{\widetilde A}$&$\begin{cases}A[\mu_0;\mu_1,\mu_2,\cdots,\mu_{r-2},\mu_{r-1},\mu_r]=[\mu_1;\mu_0,\mu_2,\cdots,\mu_{r-2},\mu_r,\mu_{r-1}]\\\tilde A[\mu_0;\mu_1,\mu_2,\cdots,\mu_{r-2},\mu_{r-1},\mu_r]=[\mu_r;\mu_{r-1},\mu_{r-2},\cdots,\mu_2,\mu_1,\mu_0]\end{cases}$\\
$D_{r=2R+1}$&$\mbb Z_4$&$A[\mu_0;\mu_1,\mu_2,\cdots,\mu_{r-2},\mu_{r-1},\mu_r]=[\mu_{r-1};\mu_r,\mu_{r-2},\cdots,\mu_2,\mu_1,\mu_0]$\\
$E_6$&$\mbb Z_3$&$A[\mu_0;\mu_1,\mu_2,\mu_3,\mu_4,\mu_5,\mu_6]=[\mu_1;\mu_5,\mu_4,\mu_3,\mu_6,\mu_0,\mu_2]$\\
$E_7$&$\mbb Z_2$&$A[\mu_0;\mu_1,\mu_2,\mu_3,\mu_4,\mu_5,\mu_6,\mu_7]=[\mu_6;\mu_5,\mu_4,\mu_3,\mu_2,\mu_1,\mu_0,\mu_7]$
\end{tabular}.
\end{center}
\caption{Action of outer automorphism}\label{outerauto}
\end{table}
\hspace{-17pt}Given the data, one can prove the adjoints are invariant under the center symmetries of $ABCDE$ type models.

\section{Some $S$-matrices}\label{Smatrix}
In this appendix, we list some $S$-matrices. (The results for $E$ type have not appeared in literature before to the best of our knowledge.) Some matrices are too large to write. (For instance, $\wf{so}(12)_3$ has $32\times32$ matrix.) For those cases, we instead write submatrices corresponding to the center symmetries. We give those for generic level $k$. The bases of the $S$-matrices are given in `dictionary order.' For example, in case of $\wf{su}(3)_1$, the bases are ordered as $\{\ho_0,\ho_1,\ho_2\}$.

There are various ways to compute $S$-matrices. The first direct method is to use the definition\footnote{Note that we are following the convention in mathematics. In physics literature, it is common to take dual of one object:
\[ \left(\widetilde S_\text{top}^\text{physics}\right)_{ij}=\sum_k{N_{ij^*}}^k\frac{e^{2\pi ih_k}}{e^{2\pi ih_i}e^{2\pi ih_{j^*}}}d_k. \]
We stick to the math convention.}
\[ \left(\widetilde S_\text{top}\right)_{ij}=\sum_k{N_{ij}}^k\frac{e^{2\pi ih_k}}{e^{2\pi ih_i}e^{2\pi ih_j}}d_k, \]
where $h_j$ and $d_j$ are conformal dimension and quantum dimension of an object $j$, respectively. This computation can be executed if fusion coefficients ${N_{ij}}^k$ (and $h_j,d_j$) are known. The fusion rules can be computed with the help of, say, LieART \cite{LieART} and Weyl reflections. Another method is to employ the Kac-Peterson formula \cite{KP84} or (14.217) of the yellow book \cite{FMS}. The third way is to simply use available software. (For example, we know one called KAC \cite{KAC}.) We employ all these available methods at our disposal.

We use these $S$-matrices to constrain RG flows. We also use them in section \ref{massive} to fix sign of relevant couplings and ground state degeneracies in massive scenarios.

\subsection{$\wf{su}(r+1)_k$}
\begin{equation}
\begin{split}
    \widetilde S_{\wf{su}(3)_1}&=\begin{pmatrix}1&1&1\\1&\omega^2&\omega\\1&\omega&\omega^2\end{pmatrix},\\
    \widetilde S_{\wf{su}(3)_2}&=\begin{pmatrix}1&\zeta&\zeta&1&\zeta&1\\\zeta&e^{-\frac{\pi i}3}&e^{\frac{\pi i}3}&\omega^2\zeta&-1&\omega\zeta\\\zeta&e^{\frac{\pi i}3}&e^{-\frac{\pi i}3}&\omega\zeta&-1&\omega^2\zeta\\1&\omega^2\zeta&\omega\zeta&\omega&\zeta&\omega^2\\\zeta&-1&-1&\zeta&-1&\zeta\\1&\omega\zeta&\omega^2\zeta&\omega^2&\zeta&\omega\end{pmatrix},\\
    \widetilde S_{\wf{su}(4)_k}\Big|_{\mbb Z_4}&=\begin{pmatrix}1&1&1&1\\1&e^{-\frac{k\pi i}2}&(-1)^k&e^{\frac{k\pi i}2}\\1&(-1)^k&1&(-1)^k\\1&e^{\frac{k\pi i}2}&(-1)^k&e^{-\frac{k\pi i}2}\end{pmatrix},\\
    \widetilde S_{\wf{su}(5)_k}\Big|_{\mbb Z_5}&=\begin{pmatrix}1&1&1&1&1\\1&e^{-\frac{2k\pi i}5}&e^{-\frac{4k\pi i}5}&e^{\frac{4k\pi i}5}&e^{\frac{2k\pi i}5}\\1&e^{-\frac{4k\pi i}5}&e^{\frac{2k\pi i}5}&e^{-\frac{2k\pi i}5}&e^{\frac{4k\pi i}5}\\1&e^{\frac{4k\pi i}5}&e^{-\frac{2k\pi i}5}&e^{\frac{2k\pi i}5}&e^{-\frac{4k\pi i}5}\\1&e^{\frac{2k\pi i}5}&e^{\frac{4k\pi i}5}&e^{-\frac{4k\pi i}5}&e^{-\frac{2k\pi i}5}\end{pmatrix},
\end{split}\label{Ssu}
\end{equation}
where $\omega:=e^{\frac{2\pi i}3},\zeta:=\frac{1+\sqrt5}2$.

\subsection{$\wf{so}(2r+1)_k$}
\begin{equation}
\begin{split}
    \widetilde S_{\wf{so}(7)_1}&=\begin{pmatrix}1&1&\sqrt2\\1&1&-\sqrt2\\\sqrt2&-\sqrt2&0\end{pmatrix},\\
    \widetilde S_{\wf{so}(7)_2}&=\begin{pmatrix}1&2&\sqrt7&1&\sqrt7&2&2\\2&4\cos\frac{2\pi}7&0&2&0&-4\cos\frac{3\pi}7&-4\cos\frac\pi7\\\sqrt7&0&\sqrt7&-\sqrt7&-\sqrt7&0&0\\1&2&-\sqrt7&1&-\sqrt7&2&2\\\sqrt7&0&-\sqrt7&-\sqrt7&\sqrt7&0&0\\2&-4\cos\frac{3\pi}7&0&2&0&-4\cos\frac\pi7&4\cos\frac{2\pi}7\\2&-4\cos\frac\pi7&0&2&0&4\cos\frac{2\pi}7&-4\cos\frac{3\pi}7\end{pmatrix}.
\end{split}\label{Ssoodd}
\end{equation}

\subsection{$\wf{sp}(2r)_k$}
\begin{equation}
\begin{split}
    \widetilde S_{\wf{sp}(6)_1}&=\begin{pmatrix}1&\zeta&\zeta&1\\\zeta&1&-1&-\zeta\\\zeta&-1&-1&\zeta\\1&-\zeta&\zeta&-1\end{pmatrix},\\
    \widetilde S_{\wf{sp}(2r)_k}\Big|_{\mbb Z_2}&=\begin{pmatrix}1&1\\1&(-1)^{kr}\end{pmatrix}.
\end{split}\label{Ssp}
\end{equation}

\subsection{$\wf{so}(2r)_k$}
\begin{equation}
\begin{split}
    \widetilde S_{\wf{so}(12)_1}&=\begin{pmatrix}1&1&1&1\\1&1&-1&-1\\1&-1&-1&1\\1&-1&1&-1\end{pmatrix},\\
    \widetilde S_{\wf{so}(4R)_k}\Big|_{\mbb Z_2\times\mbb Z_2}&=\begin{pmatrix}1&1&1&1\\1&1&(-1)^k&(-1)^k\\1&(-1)^k&(-1)^{kR}&(-1)^{k(1-R)}\\1&(-1)^k&(-1)^{k(1-R)}&(-1)^{kR}\end{pmatrix},\\
    \widetilde S_{\wf{so}(10)_1}&=\begin{pmatrix}1&1&1&1\\1&1&-1&-1\\1&-1&i&-i\\1&-1&-i&i\end{pmatrix},\\
    \widetilde S_{\wf{so}(4R+2)_k}\Big|_{\mbb Z_4}&=\begin{pmatrix}1&1&1&1\\1&1&(-1)^k&(-1)^k\\1&(-1)^k&e^{\frac{\pi i}2k(1-2R)}&e^{-\frac{\pi i}2k(1+2R)}\\1&(-1)^k&e^{-\frac{\pi i}2k(1+2R)}&e^{\frac{\pi i}2k(1-2R)}\end{pmatrix}.
\end{split}\label{Ssoeven}
\end{equation}

\subsection{$(\wf{e_6})_k$}
\begin{equation}
\scalebox{0.85}{$\begin{split}
    \widetilde S_{(\wf{e_6})_1}&=\begin{pmatrix}1&1&1\\1&e^{2\pi i/3}&e^{4\pi i/3}\\1&e^{4\pi i/3}&e^{2\pi i/3}\end{pmatrix},\\
    \widetilde S_{(\wf{e_6})_2}&=\begin{pmatrix}1&\frac{\sin\frac{3\pi}7}{\sin\frac\pi7}&\frac{\sin\frac{3\pi}7}{\sin\frac\pi7}&1&\frac{\sin\frac{3\pi}7}{\sin\frac\pi7}&2\cos\frac\pi7&2\cos\frac\pi7&1&2\cos\frac\pi7\\\frac{\sin\frac{3\pi}7}{\sin\frac\pi7}&2e^{\frac{\pi i}3}\cos\frac\pi7&2e^{-\frac{\pi i}3}\cos\frac\pi7&e^{\frac{2\pi i}3}\frac{\sin\frac{3\pi}7}{\sin\frac\pi7}&-2\cos\frac\pi7&e^{\frac{2\pi i}3}&e^{-\frac{2\pi i}3}&e^{-\frac{2\pi i}3}\frac{\sin\frac{3\pi}7}{\sin\frac\pi7}&1\\\frac{\sin\frac{3\pi}7}{\sin\frac\pi7}&2e^{-\frac{\pi i}3}\cos\frac\pi7&2e^{\frac{\pi i}3}\cos\frac\pi7&e^{-\frac{2\pi i}3}\frac{\sin\frac{3\pi}7}{\sin\frac\pi7}&-2\cos\frac\pi7&e^{-\frac{2\pi i}3}&e^{\frac{2\pi i}3}&e^{\frac{2\pi i}3}\frac{\sin\frac{3\pi}7}{\sin\frac\pi7}&1\\1&e^{\frac{2\pi i}3}\frac{\sin\frac{3\pi}7}{\sin\frac\pi7}&e^{-\frac{2\pi i}3}\frac{\sin\frac{3\pi}7}{\sin\frac\pi7}&e^{-\frac{2\pi i}3}&\frac{\sin\frac{3\pi}7}{\sin\frac\pi7}&2e^{-\frac{2\pi i}3}\cos\frac\pi7&2e^{\frac{2\pi i}3}\cos\frac\pi7&e^{\frac{2\pi i}3}&2\cos\frac\pi7\\\frac{\sin\frac{3\pi}7}{\sin\frac\pi7}&-2\cos\frac\pi7&-2\cos\frac\pi7&\frac{\sin\frac{3\pi}7}{\sin\frac\pi7}&-2\cos\frac\pi7&1&1&\frac{\sin\frac{3\pi}7}{\sin\frac\pi7}&1\\2\cos\frac\pi7&e^{\frac{2\pi i}3}&e^{-\frac{2\pi i}3}&2e^{-\frac{2\pi i}3}\cos\frac\pi7&1&e^{\frac{\pi i}3}\frac{\sin\frac{3\pi}7}{\sin\frac\pi7}&e^{-\frac{\pi i}3}\frac{\sin\frac{3\pi}7}{\sin\frac\pi7}&2e^{\frac{2\pi i}3}\cos\frac\pi7&-\frac{\sin\frac{3\pi}7}{\sin\frac\pi7}\\2\cos\frac\pi7&e^{-\frac{2\pi i}3}&e^{\frac{2\pi i}3}&2e^{\frac{2\pi i}3}\cos\frac\pi7&1&e^{-\frac{\pi i}3}\frac{\sin\frac{3\pi}7}{\sin\frac\pi7}&e^{\frac{\pi i}3}\frac{\sin\frac{3\pi}7}{\sin\frac\pi7}&2e^{-\frac{2\pi i}3}\cos\frac\pi7&-\frac{\sin\frac{3\pi}7}{\sin\frac\pi7}\\1&e^{-\frac{2\pi i}3}\frac{\sin\frac{3\pi}7}{\sin\frac\pi7}&e^{\frac{2\pi i}3}\frac{\sin\frac{3\pi}7}{\sin\frac\pi7}&e^{\frac{2\pi i}3}&\frac{\sin\frac{3\pi}7}{\sin\frac\pi7}&2e^{\frac{2\pi i}3}\cos\frac\pi7&2e^{-\frac{2\pi i}3}\cos\frac\pi7&e^{-\frac{2\pi i}3}&2\cos\frac\pi7\\2\cos\frac\pi7&1&1&2\cos\frac\pi7&1&-\frac{\sin\frac{3\pi}7}{\sin\frac\pi7}&-\frac{\sin\frac{3\pi}7}{\sin\frac\pi7}&2\cos\frac\pi7&-\frac{\sin\frac{3\pi}7}{\sin\frac\pi7}\end{pmatrix},\\
    \widetilde S_{(\wf{e_6})_k}\Big|_{\mbb Z_3}&=\begin{pmatrix}1&1&1\\1&e^{\frac{2\pi ik}3}&e^{-\frac{2\pi ik}3}\\1&e^{-\frac{2\pi ik}3}&e^{\frac{2\pi ik}3}\end{pmatrix}.
\end{split}$}\label{Se6}
\end{equation}
The last expression has an implication to three-dimensional physics. Recall the correspondence between the two-dimensional WZW theory and three-dimensional Chern-Simons (CS) theory. Since the $S$-matrix is the expectation values of linked Wilson loops in CS theories \cite{W89}, nontrivial phases mean an anomaly in the one-form symmetry \cite{GKSW14}. The last formula tells us the $(E_6)_k$ CS theory has an anomalous $\mbb Z_3$ one-form symmetry iff $k\notin3\mbb Z$. This is consistent with the presence of invariant boundary state for $k\in3\mbb Z$ \cite{KY19}.

\subsection{$(\wf{e_7})_k$}
\begin{equation}
\scalebox{0.83}{$\begin{split}
    \widetilde S_{(\wf{e_7})_1}&=\begin{pmatrix}1&1\\1&-1\end{pmatrix},\\
    \widetilde S_{(\wf{e_7})_3}&=\begin{pmatrix}1&\frac{3+\sqrt{21}}2&\frac{3+\sqrt{21}}2&\frac{7+\sqrt{21}}2&\frac{3+\sqrt{21}}2&\frac{5+\sqrt{21}}2&\frac{7+\sqrt{21}}2&\frac{3+\sqrt{21}}2&\frac{3+\sqrt{21}}2&\frac{3+\sqrt{21}}2&1&\frac{5+\sqrt{21}}2\\
    \frac{3+\sqrt{21}}2&a&b&0&-a&\frac{3+\sqrt{21}}2&0&c&-c&-b&-\frac{3+\sqrt{21}}2&-\frac{3+\sqrt{21}}2\\
    \frac{3+\sqrt{21}}2&b&c&0&b&-\frac{3+\sqrt{21}}2&0&-a&-a&c&\frac{3+\sqrt{21}}2&-\frac{3+\sqrt{21}}2\\
    \frac{7+\sqrt{21}}2&0&0&-\frac{7+\sqrt{21}}2&0&\frac{7+\sqrt{21}}2&-\frac{7+\sqrt{21}}2&0&0&0&\frac{7+\sqrt{21}}2&\frac{7+\sqrt{21}}2\\
    \frac{3+\sqrt{21}}2&-a&b&0&-a&-\frac{3+\sqrt{21}}2&0&c&c&b&\frac{3+\sqrt{21}}2&-\frac{3+\sqrt{21}}2\\
    \frac{5+\sqrt{21}}2&\frac{3+\sqrt{21}}2&-\frac{3+\sqrt{21}}2&\frac{7+\sqrt{21}}2&-\frac{3+\sqrt{21}}2&-1&-\frac{7+\sqrt{21}}2&-\frac{3+\sqrt{21}}2&\frac{3+\sqrt{21}}2&\frac{3+\sqrt{21}}2&-\frac{5+\sqrt{21}}2&1\\
    \frac{7+\sqrt{21}}2&0&0&-\frac{7+\sqrt{21}}2&0&-\frac{7+\sqrt{21}}2&\frac{7+\sqrt{21}}2&0&0&0&-\frac{7+\sqrt{21}}2&\frac{7+\sqrt{21}}2\\
    \frac{3+\sqrt{21}}2&c&-a&0&c&-\frac{3+\sqrt{21}}2&0&c&c&-a&\frac{3+\sqrt{21}}2&-\frac{3+\sqrt{21}}2\\
    \frac{3+\sqrt{21}}2&-c&-a&0&c&\frac{3+\sqrt{21}}2&0&c&-c&a&-\frac{3+\sqrt{21}}2&-\frac{3+\sqrt{21}}2\\
    \frac{3+\sqrt{21}}2&-b&c&0&b&\frac{3+\sqrt{21}}2&0&-a&a&-c&-\frac{3+\sqrt{21}}2&-\frac{3+\sqrt{21}}2\\
    1&-\frac{3+\sqrt{21}}2&\frac{3+\sqrt{21}}2&\frac{7+\sqrt{21}}2&\frac{3+\sqrt{21}}2&-\frac{5+\sqrt{21}}2&-\frac{7+\sqrt{21}}2&\frac{3+\sqrt{21}}2&-\frac{3+\sqrt{21}}2&-\frac{3+\sqrt{21}}2&-1&\frac{5+\sqrt{21}}2\\
    \frac{5+\sqrt{21}}2&-\frac{3+\sqrt{21}}2&-\frac{3+\sqrt{21}}2&\frac{7+\sqrt{21}}2&-\frac{3+\sqrt{21}}2&1&\frac{7+\sqrt{21}}2&-\frac{3+\sqrt{21}}2&-\frac{3+\sqrt{21}}2&-\frac{3+\sqrt{21}}2&\frac{5+\sqrt{21}}2&1\end{pmatrix},\\
    \widetilde S_{(\wf{e_7})_k}\Big|_{\mbb Z_2}&=\begin{pmatrix}1&1\\1&(-1)^k\end{pmatrix},
\end{split}$}\label{Se7}
\end{equation}
where
\begin{align*}
    a&=-\frac{3+\sqrt{21}}2+\cos\frac{5\pi}7+\frac{7+\sqrt{21}}2\cos\frac\pi{21}+\frac{3+\sqrt{21}}2\cos\frac\pi7,\\
    b&=\frac{3+\sqrt{21}}2\cos\frac{2\pi}7-\frac{5+\sqrt{21}}2\cos\frac{3\pi}7+\frac{7+\sqrt{21}}2\cos\frac{2\pi}{21},\\
    c&=\frac{5+\sqrt{21}}2\cos\frac{2\pi}7+\frac{7+\sqrt{21}}2\cos\frac{8\pi}{21}-\frac{3+\sqrt{21}}2\cos\frac\pi7.
\end{align*}
Similarly to the previous subsection, the last formula says the following; the $(E_7)_k$ CS has an anomalous $\mbb Z_2$ one-form symmetry iff $k\notin2\mbb Z$. This is again consistent with the invariant boundary condition $k\in2\mbb Z$ \cite{KY19}.


\begin{thebibliography}{30}
\bibitem{tH79}
  G.~'t Hooft,
  ``Naturalness, chiral symmetry, and spontaneous chiral symmetry breaking,''
  NATO Sci. Ser. B \textbf{59}, 135-157 (1980)
  doi:10.1007/978-1-4684-7571-5\_9
  %827 citations counted in INSPIRE as of 27 Aug 2022
\bibitem{GKKS17}
  D.~Gaiotto, A.~Kapustin, Z.~Komargodski and N.~Seiberg,
  ``Theta, Time Reversal, and Temperature,''
  JHEP \textbf{05}, 091 (2017)
  doi:10.1007/JHEP05(2017)091
  [arXiv:1703.00501 [hep-th]].
  %287 citations counted in INSPIRE as of 13 Oct 2022
\bibitem{GKSW14}
  D.~Gaiotto, A.~Kapustin, N.~Seiberg and B.~Willett,
  ``Generalized Global Symmetries,''
  JHEP \textbf{02}, 172 (2015)
  doi:10.1007/JHEP02(2015)172
  [arXiv:1412.5148 [hep-th]].
  %705 citations counted in INSPIRE as of 13 Oct 2022
\bibitem{MS1}
  G.~W.~Moore and N.~Seiberg,
  ``Classical and Quantum Conformal Field Theory,'' Commun. Math. Phys. \textbf{123}, 177 (1989) doi:10.1007/BF01238857
  %761 citations counted in INSPIRE as of 27 Aug 2021
\bibitem{MS2}
  G.~W.~Moore and N.~Seiberg,
  ``LECTURES ON RCFT,'' RU-89-32.
  %16 citations counted in INSPIRE as of 27 Aug 2021
\bibitem{BT17}
  L.~Bhardwaj and Y.~Tachikawa,
  ``On finite symmetries and their gauging in two dimensions,''
  JHEP \textbf{03}, 189 (2018)
  doi:10.1007/JHEP03(2018)189
  [arXiv:1704.02330 [hep-th]].
  %93 citations counted in INSPIRE as of 25 Aug 2022
\bibitem{ENO}
  P.~Etingof, D.~Nikshych, V.~Ostrik, ``On fusion categories,'' [arXiv:math/0203060 [math.QA]].
\bibitem{G12}
  D.~Gaiotto,
  ``Domain Walls for Two-Dimensional Renormalization Group Flows,''
  JHEP \textbf{12}, 103 (2012)
  doi:10.1007/JHEP12(2012)103
  [arXiv:1201.0767 [hep-th]].
  %70 citations counted in INSPIRE as of 25 Aug 2022
\bibitem{KK-ARG}
  K.~Kikuchi,
  ``Axiomatic rational RG flow,''
  [arXiv:2209.00016 [hep-th]].
  %0 citations counted in INSPIRE as of 14 Oct 2022
\bibitem{KK21}
  K.~Kikuchi,
  ``Symmetry enhancement in RCFT,''
  [arXiv:2109.02672 [hep-th]].
  %12 citations counted in INSPIRE as of 14 Oct 2022
\bibitem{KKSUSY}
  K.~Kikuchi,
  ``Emergent SUSY in two dimensions,''
  [arXiv:2204.03247 [hep-th]].
  %9 citations counted in INSPIRE as of 14 Oct 2022
\bibitem{KK22II}
  K.~Kikuchi,
  ``Symmetry enhancement in RCFT II,''
  [arXiv:2207.06433 [hep-th]].
  %4 citations counted in INSPIRE as of 14 Oct 2022
\bibitem{KK22free}
  K.~Kikuchi,
  ``Emergent symmetry and free energy,''
  [arXiv:2207.10095 [hep-th]].
  %1 citations counted in INSPIRE as of 14 Oct 2022
\bibitem{NK22}
  Y.~Nakayama and K.~Kikuchi,
  ``The fate of non-supersymmetric Gross-Neveu-Yukawa fixed point in two dimensions,''
  [arXiv:2212.06342 [hep-th]].
  %0 citations counted in INSPIRE as of 14 Dec 2022
\bibitem{Z86}
  A.~B.~Zamolodchikov,
  ``Irreversibility of the Flux of the Renormalization Group in a 2D Field Theory,''
  JETP Lett. \textbf{43}, 730-732 (1986)
  %1569 citations counted in INSPIRE as of 24 Dec 2022
\bibitem{KP05}
  A.~Kitaev and J.~Preskill,
  ``Topological entanglement entropy,''
  Phys. Rev. Lett. \textbf{96}, 110404 (2006)
  doi:10.1103/PhysRevLett.96.110404
  [arXiv:hep-th/0510092 [hep-th]].
  %1087 citations counted in INSPIRE as of 10 Jul 2022
\bibitem{LW05}
  M.~Levin and X.~G.~Wen,
  ``Detecting Topological Order in a Ground State Wave Function,''
  Phys. Rev. Lett. \textbf{96}, 110405 (2006)
  doi:10.1103/PhysRevLett.96.110405
  [arXiv:cond-mat/0510613 [cond-mat.str-el]].
  %948 citations counted in INSPIRE as of 20 Jul 2022
\bibitem{DDT00}
  P.~Dorey, C.~Dunning and R.~Tateo,
  ``New families of flows between two-dimensional conformal field theories,''
  Nucl. Phys. B \textbf{578}, 699-727 (2000)
  doi:10.1016/S0550-3213(00)00185-1
  [arXiv:hep-th/0001185 [hep-th]].
  %17 citations counted in INSPIRE as of 19 Jul 2022
\bibitem{RV98}
  P.~Ruelle and O.~Verhoeven,
  ``Discrete symmetries of unitary minimal conformal theories,'' Nucl. Phys. B \textbf{535}, 650-680 (1998) doi:10.1016/S0550-3213(98)00639-7
  [arXiv:hep-th/9803129 [hep-th]].
  %16 citations counted in INSPIRE as of 12 Jun 2022
\bibitem{BBLKS13}
  M.~Beria, G.~P.~Brandino, L.~Lepori, R.~M.~Konik and G.~Sierra,
  ``Truncated Conformal Space Approach for Perturbed Wess-Zumino-Witten $SU(2)_k$ Models,''
  Nucl. Phys. B \textbf{877}, 457-483 (2013)
  doi:10.1016/j.nuclphysb.2013.10.005
  [arXiv:1301.0084 [hep-th]].
  %20 citations counted in INSPIRE as of 28 Nov 2022
\bibitem{KPTT15}
  R.~M.~Konik, T.~P\'almai, G.~Tak\'acs and A.~M.~Tsvelik,
  ``Studying the perturbed Wess\textendash{}Zumino\textendash{}Novikov\textendash{}Witten SU(2)$_k$ theory using the truncated conformal spectrum approach,''
  Nucl. Phys. B \textbf{899}, 547-569 (2015)
  doi:10.1016/j.nuclphysb.2015.08.016
  [arXiv:1505.03860 [cond-mat.str-el]].
  %21 citations counted in INSPIRE as of 28 Nov 2022
\bibitem{YZ89}
  V.~P.~Yurov and A.~B.~Zamolodchikov,
  ``TRUNCATED CONFORMAL SPACE APPROACH TO SCALING LEE-YANG MODEL,''
  Int. J. Mod. Phys. A \textbf{5}, 3221-3246 (1990)
  doi:10.1142/S0217751X9000218X
  %209 citations counted in INSPIRE as of 04 Mar 2022
\bibitem{WHZ03}
  C.~Wu, J.~p.~Hu and S.~c.~Zhang,
  ``Exact SO(5) Symmetry in spin 3/2 fermionic system,''
  Phys. Rev. Lett. \textbf{91}, 186402 (2003)
  doi:10.1103/PhysRevLett.91.186402
  [arXiv:cond-mat/0302165 [cond-mat.str-el]].
  %112 citations counted in INSPIRE as of 29 Nov 2022
\bibitem{C17}
  J.~Cardy,
  ``Bulk Renormalization Group Flows and Boundary States in Conformal Field Theories,''
  SciPost Phys. \textbf{3}, no.2, 011 (2017) doi:10.21468/SciPostPhys.3.2.011 [arXiv:1706.01568 [hep-th]].
  %17 citations counted in INSPIRE as of 28 Jul 2021
\bibitem{L15}
  P.~Lecheminant,
  ``Massless renormalization group flow in SU(N)$_k$ perturbed conformal field theory,''
  Nucl. Phys. B \textbf{901}, 510-525 (2015)
  doi:10.1016/j.nuclphysb.2015.11.004
  [arXiv:1509.01680 [cond-mat.str-el]].
  %18 citations counted in INSPIRE as of 11 Dec 2022
\bibitem{NY17}
  T.~Numasawa and S.~Yamaguchi,
  ``Mixed global anomalies and boundary conformal field theories,''
  JHEP \textbf{11}, 202 (2018)
  doi:10.1007/JHEP11(2018)202
  [arXiv:1712.09361 [hep-th]].
  %15 citations counted in INSPIRE as of 11 Oct 2022
\bibitem{TS18}
  Y.~Tanizaki and T.~Sulejmanpasic,
  ``Anomaly and global inconsistency matching: $\theta$-angles, $SU(3)/U(1)^2$ nonlinear sigma model, $SU(3)$ chains and its generalizations,''
  Phys. Rev. B \textbf{98}, no.11, 115126 (2018)
  doi:10.1103/PhysRevB.98.115126
  [arXiv:1805.11423 [cond-mat.str-el]].
  %74 citations counted in INSPIRE as of 19 Nov 2022
\bibitem{YHO18}
  Y.~Yao, C.~T.~Hsieh and M.~Oshikawa,
  ``Anomaly matching and symmetry-protected critical phases in $SU(N)$ spin systems in 1+1 dimensions,''
  Phys. Rev. Lett. \textbf{123}, no.18, 180201 (2019)
  doi:10.1103/PhysRevLett.123.180201
  [arXiv:1805.06885 [cond-mat.str-el]].
  %52 citations counted in INSPIRE as of 11 Dec 2022
\bibitem{H82}
  F.~D.~M.~Haldane,
  ``Continuum dynamics of the 1-D Heisenberg antiferromagnetic identification with the O(3) nonlinear sigma model,''
  Phys. Lett. A \textbf{93}, 464-468 (1983)
  doi:10.1016/0375-9601(83)90631-X
  %664 citations counted in INSPIRE as of 23 Dec 2022
\bibitem{H83}
  F.~D.~M.~Haldane,
  ``Nonlinear field theory of large spin Heisenberg antiferromagnets. Semiclassically quantized solitons of the one-dimensional easy Axis Neel state,''
  Phys. Rev. Lett. \textbf{50}, 1153-1156 (1983)
  doi:10.1103/PhysRevLett.50.1153
  %775 citations counted in INSPIRE as of 23 Dec 2022
\bibitem{FO15}
  S.~C.~Furuya and M.~Oshikawa,
  ``Symmetry Protection of Critical Phases and a Global Anomaly in $1+1$ Dimensions,''
  Phys. Rev. Lett. \textbf{118}, no.2, 021601 (2017)
  doi:10.1103/PhysRevLett.118.021601
  [arXiv:1503.07292 [cond-mat.stat-mech]].
  %60 citations counted in INSPIRE as of 11 Dec 2022
\bibitem{LS21}
  Y.~H.~Lin and S.~H.~Shao,
  ``$\mathbb{Z}_N$ symmetries, anomalies, and the modular bootstrap,''
  Phys. Rev. D \textbf{103}, no.12, 125001 (2021)
  doi:10.1103/PhysRevD.103.125001
  [arXiv:2101.08343 [hep-th]].
  %25 citations counted in INSPIRE as of 11 Dec 2022
\bibitem{GK94}
  D.~Gepner and A.~Kapustin,
  ``On the classification of fusion rings,''
  Phys. Lett. B \textbf{349}, 71-75 (1995)
  doi:10.1016/0370-2693(95)00172-H
  [arXiv:hep-th/9410089 [hep-th]].
  %15 citations counted in INSPIRE as of 03 Jul 2022
\bibitem{KY19}
  K.~Kikuchi and Y.~Zhou,
  ``Two-dimensional Anomaly, Orbifolding, and Boundary States,''
  [arXiv:1908.02918 [hep-th]].
  %7 citations counted in INSPIRE as of 13 Dec 2022
\bibitem{CW20}
  M.~Cheng and D.~J.~Williamson,
  ``Relative anomaly in ( 1+1 )d rational conformal field theory,''
  Phys. Rev. Res. \textbf{2}, no.4, 043044 (2020) doi:10.1103/PhysRevResearch.2.043044 [arXiv:2002.02984 [cond-mat.str-el]].
  %5 citations counted in INSPIRE as of 17 Nov 2021
\bibitem{LHYO22}
  L.~Li, C.~T.~Hsieh, Y.~Yao and M.~Oshikawa,
 ``Boundary conditions and anomalies of conformal field theories in 1+1 dimensions,''
  [arXiv:2205.11190 [hep-th]].
  %1 citations counted in INSPIRE as of 12 Dec 2022
\bibitem{V89}
  C.~Vafa,
  ``Quantum Symmetries of String Vacua,''
  Mod. Phys. Lett. A \textbf{4}, 1615 (1989)
  doi:10.1142/S0217732389001842
  %98 citations counted in INSPIRE as of 24 Dec 2022
\bibitem{T18}
  Y.~Tachikawa, ``Topological Phases and Relativistic QFTs,'' https://member.ipmu.jp/yuji.tachikawa/lectures/2018-cern-rikkyo/.
\bibitem{FMS}
  P.~Di Francesco, P.~Mathieu and D.~Senechal,
  ``Conformal Field Theory,''
  doi:10.1007/978-1-4612-2256-9
  %%CITATION = doi:10.1007/978-1-4612-2256-9;%%
  %161 citations counted in INSPIRE as of 16 Feb 2019
\bibitem{LieART}
  R.~Feger and T.~W.~Kephart,
  ``LieART\textemdash{}A Mathematica application for Lie algebras and representation theory,''
  Comput. Phys. Commun. \textbf{192}, 166-195 (2015)
  doi:10.1016/j.cpc.2014.12.023
  [arXiv:1206.6379 [math-ph]].
  %156 citations counted in INSPIRE as of 16 Dec 2022
\bibitem{KP84}
  V.~G.~Kac and D.~H.~Peterson,
  ``Infinite dimensional Lie algebras, theta functions and modular forms,''
  Adv. Math. \textbf{53}, 125-264 (1984)
  doi:10.1016/0001-8708(84)90032-X
  %459 citations counted in INSPIRE as of 17 Dec 2022
\bibitem{KAC}
  ``Komputations with Algebras and Currents,'' https://www.nikhef.nl/~t58/Site/Kac.html
\bibitem{W89}
  E.~Witten,
  ``Quantum Field Theory and the Jones Polynomial,''
  Commun. Math. Phys. \textbf{121}, 351-399 (1989)
  doi:10.1007/BF01217730
  %3402 citations counted in INSPIRE as of 16 Dec 2022
\end{thebibliography}
\end{document}